\documentclass[article]{JHEP3}

\usepackage{amsmath,amssymb}
\usepackage[dvips]{graphics}
\usepackage{epsfig}
\usepackage{psfrag,comment}
\usepackage{graphicx}

\newcommand{\CB}{{\cal B}}
\newcommand{\CC}{{\cal C}}
\newcommand{\CE}{{\cal E}}
\newcommand{\CF}{{\cal F}}
\newcommand{\CG}{{\cal G}}
\newcommand{\CH}{{\cal H}}

\newcommand{\CL}{{\cal L}}

\newcommand{\CN}{{\cal N}}
\newcommand{\CO}{{\cal O}}

\newcommand{\CR}{{\cal R}}

\newcommand{\CW}{{\cal W}}
\newcommand{\CX}{{\cal X}}

\newcommand{\CZ}{{\cal Z}}

\def\BN{{\mathbb N}}
\def\BZ{{\mathbb Z}}
\def\BR{{\mathbb R}}
\def\BC{{\mathbb C}}
\def\BP{{\mathbb P}}

\def\BS{{\mathbb S}}

\newcommand{\be}{\begin{equation}}
\newcommand{\ee}{\end{equation}}
\newcommand{\ba}{\begin{aligned}}
\newcommand{\ea}{\end{aligned}}
\newcommand{\bea}{\begin{eqnarray}}
\newcommand{\eea}{\end{eqnarray}}
\newcommand{\bean}{\begin{eqnarray*}}
\newcommand{\eean}{\end{eqnarray*}}

\newcommand{\p}{\partial}

\def\r{\right\rangle}
\def\a{\alpha'}
\def\1{\mathbf{1}}
\def\0{|\1\r}
\def\im{{\mathbb{I}}{\mathrm{m}}}
\def\re{{\mathbb{R}}{\mathrm{e}}}

\newcommand{\tr}{{\rm Tr}}
\newcommand{\rme}{{\rm e}}
\newcommand{\rmi}{{\rm i}}
\newcommand{\rmd}{{\rm d}}
\newcommand{\rd}{{\rm d}}

\def\XXint#1#2#3{{\setbox0=\hbox{$#1{#2#3}{\int}$}
     \vcenter{\hbox{$#2#3$}}\kern-.5\wd0}}

\newdimen\tableauside\tableauside=1.0ex
\newdimen\tableaurule\tableaurule=0.4pt
\newdimen\tableaustep
\def\phantomhrule#1{\hbox{\vbox to0pt{\hrule height\tableaurule width#1\vss}}}
\def\phantomvrule#1{\vbox{\hbox to0pt{\vrule width\tableaurule height#1\hss}}}
\def\sqr{\vbox{%
  \phantomhrule\tableaustep
  \hbox{\phantomvrule\tableaustep\kern\tableaustep\phantomvrule\tableaustep}%
  \hbox{\vbox{\phantomhrule\tableauside}\kern-\tableaurule}}}
\def\squares#1{\hbox{\count0=#1\noindent\loop\sqr
  \advance\count0 by-1 \ifnum\count0>0\repeat}}
\def\tableau#1{\vcenter{\offinterlineskip
  \tableaustep=\tableauside\advance\tableaustep by-\tableaurule
  \kern\normallineskip\hbox
    {\kern\normallineskip\vbox
      {\gettableau#1 0 }%
     \kern\normallineskip\kern\tableaurule}%
  \kern\normallineskip\kern\tableaurule}}
\def\gettableau#1{\ifnum#1=0\let\next=\null\else
\squares{#1}\let\next=\gettableau\fi\next}
\preprint{
{\small{\textsf{CERN-PH-TH/2009-130}}}}

\title{Borel and Stokes Nonperturbative Phenomena in Topological String Theory and $c=1$ Matrix Models}

\author{Sara Pasquetti$^1$ and Ricardo Schiappa$^2$
\\
$^1$Theory Division, Department of Physics, CERN,\\
CH--1211 Gen\`eve 23, Switzerland\\
\\
$^2$CAMGSD, Departamento de Matem\'atica, Instituto Superior T\'ecnico,\\ 
Av. Rovisco Pais 1, 1049--001 Lisboa, Portugal\\
\\
\email{sara.pasquetti@cern.ch}, \quad
\email{schiappa@math.ist.utl.pt}

}

\abstract{
We address the nonperturbative structure of topological strings and $c=1$ matrix models, focusing on understanding the nature of instanton effects alongside with exploring their relation to the large--order behavior of the $1/N$ expansion. We consider the Gaussian, Penner and Chern--Simons matrix models, together with their holographic duals, the $c=1$ minimal string at self--dual radius and topological string theory on the resolved conifold. We employ Borel analysis to obtain the exact all--loop multi--instanton corrections to the free energies of the aforementioned models, and show that the leading poles in the Borel plane control the large--order behavior of perturbation theory. We understand the nonperturbative effects in terms of the Schwinger effect and provide a semiclassical picture in terms of eigenvalue tunneling between critical points of the multi--sheeted matrix model effective potentials. In particular, we relate instantons to Stokes phenomena via a hyperasymptotic analysis, providing a smoothing of the nonperturbative ambiguity. Our predictions for the multi--instanton expansions are confirmed within the trans--series set--up, which in the double--scaling limit describes nonperturbative corrections to the Toda equation. Finally, we provide a spacetime realization of our nonperturbative corrections in terms of toric D--brane instantons which, in the double--scaling limit, precisely match D--instanton contributions to $c=1$ minimal strings.
}


\begin{document}



\vfill

\eject

\section{Introduction and Summary}

The nonperturbative realm of quantum field and string theories has often been a source of many new results and surprises. Another recurrent topic of great interest over the past decades has been the large $N$ approximation, lately relating gauge and string theories in a nonperturbative fashion. Of particular interest to us in this work is the case of the perturbative $1/N$ expansion of hermitian matrix models, whose nonperturbative corrections are exponentially suppressed as $\exp \left( - N \right)$. In the double--scaling limit these models describe noncritical or minimal (super)string theories, and the nonperturbative structure of the matrix model is related to that of the corresponding string theory \cite{fgz93}: the $\exp \left( - N \right)$ contributions are instanton effects in the matrix model \cite{d91, d92} and they are interpreted as D--brane configurations in the string theoretic description \cite{s90, p94, akk03}. Of course the study of matrix models is not confined to the vicinity of their critical points and one may also study nonperturbative effects away from the double--scaling limit. The interesting point is that off--critical matrix models may be dual to topological string theories. For instance, this happens in the case first suggested by \cite{dv02}, where some off--critical matrix models describe the topological string B--model on certain non--compact Calabi--Yau (CY) backgrounds, with the string genus expansion (in powers of the string coupling, $g_s$) being identified with the $1/N$ matrix model expansion; and it is also the case for topological strings with a Chern--Simons dual, first studied in \cite{m02, akmv02}. This turns out to be a more general statement, as it was later shown in \cite{m06, bkmp07, bkmp08} that topological string theory on mirrors of toric manifolds also enjoys a dual holographic description in terms of off--critical matrix models. It is thus evident that fully understanding the nonperturbative structure of matrix models, both at and off criticality, will have many applications in both minimal and topological string theories.

Recently \cite{msw07, m08, msw08} there has been significative progress in understanding and in quantitatively computing nonperturbative effects in matrix models away from criticality. In \cite{msw07}, and building upon double--scaled results \cite{d91, d92, hhikkmt04, iy05}, off--critical saddle--point techniques were developed in order to compute instanton amplitudes (up to two loops) in terms of spectral curve geometrical data. This work focused upon one--instanton contributions in one--cut models, and in \cite{msw08} an extension to multi--instanton contributions, again in one--cut models, was obtained, starting from a two--cut analysis. Extensive checks of the nonperturbative proposals in these papers were also performed, by matching against the large--order behavior of the $1/N$ expansion. Another approach to multi--instanton amplitudes was developed in \cite{m08}, this time around based on orthogonal polynomial methods, via the use of trans--series solutions to the string equations \cite{biz80}. Further progress along these lines recently led to the proposal of \cite{em08}, where a proper nonperturbative definition of a modular--invariant holomorphic partition function was presented, which was also shown to be manifestly background independent. Remarkably, many of the results uncovered in \cite{msw07, m08, msw08} appear to extend beyond the context of matrix models; \textit{e.g.}, in cases where the theory is controlled by a finite--difference equation---such as the string equation \cite{biz80} for matrix models---it is possible to compute nonperturbative effects and relate them to the large--order behavior of the theory. This is the case of Hurwitz theory \cite{msw07}, which is controlled by a Toda--like equation, and also the case of topological strings on the background considered in \cite{cgmps06}\footnote{A local CY threefold given by a bundle over a two--sphere, $X_p = \CO (p-2) \oplus \CO (-p) \to \BP^1$, $p \in \BZ$, which may be regarded as a quantum group deformation of Hurwitz theory; see \cite{cgmps06} for further details.}. However, all models considered in the aforementioned articles lie in the universality class of two--dimensional gravity, with $c=0$, and methods that have been worked out in this case cannot be applied in a straightforward fashion to the case of topological strings in the universality class of $c=1$. In view of this, it is necessary to develop new techniques in order to approach nonperturbative effects in models which belong to the universality class  of the $c=1$ string at the self--dual radius.

Let us be a bit more specific about the nature of the string perturbative expansion and the type of nonperturbative contributions we shall be looking for. Topological strings, much like physical string theory, are perturbatively defined in terms of two couplings, $\a$ and $g_s$, as\footnote{Recall that in the A--model the $\{t_i\}$ are K\"ahler parameters while in the B--model they are complex parameters.}
\be\label{freegt}
F (g_s; \{t_i\}) = \sum_{g=0}^{+\infty} g_s^{2g-2} F_g (t_i),
\ee
\noindent
where $F = \log Z$ is the free energy and $Z$ the partition function, and where the fixed genus free energies $F_g (t_i)$ are themselves perturbatively expanded in $\a$. In some sense the $\a$ expansion is the milder one: it has finite convergence radius, with this radius given by the critical value of the K\"ahler parameters where one reaches a conifold point in moduli space. As it turns out, the problem of finding a nonperturbative formulation of the A--model free energy, in $\a$, may be reduced to that of solving the mirror B--model description, where topological string amplitudes become exact in $\a$. In this way, the A--model solution is found by translating B--model amplitudes back to the A--model, by means of the mirror map. This topic has been extensively studied in the literature and we refer the reader to the recent developments \cite{bkmp07, bkmp08} and references therein.

The situation gets more complicated as one tries to go beyond perturbation theory in $g_s$. In this case, one is immediately faced with the familiar string theoretic large--order behavior $F_g \sim (2g)!$ rendering (\ref{freegt}) as an asymptotic expansion \cite{s90}. In this case, one expects nonperturbative corrections of order $\sim \exp \left( - 1/g_s \right)$, and an adequate nonperturbative formulation of the theory must encode all these corrections. As described above, there are certain cases---such as the backgrounds considered by Dijkgraaf and Vafa \cite{dv02}, or models with a dual Chern--Simons interpretation---where topological strings have a holographic matrix model description, with the matrix model large $N$ expansion reproducing the topological string genus expansion. In these set of backgrounds one would be tempted to use the finite $N$ matrix model free energy as the $g_s$ nonperturbative definition of topological string theory\footnote{Other nonperturbative completions, provided by a holographic dual, have been proposed in \cite{osv04}.}. In order to establish this result, one must first understand how the finite $N$ matrix model would encompass all nonperturbative contributions $\sim \exp \left( - 1/g_s \right)$. This situation is clear for minimal strings, realized in the double--scaling limit of hermitian matrix models: the nonperturbative effects associated to the asymptotic nature of the genus expansion are implemented via eigenvalue tunneling effects in the dual matrix model, and are interpreted in the continuum formulation in terms of Liouville branes in spacetime \cite{fgz93, akk03}. For topological strings, a similar understanding has been achieved in the case of the local curve \cite{m06, msw07}, where a matrix model description is available \cite{cgmps06}. In this case, the nonperturbative effects associated to the asymptotic behavior, or large--order behavior, have again been matched to instantons arising from matrix eigenvalue tunneling, and a spacetime interpretation in terms of domain walls has been provided \cite{msw07}.

However, there are several cases where this paradigm seems not to apply, at least not in a straightforward fashion. It is our goal to address such issues in the present work in the prototypical example of the resolved conifold, but also encompassing matrix models in the $c=1$ universality class. Topological strings on the resolved conifold are holographically described by the Chern--Simons matrix model, but there are now no obvious instantons associated to eigenvalue tunneling as the Chern--Simons potential has no local maxima outside of the cut, where the eigenvalue instantons could tunnel to. This problem, which was not an issue in any of the previously mentioned examples, also appears in other matrix models, such as the Gaussian and Penner models; all of them in the $c=1$ universality class. One may then ask where do nonperturbative corrections arise from, or what exactly controls the large--order behavior of the $1/N$ perturbative expansion in these models. We shall answer these questions in this paper. One way out is to directly compute the (would--be) instanton action that controls the large--order behavior of the perturbative expansion, by means of a standard Borel analysis (see, \textit{e.g.}, \cite{z81}). At first this may look like a formidable task, as one may expect the topological string genus expansion to be rather complicated, not amenable to a Borel transform. Happily, the free energies of all cases we consider enjoy a Gopakumar--Vafa (GV) integral representation \cite{gv98, gv98a} which allows for an exact location of the singularities in the Borel complex plane controlling the divergence of the asymptotic perturbative series, \textit{i.e.}, the instanton action \cite{z81}. This is the topological string generalization of a celebrated $c=1$ string result \cite{gm90}. Interestingly enough, this integral representation may also be regarded as an one--loop Schwinger integral \cite{gv98a}, thus providing a spacetime interpretation of these nonperturbative effects; as already pointed out in \cite{gv98a} they control the pair--production rate of BPS bound states. As we shall later see, these results---which we further identify as Stokes phenomena of the finite $N$ partition function---will also allow us to explain the nonperturbative contributions as one--eigenvalue effects in the matrix model picture. We find, from a saddle--point analysis, that the $c=1$ nonperturbative effects arise due to the multi--valued structure of the effective potential (as preliminarily suggested in \cite{m05}); a different picture from that of matrix models in the universality class of 2d gravity plus matter, where the one--eigenvalue tunneling occurs from a metastable minimum to the most stable one.

This paper is organized as follows. We begin in section 2 by reviewing the main ideas behind our subsequent work. This includes the definition of the Borel transform and the relation between instantons and the large--order behavior of perturbation theory, both related to the existence of a nonperturbative ambiguity in the calculation of the free energy. In this section we also discuss the Schwinger effect, where one actually has a physical prescription to define the inverse Borel transform, which will turn out to be the case for topological strings and $c=1$ matrix models via the GV integral representation of the topological string free energy. In section 3 we then move on to presenting the matrix models we shall be focusing upon. We review some of their properties, such as their spectral curves and their perturbative genus expansions, and also obtain expressions for their exact, finite $N$ partition functions and holomorphic effective potentials, both of which play important roles in sections to come. In this section we also discuss the double--scaling limit of these models and show how they relate to FZZT branes. Section 4 presents one of the main topics in this paper, the Borel analysis of the Gaussian, Penner and Chern--Simons matrix models. We show how to obtain Schwinger--like integral representations of the free energy, via Borel resummation, and how the correct identification of the leading poles in the complex Borel plane leads to the one--instanton action in all our examples. We further show in this section that while the all--loop multi--instanton amplitudes precisely reconstruct the perturbative series, the one--instanton results control the large--order behavior of perturbation theory. We then move on to another of our main topics in section 5, namely the issue of Stokes phenomena. We recall how to obtain Stokes phenomena for integrals with saddles via hyperasymptotic analysis, and perform a detailed calculation for the Gamma function. This extends to the Barnes function and, in this way, allows us to identify instantons with Stokes phenomena as we reproduce the results we have previously found in section 4, out of hyperasymptotic analysis. In section 6 we provide a semiclassical interpretation of our instantons via eigenvalue tunneling, where this tunneling is now associated to the existence of a branched multi--sheeted structure in the relevant holomorphic effective potentials. Indeed, simple monodromy calculations reproduce our results for the multi--instanton action straight out of this interpretation. We further show in this section how to interpret our instantons in spacetime, from the point of view of ZZ branes. In section 7, we discuss the trans--series approach to $c=1$ matrix models and how it further validates our results. Finally, we conclude in section 8 with an outlook and future prospects. We also include two appendices, one dedicated to the study of the monodromy structure of the polylogarithm, and the other dedicated to the Cauchy dispersion relation, in the case of more general topological string theories than the ones we address in this paper.

\section{Asymptotic Series, Large Order and Topological Strings}\label{sec:ds}

We start by reviewing some useful facts concerning asymptotic series, the relation of their large--order behavior to nonperturbative effects, as described by instantons or by the Schwinger effect, and put them in the context of topological string theory as we wish to study in the present work. For an introduction to these topics with applications in quantum mechanics and quantum field theory, we refer the reader to \cite{z81} and references therein.

Let us consider the perturbative expansion of some function, $F(z)$, with $z$ the specific perturbative expansion parameter,
\be\label{d}
F(z) \sim \sum_{n=0}^{+\infty} F_n\, z^{n}. 
\ee
\noindent
In many interesting examples one may infer that, at large $n$, the coefficients behave as $F_n \sim \left( \beta n \right)!$, thus rendering the series divergent. As an approximation to the function $F(z)$, the asymptotic series (\ref{d}) must necessarily be truncated. As such, one is faced with an obvious problem: how to deal with the fact that the perturbative expansion has zero convergence radius? In particular, if we do not know the function $F(z)$, but only its asymptotic series expansion, how do we associate a value to the divergent sum? The best framework to address issues related to asymptotic series is Borel analysis. One starts by introducing the Borel transform of the asymptotic series (\ref{d}) as
\be\label{defborelt}
\CB[F](\xi) = \sum_{n=0}^{+\infty} \frac{F_n}{(\beta n)!}\, \xi^n,
\ee
\noindent
which removes the divergent part of the coefficients $F_n$ and renders $\CB[F](\xi)$ with  finite convergence radius. In particular, if $F(z)$ originally had a finite radius of convergence (\textit{i.e.}, if it was \textit{not} an asymptotic series), $\CB[F](\xi)$ would be an entire function in the Borel complex $\xi$--plane. In general, however, $\CB[F](\xi)$ will have singularities and it is crucial to locate them in the complex plane. The reason for this is simple to understand: if $\CB[F](\xi)$ has no singularities for real positive $\xi$ one may analytically continue this function on $\BR^+$ and thus define the \textit{inverse} Borel transform by means of a Laplace transform as\footnote{For simplicity, we are assuming $z \in \BR^+$ in this expression.}
\be\label{borelint}
\widetilde{F} (z) = \int_0^{+\infty} \rmd s\, \CB[F] \big( z s^\beta \big)\, \rme^{-s}.
\ee
\noindent
The function $\widetilde{F} (z)$ has, by construction, the same asymptotic expansion as $F(z)$ and may thus provide a solution to our original question; it associates a value to the divergent sum (\ref{d}).
If, however, the function $\CB[F](\xi)$ has poles or branch cuts on the real axis, things get a bit more subtle: in order to perform the integral (\ref{borelint}) one needs to choose a contour which avoids such singularities. This choice of contour naturally introduces an ambiguity (as we shall see next, a \textit{nonperturbative} ambiguity) in the reconstruction of the original function, which renders $F(z)$ \textit{non}--Borel summable\footnote{Strictly speaking, the function is said not to be Borel summable if different integration contours yield different results. It may still be the case that, in spite of having singularities in the real axis, \textit{all} alternative integration contours yield the \textit{same} result.}. As it turns out, different integration paths produce functions  with the same asymptotic behavior, but differing by exponentially suppressed terms. For instance, in the presence of a singularity at a distance $A$ from the origin, on the real axis, one may define the integral (\ref{borelint}) on contours $\CC_{\pm}$, either avoiding the singularity from above, and leading to $\widetilde{F}_+(z)$, or from below, and leading to $\widetilde{F}_-(z)$. One finds that these two functions differ by a nonperturbative term \cite{z81}
\be\label{npa}
\widetilde{F}_+(z) - \widetilde{F}_-(z) \sim \rmi\, \rme^{-\frac{A}{z^{1/\beta}}}.
\ee
\noindent
In certain cases, \textit{e.g.}, when one has a Schwinger representation for the function \cite{s51, co77}, there is a natural and rigorous way to define the integral (\ref{borelint}) on a contour which avoids the singularities, and which also allows for a physical interpretation of the nonperturbative contributions.

So far our discussion has been rather general. However, it takes no effort to figure out the physical relevance of our discussion: divergent series are almost ubiquitous in physics and appear basically each time we approach an interesting problem in perturbation theory \cite{z81}. A typical and extensively studied case in quantum mechanics is the anharmonic oscillator (see, \textit{e.g.}, \cite{bw73, cs78, z81}). Herein, the ground state energy may be computed in perturbation theory---as a power series in the quartic coupling---and one finds that it is analytic in all the (coupling constant) complex plane except for a branch cut on the negative real axis, associated to the instability of the potential which becomes unbounded for negative values of the coupling. This instability is reflected by the fact that the series is, as expected, asymptotic. In particular, one can perform a Borel analysis as above and discover that the Borel transform of the ground state energy has singularities on the positive real axis, leading to an ambiguity of order $\sim \rmi\, \rme^{-1/g}$, with $g$ the quartic coupling constant. In this simple quantum mechanical example the nonperturbative ambiguity has a clear physical interpretation: it signals the presence---at negative $g$---of \textit{instantons} mediating the decay from the unstable to the true vacuum, via tunneling under the local maximum of the potential.

What these ideas illustrate is that by means of a purely perturbative analysis, \textit{i.e.}, finding the singularities of the Borel transform of the original perturbative series, it is possible to learn about nonperturbative effects---at least the intensity of the nonperturbative ambiguity (but we shall say more on this in the following). In some examples it is possible to independently compute these nonperturbative terms directly, \textit{e.g.}, using WKB methods or computing the path integral around non--trivial (subdominant) saddle points \cite{z81}. In these examples one may then proceed in the opposite way from above and obtain information on the large--order behavior of the perturbative expansion out of the nonperturbative data. This is what we shall illustrate next.

\subsection{From Instantons to Large--Order and Back}\label{sec:lo}

In physical applications, the factor $A$ appearing in (\ref{npa}) is the one--instanton action (see, \textit{e.g.}, \cite{z81}). Let us make this relation between instantons and the large--order behavior of perturbation theory a bit more precise, as it will play a crucial role in our later analysis. Consider a quantum system whose free energy is expressed as a perturbative expansion in $g$, the coupling constant\footnote{Here and below the index $(\ell)$ labels the $\ell$--instanton sector, so that $(0)$ labels the perturbative expansion.},
\be\label{zero}
F^{(0)}(g) = \sum_{k=0}^{+\infty} f^{(0)}_k g^k.
\ee
\noindent
The series (\ref{zero}) will generically be asymptotic, with zero radius of convergence. This is naturally associated to a branch cut of $F(g)$ in the complex $g$--plane, located in the negative real axis and associated to instanton effects (just like in the anharmonic oscillator example above). The function $F(g)$ is expected to be analytic otherwise. In fact this is saying that our quantum system should actually be thought of as an asymptotic formal power series in two expansion parameters, $g$ and $\exp \left(-\frac{1}{g}\right)$, see \cite{m08} for a discussion in the matrix model context. The appropriate expansion of the free energy is thus \cite{m08}
\be\label{inst}
F(g) = \sum_{\ell=0}^{+\infty} C^{\ell} F^{(\ell)}(g), \qquad F^{(\ell)} (g) = \frac{\rmi}{g^b}\, \rme^{- \frac{\ell A}{g}} \sum_{k=0}^{+\infty} f^{(\ell)}_{k+1} g^k.
\ee
\noindent
Here, $C$ is a parameter corresponding to the nonperturbative ambiguity. Also, $A$ is the one--instanton action, $b$ a characteristic exponent and $f^{(\ell)}_k$ is the $k$--loop contribution around the $\ell$--instanton configuration. Typically, the coefficients $f^{(\ell)}_k$ are factorially divergent for any $\ell$ \cite{z81}, in which case we may think about the $(\ell+1)$--instanton sector as the nonperturbative contribution related to the asymptotic nature of the loop expansion around the $\ell$--instanton sector.

A standard procedure then relates the coefficients of the perturbative expansion around the zero--instanton sector, $f^{(0)}_{k}$, with the one--instanton free energy as follows. The discontinuity of the free energy across the branch cut (associated to the instability of the theory for negative $g$) is expressed, at first order, in terms of the leading instanton expansion (\ref{inst})
\be
{\rm Disc}~F(g) \equiv \lim_{\epsilon \to 0^+} F(g+\rmi \epsilon) - F(g-\rmi \epsilon) = 2 \rmi~\im~F(g) = F^{(1)}(g) + \cdots.
\ee
\noindent
At the same time, we may use the the Cauchy formula to write
\be\label{disc}
F(g) = \frac{1}{2\pi\rmi} \int_{-\infty}^0 \rmd w\, \frac{{\rm Disc}~F(w)}{w-g} - \oint_{(\infty)} \frac{\rmd w}{2\pi\rmi}\, \frac{F(w)}{w-g}.
\ee
\noindent
In certain situations, \textit{e.g.}, in the aforementioned anharmonic oscillator example \cite{bw73}, it is possible to show by scaling arguments that the last integral in the expression above does not contribute. In such cases, (\ref{disc}) provides a remarkable connection between perturbative and nonperturbative expansions. Using the perturbative expansion (\ref{zero}) and the leading one--instanton contribution to the discontinuity ${\rm Disc}~F(g) \sim F^{(1)}$, one may obtain from the Cauchy formula (\ref{disc}) the following large order (or large $k$) relation
\be\label{for}
f^{(0)}_k = \int_0^{+\infty} \frac{\rmd z}{2\pi\rmi}\, \frac{F^{(1)}(z)}{z^{k+1}} \sim \frac{\Gamma \left( k+b \right)}{2\pi A^{k+b}}\, \sum_{n=0}^{+\infty} \frac{\Gamma \left( k+b-n \right)}{\Gamma \left( k+b \right)}\, f^{(1)}_{n+1} A^n.
\ee
\noindent
This explicitly shows that the computation of the one--loop one--instanton partition function determines the leading order of the asymptotic expansion for the perturbative coefficients of the zero--instanton partition function. Higher loop corrections then yield the successive $\frac{1}{k}$ corrections. Furthermore, instanton corrections with action $A'>A$, where we have in mind multi--instanton corrections with action $\ell A$, $\ell \ge 2$, will yield corrections to the asymptotics of the $f^{(0)}_k$ coefficients which are exponentially suppressed in $k$.

For the cases we shall consider in this work, namely matrix models and string theory, one finds genus expansions as in (\ref{freegt}), with $F_g^{(0)}\sim (2g)!$, so that the relation (\ref{for}) gets slightly re--written as follows (see, \textit{e.g.}, \cite{msw07}). Begin with the free energy in the zero--instanton sector, $g_s^2 F^{(0)} (g_s)$. Setting $z=g_s^2$, the one--instanton path integral then yields a series of the form
\be\label{oneinstantonlargeorder}
z F^{(1)} (z) = \frac{\rmi}{z^{\frac{b}{2}}}\, \rme^{- \frac{A}{\sqrt{z}}} \sum_{g=0}^{+\infty} z^{\frac{g}{2}} F^{(1)}_{g+1} (t).
\ee
\noindent
Following a procedure analogous to the one above, where one further assumes that the standard dispersion relation (\ref{disc}) still holds, it follows for the zero--instanton sector perturbative coefficients
\be\label{lomm}
F^{(0)}_g (t) = \int_0^{+\infty} \frac{\rmd z}{2\pi\rmi}\, \frac{z F^{(1)} (z)}{z^{g+1}} \sim \frac{\Gamma \left( 2g+b \right)}{\pi A^{2g+b}}\, \sum_{h=0}^{+\infty} \frac{\Gamma \left( 2g+b-h \right)}{\Gamma \left( 2g+b \right)}\, F^{(1)}_{h+1} (t)\, A^{h}.
\ee
\noindent
Again, the computation of the one--loop one--instanton free energy determines the leading order of the asymptotic expansion for the perturbative coefficients of the zero--instanton free energy. Higher loop corrections then yield the successive $\frac{1}{2g}$ corrections. One should further notice that recently, in \cite{msw07, m08, msw08}, the relation (\ref{lomm}) has been tested in several models and rather conclusive numerical checks have confirmed its validity.

\subsection{The Schwinger Effect and a Semiclassical Interpretation}\label{sec:se}

As discussed above, a nonperturbative ambiguity---typically associated to instantons, from a physical point of view---can arise when defining the integration contour for the inverse Borel transform. However, this is not always the case: we shall now review an example where a prescription to define the inverse Borel transform naturally arises, together with a physical interpretation for the nonperturbative contributions  \cite{co77}. This is the Schwinger effect \cite{s51}.

The one--loop effective Lagrangian describing a charged scalar particle, of charge $e$ and mass $m$,  in a constant electric field, $E>0$, has an integral representation given by \cite{s51} (see, \textit{e.g.}, \cite{d04} for a recent review)
\be\label{1lo}
\CL = \frac{e^2 E^2}{16 \pi^2} \int_0^{+\infty} \frac{\rmd s}{s^2} \left( \frac{1}{\sin s}-\frac{1}{s}-\frac{s}{6} \right) \rme^{-s \frac{m^2}{e E}},
\ee
\noindent
which admits the weak coupling expansion
\be\label{exp}
\CL \sim \frac{m^4}{16 \pi^2} \sum_{n=0}^{+\infty} (-1)^n \frac{\overline{B}_{2n+4}}{(2n+4)(2n+3)(2n+2)} \left(\frac{2 e E}{m^2}\right)^{2n+4}.
\ee
\noindent
In here we used the shorthand $\overline{B}_{2n} = \frac{1-2^{2n-1}}{2^{2n-1}} B_{2n}$, with $B_{2n}$ the Bernoulli numbers. Since one may further relate the Bernoulli numbers to the Riemann zeta function via
\be
B_{2n} = (-1)^{(n+1)} \frac{2\, \zeta (2n)}{(2\pi)^{2n}}\, (2n)!
\ee
\noindent
it becomes evident that $B_{2n}$ diverges factorially fast. In this case, if one first writes the expansion (\ref{exp}) as
\be
\CL \sim \sum_{n=0}^{+\infty} a_{2n+4}\, x^{2n+4},
\ee
\noindent
with $x=\frac{2 e E}{m^2}$, it follows that at large $n$ one has $a_{2n+4} \sim (2n+1)!$ rendering this perturbative expansion asymptotic---and actually non--Borel summable as we shall see next. Indeed, computing the Borel transform it follows
\be
\CB[\CL](\xi) = \sum_{n=0}^{+\infty} \frac{a_{2n+4}}{(2n+1)!}\, \xi^{2n+4} = \frac{m^4}{16 \pi^2} \left( \frac{\xi/2}{\sin \left( \xi/2 \right)} - 1 - \frac{\left( \xi/2 \right)^2}{6} \right),
\ee
\noindent
from where one immediately notices that the Schwinger integral representation of the effective Lagrangian (\ref{1lo}) is essentially the inverse Borel transform
\be
\widetilde{\CL} (x) = \int_0^{+\infty} \frac{\rmd t}{t^3}\, \CB[\CL] \left( x t \right) \rme^{-t} = \frac{e^2 E^2}{16 \pi^2} \int_0^{+\infty} \frac{\rmd s}{s^2} \left( \frac{1}{\sin s} - \frac{1}{s} - \frac{s}{6} \right) \rme^{-s \frac{m^2}{e E}}.
\ee
\noindent
Of course so far we still have a nonperturbative ambiguity to deal with: in order to perform the integration on the real axis one still needs to specify a prescription in order to avoid the poles at $s = n \pi$, $n \in \BN$. This introduces the usual ambiguities leading to exponentially suppressed contributions to the effective Lagrangian. The novelty in this case is that there is now a natural way to address the integration avoiding the singularities in an unambiguous way \cite{co77}. 

As it turns out, the contour of integration needs to be deformed in such a way that the integral picks up the contributions of all the poles as if the real axis is approached from above, tantamount to a $+\rmi\epsilon$ prescription; and this is the requirement which is dictated by unitarity \cite{co77}. As such, one has a physical principle behind the unambiguous choice of contour. Furthermore, the Lagrangian develops an imaginary part which is simple to compute by summing residues\footnote{Notice that the residue at $s=0$ precisely vanishes.}, and which cannot be seen to any finite order in perturbation theory,
\be\label{rate}
\im\, \mathcal{L} = \frac{1}{8 \pi} \left(\frac{e E}{\pi}\right)^2\, \sum_{n=1}^{+\infty} \frac{(-1)^{n-1}}{n^2}\, \exp\left(-n \frac{\pi m^2}{eE}\right),
\ee
\noindent
an expression with an evident multi--instanton flavor \cite{kp03}, as in (\ref{inst}). Besides the appropriate, physical prescription to perform the integration, and as such unambiguously compute the nonperturbative contributions to the Lagrangian, the Schwinger effect gives us something else: a physical interpretation of this imaginary part. Indeed, the imaginary part of the effective Lagrangian (\ref{rate}) is precisely the pair--production rate, or probability per unit volume for pair creation, for scalar electrodynamics in a constant electric field \cite{s51}. In other words, the above unitary $+\rmi\epsilon$ prescription for the integration contour guarantees that this probability is a positive number between zero and one (which basically demands (\ref{rate}) to be real and positive).

Another interesting illustration of the Schwinger effect, which will be of particular relevance in our subsequent discussion on topological strings and matrix models, is the case of a constant (euclidean) self--dual electromagnetic background \cite{ds01, ds02, ds02a}, satisfying
\be
F_{\mu\nu} = \star F_{\mu\nu} \equiv \frac{1}{2} \epsilon_{\mu\nu\rho\lambda} F^{\rho\lambda}.
\ee
\noindent
Following \cite{d04}, we introduce $\CF^2 = \frac{1}{4} F_{\mu\nu} F^{\mu\nu}$ and the natural dimensionless parameter $\gamma = \frac{2 e \CF}{m^2}$. In this case, the one--loop effective Lagrangian describing a charged scalar particle is now given by \cite{s51, d04}
\be\label{sd}
\CL =\frac{e^2 \CF^2}{16 \pi^2} \int_0^{+\infty} \frac{\rmd s}{s} \left( \frac{1}{\sinh^2 s} - \frac{1}{s^2} + \frac{1}{3} \right) \rme^{- \frac{2 s}{\gamma}},
\ee
\noindent
admitting the weak coupling expansion
\be\label{sdexp}
\CL \sim - \frac{m^4}{16 \pi^2} \sum_{n=1}^{+\infty} \frac{B_{2n+2}}{2n \left( 2n+2 \right)} \left( \frac{2 e \CF}{m^2} \right)^{2n+2}.
\ee
\noindent
Notice that there are two possible self--dual backgrounds \cite{ds01, ds02, ds02a}: a
magnetic--like background with $\CF$ real, in which case (\ref{sdexp}) has an alternating sign; and an electric--like background with $\CF$ imaginary, in which case (\ref{sdexp}) is not alternating. If one carries through a Borel analysis similar to the previous one, where we studied the case of constant electric field, one may further notice that while the alternating (magnetic) series has Borel poles on the positive \textit{imaginary} axis, the non--alternating (electric) series has the Borel poles on the positive \textit{real} axis, making both situations rather distinct on what respects evaluating the inverse Borel transforms. For the non--alternating (electric) series it is also possible to see that the aforementioned unitarity prescription will pick a Borel contour leading to a nonperturbative imaginary contribution to the Lagrangian, unambiguously given by
\be\label{imlsdb}
\im\, \mathcal{L} = \frac{e m^2 \CF}{32 \pi^3}\, \sum_{n=1}^{+\infty} \left( \frac{2\pi}{n} + \frac{\gamma}{n^2} \right) \rme^{-\frac{2\pi n}{\gamma}}.
\ee
\noindent
This expression can similarly be obtained by first considering the magnetic series, reflecting the integrand in (\ref{sd}) to the negative real axis in order to obtain an integral over the entire real line, and then deforming this integration contour such that it just incloses all the poles in the positive imaginary axis (a contour surrounding $\rmi \BR^+$). The resulting integral will then produce a sum over residues which, upon ``Wick rotation'' of the dimensionless coupling $\gamma \to \rmi \bar{\gamma}$, leads to the same expression as above, (\ref{imlsdb}). As we shall see in the course of this work, this expression is also at the basis of the nonperturbative structure of topological strings and $c=1$ matrix models.

Another important feature of the Schwinger effect, that we shall further explore later on, is the fact that in the presence of a constant electric field the pair--production process can be given a semiclassical interpretation in terms of a tunneling process, where electrons of negative energy are extracted from the Dirac background by the application of the external field \cite{p72}. The motion under the potential barrier, classically forbidden, is considered for \textit{imaginary} values of time, allowing for a computation of the tunneling probability corresponding to the pair--production rate as
\be
w \sim \rme^{- 2\, \im\, S},
\ee
\noindent
where $\im\, S $ is the imaginary part of the action developed during motion under the barrier. In here, a crucial point is that a particle in a sub--barrier trajectory satisfies the classical equations of motion. One may then use standard classical mechanics of a relativistic particle in order to describe this process. Indeed, energy conservation
\be
\CE = \pm \sqrt{p^2+m^2} - e E x,
\ee
\noindent
together with the equation of motion $\partial_t p = e E$, allow for an immediate re--writing of the action (for $\CE=0$) as:
\be\label{ac}
S(p) = \frac{p}{2 e E}\, \sqrt{p^2+m^2} -\frac{m^2}{2 e E}\, \log \left( p+\sqrt{p^2+m^2} \right).
\ee
\noindent
Notice that, because of the logarithm, the action is a multi--valued function.

\FIGURE[ht]{
\label{popov}
\centering
\psfrag{EE}{$E$}
\psfrag{XX}{$x$}
\psfrag{AAA}{$A_1$}
\psfrag{BBB}{$A_2$}
\psfrag{IMP}{$\im\, p$}
\psfrag{REP}{$\re\, p$}
\psfrag{AA1}{$A_1$}
\psfrag{AA2}{$A_2$}
\psfrag{BB1}{$B$}
\psfrag{pim}{$+\rmi m$}
\psfrag{nim}{$-\rmi m$}
\epsfig{file=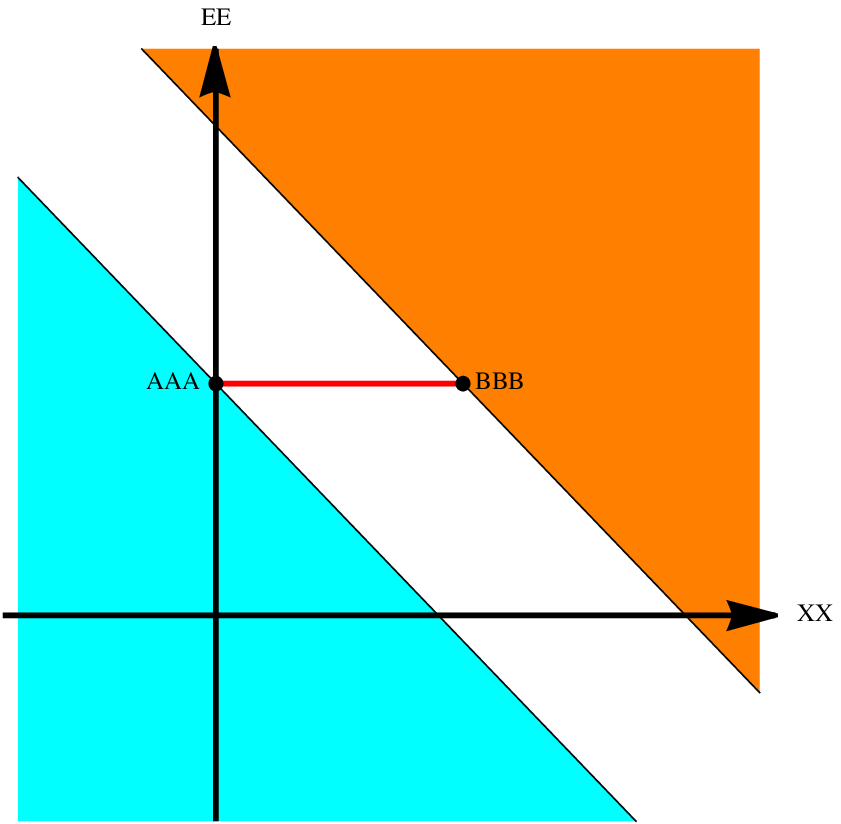, width=6cm, height=6cm}
$\qquad$
\epsfig{file=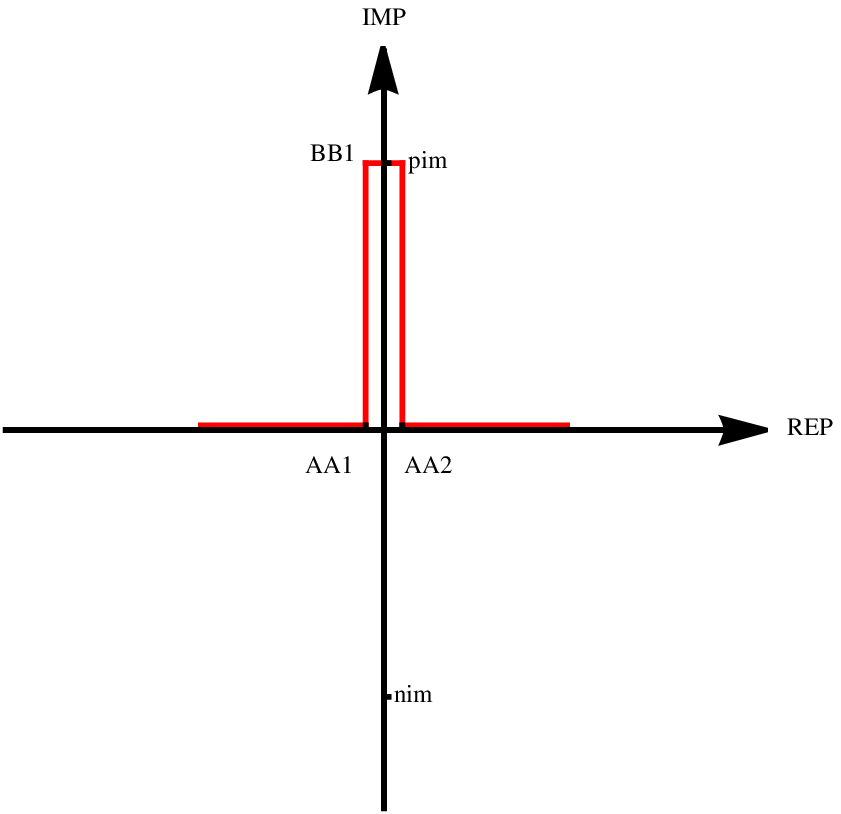, width=6cm, height=6cm}
\caption{On the left, the energy as a function of the position. The blue (orange) region corresponds to the lower (upper) continuum of particles with negative (positive) energy. The white region is classically forbidden. On the right, the variation of the imaginary momentum in the sub--barrier motion.}
}

The spectrum of possible values for the energy is displayed in Figure \ref{popov}. A potential barrier separates the lower continuum of negative--energy states (the minus sign of $\sqrt{p^2+m^2}$) from the upper continuum of positive--energy states (the plus sign of $\sqrt{p^2+m^2}$). Sub--barrier motion between points $A_1$ and $A_2$ will start at $A_1$, where $t=0=p$, and corresponds to the variation of the imaginary time/momentum along the path $A_1 B A_2$, while the real part of the energy remains constant. Indeed, at the classical turning point $B$ we will have $t = \rmi m/e E$ and $p=\rmi m$, which corresponds to a square--root branch point of the function $S(p)$. The motion ends back at $t=0=p$ in point $A_2$, as shown in Figure \ref{popov}. In this case, we see that the sub--barrier trajectory correspond to an increment of the imaginary part of the action as
\be
\Delta_{A_1BA_2}\, S = \im\, S,
\ee
\noindent
which we may compute as the \textit{shift} of the multi--valued function $S(p)$ as we move in--between the sheets of the logarithm. In fact, it is rather simple to realize that the value of $S(p)$ on a generic sheet differs from its value on the principal sheet, $S^*(p)$, by
\be
S(p) = S^*(p) + \rmi n\, \frac{\pi m^2}{e E}, \qquad n \in \BZ.
\ee
\noindent
In the illustration above we went (once) ``half way'' around the branch cut $[-\rmi m, +\rmi m]$ in which case the shift in the action is given by
\be
\Delta_{A_1BA_2}\, S = \frac{\pi m^2}{2e E}.
\ee
\noindent
For a generic sub--barrier motion, corresponding to a repeated wandering of the particle between the turning points $A_1$ and $A_2$ we may write
\be
2\, \im\, S = \oint_{\gamma_n} \sqrt{p^2+m^2} = n\, \frac{\pi m^2}{e E}, \qquad n \in \BZ,
\ee
\noindent
where $\gamma_n$ is a contour encircling $n$--times the branch cut of the action. One may notice \cite{p72} that this result is in complete agreement with the Schwinger computation result (\ref{rate}).

Naturally, this semiclassical argument may be refined in order to reproduce the pre--factors of the exponential term, and also so as to include the effect of a magnetic field. The point we wanted to make is that, from a semiclassical perspective, the instanton action describing the pair--production rate as a tunneling process may be computed via the branch cut discontinuities of the multi--valued function $S(p)$. This is a technique that we shall deploy later on in order to provide for a semiclassical interpretation of nonperturbative effects in $c=1$ matrix models in terms of eigenvalue tunneling.

In this paper we shall apply the techniques we have just described, Borel analysis, instanton calculus, and the Schwinger effect, in order to study the nonperturbative structure of topological strings and $c=1$ matrix models. As such, we now turn to topological string theory with emphasis towards the integral representation of its free energy.

\subsection{The Topological String Free Energy}\label{sec:gwt}

Asymptotic series and the Schwinger integral representation also appear in the context of topological string theory (see, \textit{e.g.}, \cite{v05} for an introduction). Let us start by describing the free energy of the A--model. The closed string sector of the A--model is a theory of maps $\phi: \Sigma_g \to \CX$ from a genus--$g$ Riemann surface, $\Sigma_g$, into a CY threefold $\CX$, which may be topologically classified by their homology class $\beta = \left[ \phi_*(\Sigma_g) \right] \in H_2 (\CX, \BZ)$. One may expand $\beta = \sum_{i=1}^{\dim H_2 (\CX, \BZ)} n_i\, [S_i]$ on a basis $[S_i]$ of $H_2(\CX, \BZ)$, with associated complexified K\"ahler parameters $t_i$.

The topological string free energy has a standard genus expansion in powers of the string coupling $g_s$, as in (\ref{freegt}), which in the large--radius phase (\textit{i.e.}, for large values of the K\"ahler parameters, in units of $\a$) becomes
\be\label{gwgenf}
F(g_s; \{ t_i \}) = \sum_{g=0}^{+\infty} g_s^{2g-2} F_g (t_i), \qquad F_g (t_i) = \sum_{\beta \ge 0} N_{g,\beta}\, Q^{\beta}.
\ee
\noindent
Here, the sum over $\beta$ is a sum over topological sectors or, equivalently, over world--sheet instantons. We have further introduced $Q_i = \rme^{-t_i}$, with $Q^{\beta}$ denoting $\prod_i Q_i^{n_i}$, and we have chosen units in which $\a = 2\pi$. The coefficients $N_{g, \beta}$ are the Gromov--Witten invariants of $\CX$, counting world--sheet instantons, \textit{i.e.}, the number of curves of genus $g$ in the two--homology class $\beta$.

The expansion in world--sheet instantons in (\ref{gwgenf}), regarded as a power series in $\rme^{-t_i}$, generically has a finite convergence radius, $t_c$, that can be estimated from the asymptotic large $\beta$ behavior of Gromov--Witten invariants \cite{bcov93}
\be\label{gwgenusg}
N_{g,\beta} \sim \beta^{(\gamma-2)(1-g)-1} \rme^{\beta t_c}, \quad \beta \to +\infty.
\ee
\noindent
In here, $\gamma$ is a critical exponent. At the critical value of the K\"ahler parameter, $t_c$, the so--called \textit{conifold point}, the geometric interpretation of the A--model large--radius phase breaks down and the topological string free energy undergoes a phase transition to a non--geometric phase, nonperturbative in $\a$. One may characterize the theory by its critical behavior at the conifold point. In particular, one can consider the following double--scaling limit
\be\label{dsl}
t \to t_c, \quad g_s \to 0, \qquad \mu = \frac{\rme^{-t_c} -\rme^{-t}}{g_s} \quad \mathrm{fixed}.
\ee
\noindent
In this case, the double--scaled free energy is universal, as first noticed in \cite{gv95}, and reads
\be\label{cone}
F_{\mathrm{DSL}} (\mu) = F_{c=1}(\mu) = \frac{1}{2} \mu^2 \log \mu - \frac{1}{12} \log \mu + \sum_{g=2}^{+\infty} \frac{B_{2g}}{2g (2g-2)} \mu^{2-2g},
\ee
\noindent
where  $F_{c=1} (\mu)$ is the all--genus free energy of the $c=1$ string at the self--dual radius (for a review on these issues see, \textit{e.g.}, \cite{k91}). The critical behavior (\ref{cone}) has been checked in many examples, such as \cite{kz99, kkv99}. Furthermore, in \cite{cgmps06}, it has been shown that certain local CYs have a critical behavior which is in the universality class of 2d quantum gravity, \textit{i.e.}, they have $\gamma = - \frac{1}{2}$. Another feature to notice is that the above genus expansion depends on the alternating Bernoulli numbers and, thus, is alternating for real $\mu$.

Of particular interest to our present work is the fact that the free energy $F_{c=1}(\mu)$ has a Schwinger--like nonperturbative integral formulation \cite{gm90, gk90}, given by
\be
F_{c=1}(\mu) = \frac{1}{4} \int_0^{+\infty} \frac{\rmd s}{s} \left(\frac{1}{\sinh^2 s} - \frac{1}{s^2} + \frac{1}{3}\right) \rme^{-s\mu},
\ee
\noindent
which coincides, after an appropriate identification of the parameters, with the one--loop effective Lagrangian for a charged particle in a constant self--dual background, (\ref{sd}). This means that $F_{c=1}(\mu)$ enjoys an asymptotic weak coupling expansion as in (\ref{sdexp}) and further develops a nonperturbative imaginary contribution akin to (\ref{imlsdb}). In this line of thought, the exploration of Schwinger--like integral representations for the free energies of topological strings and $c=1$ matrix models is one of the main topics in this paper.

\subsection{A Schwinger--Gopakumar--Vafa Integral Representation}\label{sec:gv}

As should be clear by now, Schwinger--like integral representations for the free energy are bound to play a critical role in our analysis. Happily, for topological string theory, such representations have been provided by Gopakumar and Vafa in \cite{gv98, gv98a, gv98c}. These works explored both the connection of topological strings to the physical IIA string, as well as the duality between type IIA compactified on a CY threefold, at strong coupling, and M--theory compactified on the same CY times a circle, in order to relate topological string amplitudes to the BPS structure of wrapped M2--branes and thus re--write the topological string free energy in terms of an integral representation. The final result in \cite{gv98c} for the all--genus topological string free energy, on a CY threefold $\CX$, is
\be\label{sgvir}
F_\CX (g_s; \{ t_i \} ) = \sum_{\{d_i\},r,m} n^{(d_i)}_r \left( \CX \right) \int_0^{+\infty} \frac{\rmd s}{s} \left( 2 \sin \frac{s}{2} \right)^{2r-2}\, \exp \left( - \frac{2 \pi s}{g_s} \left( d \cdot t + \rmi\, m \right) \right).
\ee
\noindent
Let us explain the diverse quantities in this expression. The integers $n^{(d_i)}_r \left( \CX \right)$ are the GV invariants of the threefold $\CX$. They depend on the K\"ahler class $d_i$ and on a spin label $r$. Later on we shall be focusing on the case where $\CX$ is the resolved conifold, for which there is only one non--vanishing integer, $n^{(1)}_0 = 1$. The combination $Z = d \cdot t + \rmi\, m$ represents the central charge of four--dimensional BPS states obtained in the following fashion \cite{gv98c}. Start with M--theory compactified on $\CX \times \BS^1$ and consider the BPS spectrum of M2--branes wrapped on cycles of the CY threefold with fixed central charge $A = d \cdot t = \sum_{i=1}^{\dim H_2 (\CX, \BZ)} d_i t_i$, with $d_i$ as above and $t_i$ the complexified K\"ahler parameters. The mass of the wrapped M2--branes is $2\pi A$. Upon reduction on $\BS^1$ each BPS state may have in addition an arbitrary (quantized) momentum $m$ around the circle, leading to BPS states of central charge $Z$ and mass $2\pi Z$. Notice that these four--dimensional BPS states contributing to the topological string free energy may be understood, from a IIA point of view, as bound states of D2 and D0--branes, and it is the physics of this system which can be related to a Schwinger--type computation and thus to the above integral representation (in fact, thanks to the $\CN=2$ supersymmetry in the problem, the Schwinger calculation one has to perform in this context turns out to be equivalent to that of a vacuum amplitude for a charged scalar field in the presence of a self--dual electromagnetic field strength, as in (\ref{sd})). Furthermore, the integer $m$, associated to the winding around $\BS^1$, counts the number of D0 branes in the D2D0 BPS bound state. This should make (\ref{sgvir}) clear.

One may also recover the perturbative genus expansion from this integral representation. Using the familiar identity
\be
\sum_{m \in \BZ} \exp \left( - 2\pi\rmi m\, \frac{s}{g_s} \right) = \sum_{n \in \BZ} \delta \left( \frac{s}{g_s} - n \right),
\ee
\noindent
with $\delta(x)$ the Dirac delta function, one may explicitly evaluate the sum over $m$ in 
(\ref{sgvir}) and thus obtain, after the trivial integration over $s$,
\be\label{GVexpansion}
F_\CX (g_s; \{t_i\}) = \sum_{r=0}^{+\infty}\, \sum_{d_i=1}^{+\infty} n_r^{(d_i)} (\CX) \sum_{n=1}^{+\infty} \frac{1}{n} \left( 2 \sin \frac{ng_s}{2} \right)^{2r-2}\, \rme^{- 2 \pi n\, d \cdot t}.
\ee
\noindent
This result expresses the topological string free energy, on a CY threefold $\CX$, in terms of the GV integer invariants \cite{gv98, gv98a, gv98c}. To be completely precise, it is important to notice that in order to recover the \textit{full} topological string free energy one still has to add to (\ref{GVexpansion}) the (alternating) constant map contribution \cite{mm98, fp98}
\be\label{fpK}
F_{\mathrm{K}} (g_s) = \sum_{g=0}^{+\infty} g_s^{2g-2}\, \chi_{\mathrm{K}} (\CX)\, \frac{(-1)^{g} \left| B_{2g} B_{2g-2} \right|}{4 g \left( 2g-2 \right) \left( 2g-2 \right)!},
\ee
\noindent
where $\chi_{\mathrm{K}} (\CX) = 2 \left( h^{1,1}-h^{2,1} \right)$ is the Euler characteristic of $\CX$. This term can also be given a Schwinger--like integral representation. From the point of view of the duality between type IIA and M--theory, this amounts to considering only the contribution arising from the D0--branes, or Kaluza--Klein modes. The result is \cite{gv98a}
\be\label{1loopk}
F_{\mathrm{K}} (g_s) = \frac{1}{8} \chi_{\mathrm{K}} (\CX) \sum_{m \in \BZ} \int_0^{+\infty} \frac{\rmd s}{s}\, \frac{1}{\sinh^2 \left( \frac{s}{2} \right)}\, \rme^{-2 \pi \rmi m\, \frac{s}{g_s}} = \frac{1}{8} \chi_{\mathrm{K}} (\CX) \sum_{n=1}^{+\infty} \frac{1}{n}\, \frac{1}{\sinh^2 \left( \frac{n g_s}{2} \right)}.
\ee

In this paper we shall mainly consider the resolved conifold, a toric CY threefold for which $\dim H_2 (\CX, \BZ) = 1$ and thus the only non--vanishing integer GV invariant is $n_0^{(1)} = 1$. In this case, the GV integral representation (\ref{sgvir}) immediately yields
\be
F_\CX (g_s; t) = \frac{1}{4} \sum_{m \in \BZ} \int_0^{+\infty} \frac{\rmd s}{s}\, \frac{1}{\sin^2 \left( \frac{s}{2} \right)}\, \rme^{- \frac{2 \pi s}{g_s} \left( t + \rmi\, m \right)},
\ee
\noindent
an expression which carries a Schwinger flavor, as we have seen above. It is also important to point out that the case of $r=0$ is the only one in which the integrand of the GV integral representation will have ``interesting'' poles, \textit{i.e.}, poles of the sine function on the real axis. When $r>0$ the only poles of the integrand will be at zero and $\infty$ in the Borel complex plane. So, in particular, when studying more complicated CY threefolds where there is a sum over $r \ge 0$, it will always be the contribution from GV invariants with $r=0$ which will be the most relevant for the Schwinger analysis we shall carry through later in the paper and, as such, the case of the resolved conifold is a prototypical example for those situations. From the previous expression it is also simple to obtain the perturbative expansion, by summing over $m$ as previously described, and one obtains
\be
F_\CX (g_s; t) = \frac{1}{4} \sum_{n=1}^{+\infty} \frac{1}{n}\, \frac{1}{\sin^2 \left( \frac{ng_s}{2} \right)}\, \rme^{- 2 \pi n\, t}.
\ee
\noindent
By carrying through this sum, expanding in powers of $g_s$, and adding the constant map contribution, one finally obtains the resolved conifold genus expansion as
\be\label{rc}
F_g (t) = \frac{(-1)^g |B_{2g} B_{2g-2}|}{2g \left( 2g-2 \right) \left( 2g-2 \right)!} + \frac{|B_{2g}|}{2g \left( 2g-2 \right)!}\, \mathrm{Li}_{3-2g} \left( \rme^{-t} \right).
\ee 
\noindent
with $\mathrm{Li}_p (x)$ the polylogarithm function. We shall later see how a Borel analysis allows for a nonperturbative completion of this expansion and moreover how to relate this nonperturbative completion to the large--order behavior of the above genus expansion.

One final word pertains to the matrix model description of strings on the resolved conifold via a large $N$ duality. It was shown in \cite{gv98b} that there is a duality between closed and open topological A--model string theory on, respectively, the resolved and the deformed conifold; two smooth manifolds related to the same singular geometry. In the resolved conifold case the conifold singularity is removed by blowing up a two--sphere around the singularity; while in the deformed conifold case the conifold singularity is removed by growing a three--sphere around it, which is also a Lagrangian sub--manifold thus providing boundary conditions for open strings. As it turns out, the full open topological string field theory in this latter background, $T^*\BS^3$, where we wrap $N$ D--branes on the Lagrangian sub--manifold base, $\BS^3$, reduces to SU$(N)$ Chern--Simons gauge theory on $\BS^3$ \cite{w92}, whose partition function further admits a matrix model description \cite{m02}. The matrix model in question, which we shall review in the next section, has a potential with a single minimum and no local maxima. In this paper we refer to this type of matrix models (which will also include the Gaussian and Penner cases) as $c=1$ matrix models since, as we shall see, they all admit a very natural double--scaling limit to the $c=1$ string at self--dual radius. Notice that $c=1$ matrix models do not belong to the class of matrix models for which the off--critical instanton analysis has been carried out so far. Because understanding nonperturbative corrections to the topological string free energy on the resolved conifold is undissociated from understanding nonperturbative corrections to $c=1$ matrix models, we shall consider this latter case more broadly in order to shed full light on this class of instanton phenomena. As such, $c=1$ matrix models is the subject we shall turn to next.

\section{$c=1$ Matrix Models and Topological String Theory}

We shall now introduce three distinct matrix models, all in the universality class of the $c=1$ string, and which will be the main focus of our subsequent discussion. As mentioned in the previous section, one of these models is the one describing Chern--Simons gauge theory on $\BS^3$, known as the Stieltjes--Wigert matrix model. Another interesting, and rather elementary, matrix model is the Gaussian model. Yet, we shall find that it already displays many features that will also appear for the resolved conifold. Finally, we also address the Penner matrix model, first introduced to study the orbifold Euler characteristic of the moduli space of punctured Riemann surfaces. These three models have been extensively studied in the literature and in the present section we will mostly gather some general facts necessary to obtain their topological large $N$ expansions and their holomorphic effective potentials. Then, in the following section, we shall analyze their large $N$ asymptotic expansions from the point of view of Borel analysis.

Let us begin by recalling some basic notions about matrix models (see, \textit{e.g.}, \cite{bipz78, biz80, fgz93, m04}). The hermitian $N \times N$ one--matrix model partition function is
\be
Z = \frac{1}{{\mathrm{vol}} \left( \mathrm{U}(N) \right)} \int \rmd M\, \rme^{- \frac{1}{g_s}\, \tr\, V(M)},
\ee
\noindent
with ${\mathrm{vol}} \left( \mathrm{U}(N) \right)$ the usual volume factor of the gauge group. In the eigenvalue diagonal gauge this becomes
\be
Z = \frac{1}{N!} \int \prod_{i=1}^N \left( \frac{\rmd \lambda_i}{2\pi} \right) \Delta^2(\lambda)\, \rme^{- \frac{1}{g_s} \sum_{i=1}^N V(\lambda_i)}, 
\ee
\noindent
where $\Delta(\lambda) = \prod_{i<j} \left( \lambda_i - \lambda_j \right)$ is the Vandermonde determinant. The free energy of the matrix model is then defined as usual $F = \log Z$ and, in the large $N$ limit, it has a perturbative genus expansion
\be
F = \sum_{g=0}^{+\infty} g_s^{2g-2} F_g(t),
\ee
\noindent
with $t = N g_s$ the 't~Hooft coupling. Multi--trace correlation functions in the matrix model may be obtained from their generating functions, the \textit{connected} correlation functions defined by
\be\label{obj}
W_h (z_1,\ldots, z_h) = \left\langle \tr\, \frac{1}{z_1-M} \cdots \tr\, \frac{1}{z_h-M} \right\rangle_{(c)} = \sum_{g=0}^{+\infty} g_s^{2g+h-2} W_{g,h} (z_1, \ldots, z_h;t).
\ee
\noindent
In particular, the generator of single--trace correlation functions is $W_1 (z) = N \omega(z)$ where $\omega (z)$ is the resolvent, \textit{i.e.}, the Hilbert transform of the eigenvalue density $\rho(\lambda)$ characterizing the saddle--point associated to the matrix model large $N$ limit. In the most general case, this saddle--point is such that $\rho(\lambda)$ has support $\CC$, with $\CC$ a multi--cut region given by an union of $s$ intervals $\CC_i$. At large $N$, the eigenvalues condense on these intervals $\CC_i$ in the complex plane and one may interpret them geometrically as branch cuts of a spectral curve which, in the hermitian one--matrix model, would be a hyperelliptic Riemann surface corresponding to a double--sheet covering of the complex plane $\BC$, with the two sheets sewed together by the cuts $\CC_i$.

The spectral curve, to be denoted by $y(z)$, may be written in terms of the genus zero resolvent which, for a generic one--cut solution with $\CC=[a,b]$, is given by the \textit{ans\"atz}
\be\label{genus0resolvent}
\omega_0(z) = \frac{1}{2t} \oint_{\CC} \frac{\rmd w}{2\pi\rmi}\, \frac{V'(w)}{z-w}\, \sqrt{\frac{(z-a)(z-b)}{(w-a)(w-b)}},
\ee
\noindent
where one still has to impose that $\omega_0(z) \sim \frac{1}{z}$ as $z \to + \infty$,
in order to fix the position of the cut endpoints\footnote{This boundary condition states that the eigenvalue density is normalized to one in the cut, $\int_\CC \rmd\lambda\, \rho(\lambda) = 1$.}. The spectral curve is then defined as
\be
y(z) = V'(z) - 2t\, \omega_0(z) \equiv M(z)\, \sqrt{(z-a)(z-b)}.
\ee

For future reference, it is also useful to define the holomorphic effective potential, defined as the line integral of the one--form $y(z)\, \rmd z$ along the spectral curve,
\be
V_{\mathrm{h;eff}}(\lambda) =  \int^\lambda_a \rmd z\, y(z),
\ee
\noindent
which appears at leading order in the large $N$ expansion of the matrix integral as
\be
Z \sim \int \prod_{i=1}^N \rmd \lambda_i\, \exp \left( - \frac{1}{g_s} \sum_{i=1}^N V_{\mathrm{h;eff}} (\lambda_i) + \cdots \right).
\ee
\noindent
Because the real part of the spectral curve relates to the force exerted on a given eigenvalue, it turns out that the effective potential $V_{\mathrm{eff}} (z) = \re\, V_{\mathrm{h;eff}} (z)$ is constant inside the cut $\mathcal{C}$, \textit{i.e.}, inside the cut the eigenvalues are free. The imaginary part of the spectral curve, on the other hand, relates to the eigenvalue density as $\im\, y(z) = 2\pi t\, \rho(z)$, thus implying that the imaginary part of $V_{\mathrm{h;eff}} (z)$ is zero outside the cut and monotonic inside. These two conditions guarantee that the eigenvalue density is real with support on $\CC$. Furthermore, as should be clear from the expression above, if the matrix integral $Z$ is to be convergent a careful choice of integration contour for the eigenvalues has to be made based also on the properties of the holomorphic effective potential \cite{d91, d92}. In particular, this contour may be analytically continued to any contour which includes the cut $\CC$ and does not cross any region where $V_{\mathrm{eff}} (z) = \re\, V_{\mathrm{h;eff}} (z) < 0$, thus guaranteeing global stability of the saddle--point configuration and convergence of the matrix integral (as $\re\, V_{\mathrm{h;eff}}(\lambda) \to + \infty$ at the endpoints of the integration contour). These properties of $V_{\mathrm{h;eff}} (z)$ ensure that, in the large $N$ limit, the matrix integral can be evaluated with the steepest--descendent method \cite{d91, d92}.

There are many ways to solve matrix models. In particular, \cite{eo07} proposed a recursive method for computing the connected correlation functions (\ref{obj}) and the genus--$g$ free energies, $F_g (t)$, entirely in terms of the spectral curve. This recursive method, sometimes denoted by the \textit{topological recursion}, appears to be extremely general and applies beyond the context of matrix models; see \cite{eo08} for a review. For our purposes of computing the genus expansion of the free energy one of the most efficient and simple methods is that of orthogonal polynomials \cite{biz80}, which we now briefly introduce. If one regards
\be
\rmd\mu (z) \equiv \rme^{- \frac{1}{g_s} V(z)}\, \frac{\rmd z}{2\pi}
\ee
\noindent
as a positive--definite measure in $\BR$, it is immediate to introduce orthogonal polynomials, $\{ p_n (z) \}$, with respect to this measure as
\be\label{op}
\int_{\BR} \rmd\mu(z)\, p_n(z) p_m(z) = h_n \delta_{nm}, \qquad n \ge 0,
\ee
\noindent
where one further normalizes $p_n (z)$ such that $p_n (z) = z^n + \cdots$. Further noticing that the Vandermonde determinant is $\Delta(\lambda) = \det p_{j-1} (\lambda_i)$, the one--matrix model partition function may be computed as
\be\label{zop}
Z_N = \prod_{n=0}^{N-1} h_n = h_0^N \prod_{n=1}^N r_n^{N-n},
\ee
\noindent
where we have defined $r_n = \frac{h_n}{h_{n-1}}$ for $n \ge 1$. These coefficients also appear in the recursion relations of the orthogonal polynomials,
\be\label{oprecursion}
p_{n+1} (z) = \left( z+s_n \right) p_n (z) - r_n\, p_{n-1} (z).
\ee
\noindent
In the large $N$ limit the recursion coefficients approach a continuous function $r_n \to R(x)$, with $x = \frac{n}{N} \in [0,1]$, and one may proceed to compute the genus expansion of $\log Z$ by making use of the Euler--MacLaurin formula; see \cite{biz80, m04} for details.

\subsection{The Gaussian Matrix Model}

Let us first focus on the Gaussian matrix model, defined by the potential $V_{\mathrm{G}} (z) = \frac{1}{2} z^2$. This case is rather simple as the matrix integral can be straightforwardly evaluated via gaussian integration, and the volume of the compact unitary group follows by a theorem of Macdonald \cite{m80} as
\be
{\mathrm{vol}} \left( \mathrm{U}(N) \right) = \frac{\left( 2\pi \right)^{\frac{1}{2}N(N+1)}}{G_2 (N+1)},
\ee
\noindent
where $G_2(z)$ is the Barnes function, $G_2(z+1) = \Gamma(z) G_2 (z)$. The Gaussian partition function thus reads
\be
Z_{\mathrm{G}} = \frac{g_s^{\frac{N^2}{2}}}{(2\pi)^{\frac{N}{2}}}\, G_2 (N+1).
\ee
\noindent
The same result can be obtained with orthogonal polynomials. With respect to the Gaussian measure $\rmd \mu(x) = \rme^{-x^2} \rmd x$ one finds Hermite polynomials, $H_n (x)$, and for the Gaussian matrix model it follows
\be
p_n(z) = \left( \frac{g_s}{2} \right)^{\frac{n}{2}} H_n \left( \frac{z}{\sqrt{2 g_s}} \right), \qquad h^{\mathrm{G}}_n = g_s^n n! \sqrt{\frac{g_s}{2\pi}},
\ee
\noindent
indeed reproducing the expected result for the partition function as
\be\label{opgaussian}
Z = \prod_{n=0}^{N-1} h^{\mathrm{G}}_n = \frac{g_s^{\frac{N^2}{2}}}{(2\pi)^{\frac{N}{2}}} \prod_{n=0}^{N-1} n! = Z_{\mathrm{G}},
\ee
\noindent
where we have also used that $G_2 (N+1) = \prod_{n=0}^{N-1} n!$. The asymptotic genus expansion of the Gaussian free energy $F_{\mathrm{G}} = \log Z_{\mathrm{G}}$ simply follows from the asymptotic expansion of the logarithm of the Barnes function and one obtains
\bea
F_0^{\mathrm{G}} (t) &=& \frac{1}{2} t^2 \left( \log t - \frac{3}{2} \right), \\
F_1^{\mathrm{G}} (t) &=& - \frac{1}{12} \log t + \zeta'(-1), \\
F_g^{\mathrm{G}} (t) &=& \frac{B_{2g}}{2g(2g-2)}\, t^{2-2g}, \qquad g \ge 2,
\label{gs}
\eea
\noindent
where $\zeta (z)$ is the Riemann zeta function. One immediately notices that all free energies with $g \ge 1$ diverge when $t \to 0$. It is then quite obvious to consider the double--scaling limit, approaching the critical point $t_c = 0$, as
\be
t \to 0, \quad g_s \to 0, \qquad \mu = \frac{t-t_c}{g_s} \quad \mathrm{fixed},
\ee
\noindent
in order to obtain the $c=1$ string at self--dual radius behavior
\be
g_s^{2g-2}\, F_g^{\mathrm{G}} (t) \to \frac{B_{2g}}{2g(2g-2)}\, \mu^{2-2g}, \qquad g \ge 2.
\ee

Finally, it is very simple to compute the one--form on the spectral curve of the Gaussian model
\be
y(z)\, \rmd z = \sqrt{z^2 - 4t}\, \rmd z,
\ee
\noindent
as well as the holomorphic effective potential
\be\label{veg}
V_{\mathrm{h;eff}}^{\mathrm{G}} (z) = \frac{1}{2}\, z \sqrt{z^2 - 4t} -  2t \log \left( \frac{z + \sqrt{z^2 - 4t}}{2 \sqrt{t}} \right),
\ee
\noindent
where we have normalized the result such that $V_{\mathrm{h;eff}}^{\mathrm{G}} (b=2\sqrt{t}) = 0$. In Figures \ref{galgcurve} and \ref{gstokescurve} we plot the Gaussian algebraic curve for different values of $t$, as well as the real value of the holomorphic effective potential in the complex plane. We notice that, with an appropriate identification of parameters, the Gaussian holomorphic effective potential coincides with the action associated to the semiclassical Schwinger effect, (\ref{ac}). In the following we shall further comment about this interesting coincidence.

\FIGURE[ht]{
\label{galgcurve}
\centering
\epsfig{file=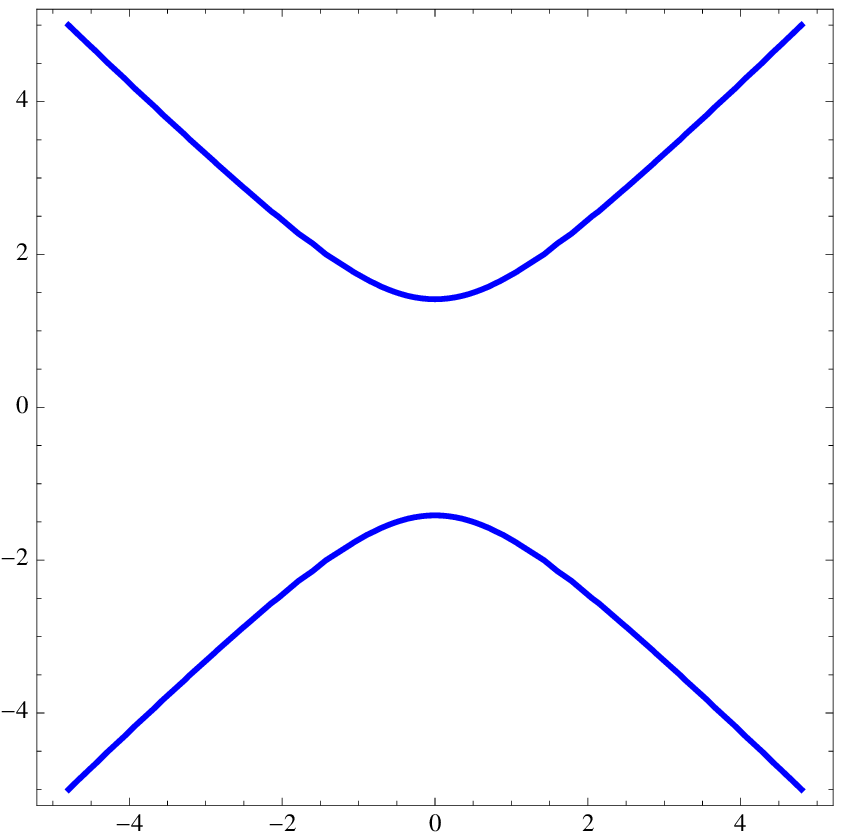, width=3cm, height=3cm}
\epsfig{file=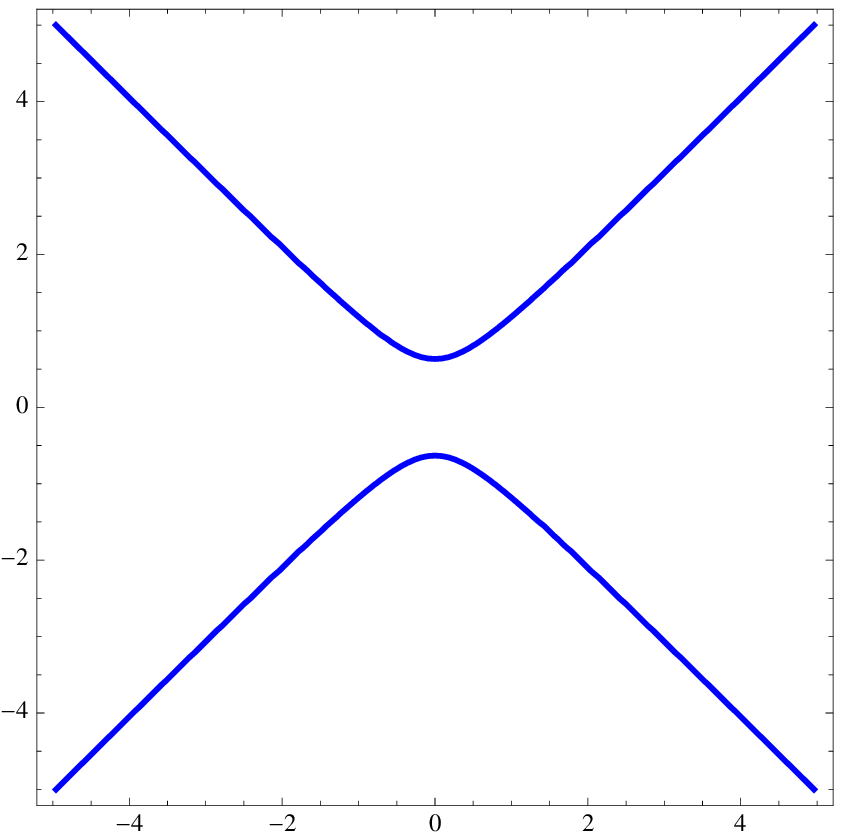, width=3cm, height=3cm}
\epsfig{file=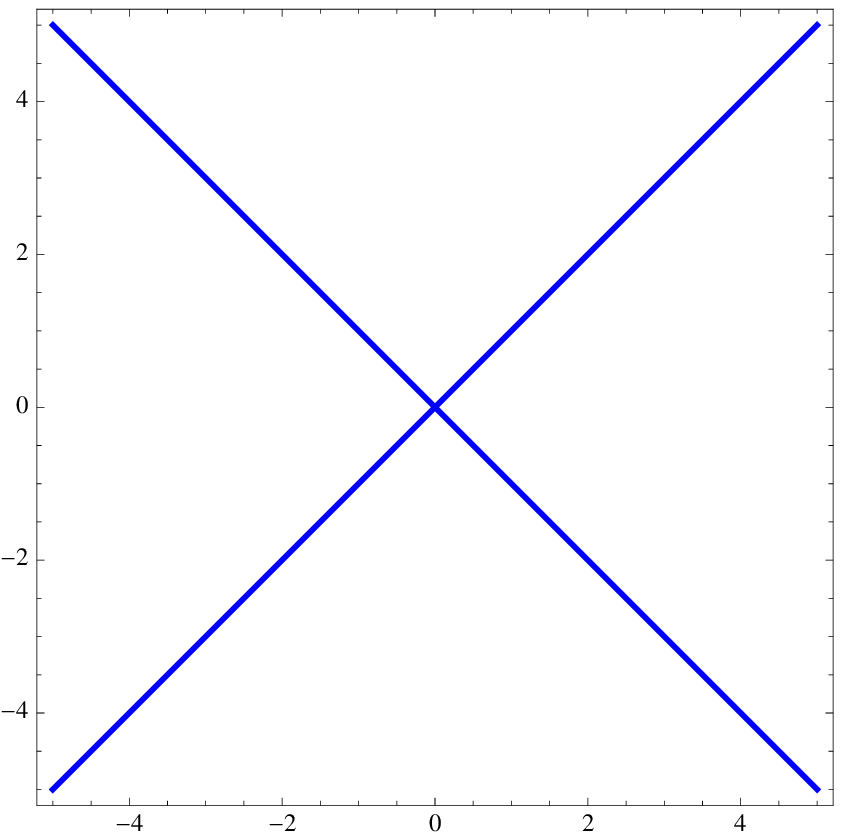, width=3cm, height=3cm}
\epsfig{file=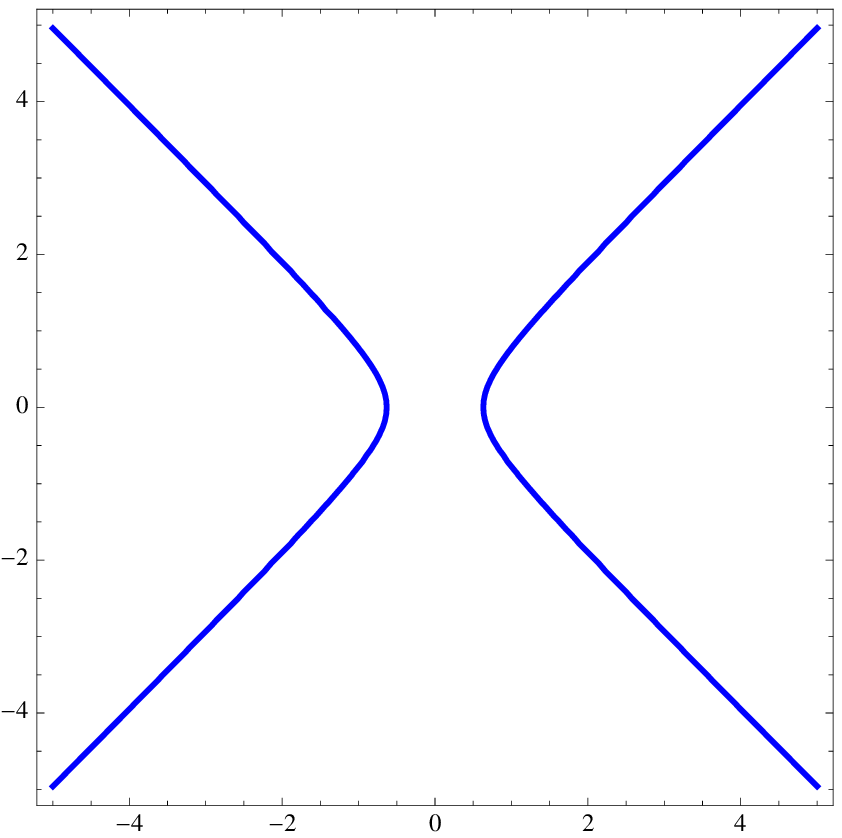, width=3cm, height=3cm}
\epsfig{file=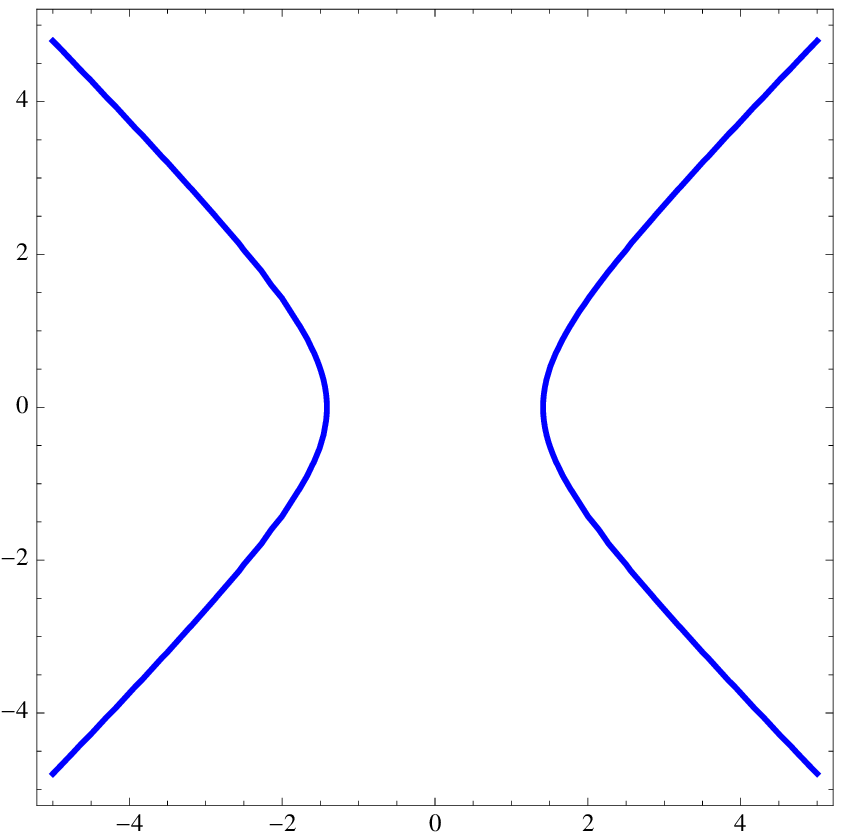, width=3cm, height=3cm}
\caption{The Gaussian algebraic curve for values of $t = -0.5, -0.1, 0, +0.1, +0.5$, from left to right, respectively. Notice that the algebraic curve is singular for $t=0$.}
}
\FIGURE[ht]{
\label{gstokescurve}
\centering
\epsfig{file=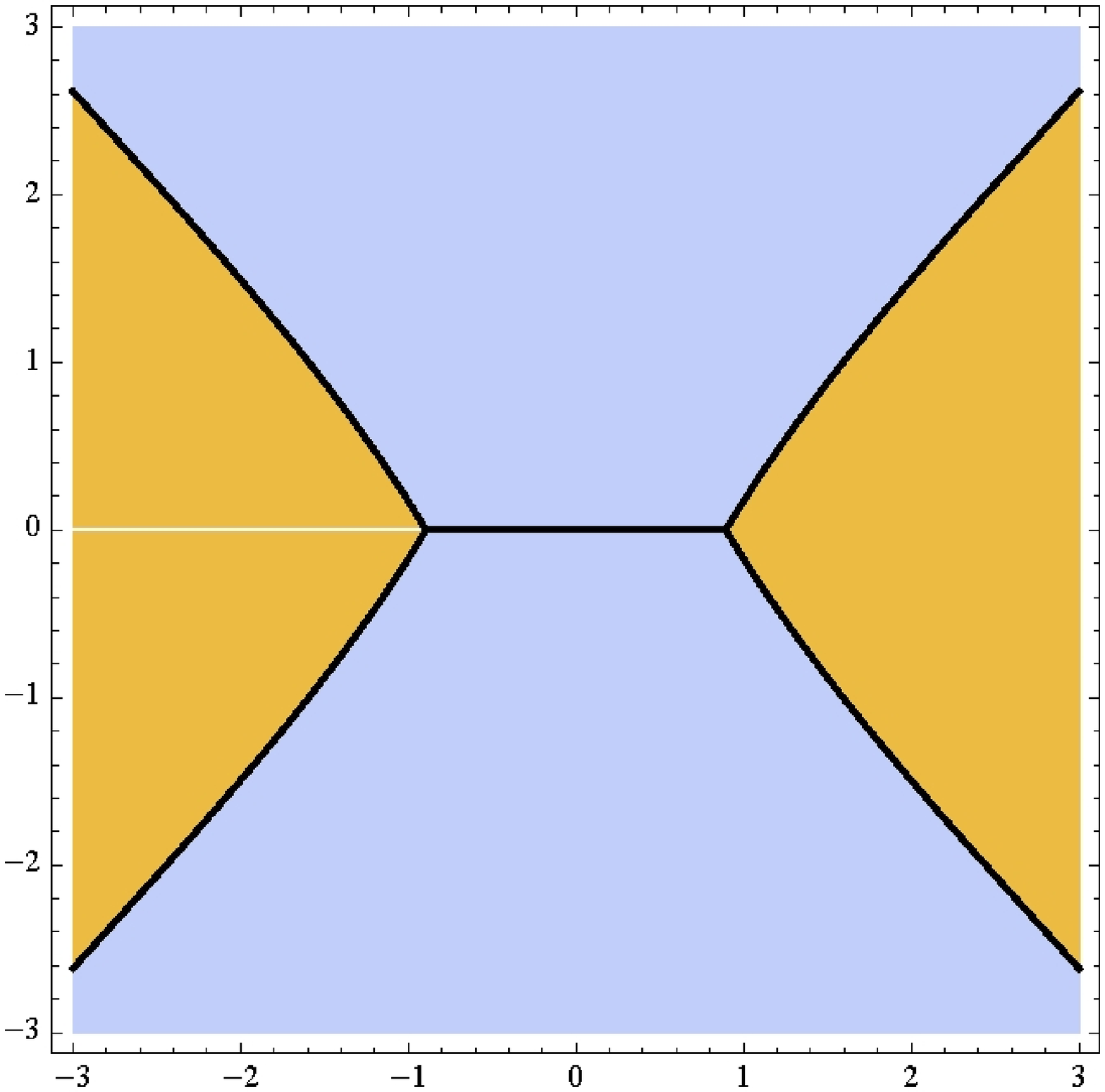, width=9.5cm, height=9.5cm}
\caption{The real part of the Gaussian holomorphic effective potential in the complex $z$--plane, for $t=0.2$. In blue is the region $\re\, V_{\mathrm{h;eff}}^{\mathrm{G}} (z) < 0$, in orange the region $\re\, V_{\mathrm{h;eff}}^{\mathrm{G}} (z) > 0$, and the black lines correspond to the Stokes lines $\re\, V_{\mathrm{h;eff}}^{\mathrm{G}} (z) = 0$ (which also include the cut of the spectral curve). The white cut corresponds to the logarithmic branch cut.}
}

\subsection{The Penner Matrix Model}

The second example we wish to address is the Penner matrix model \cite{p88}. First introduced to study the orbifold Euler characteristic of the moduli space of Riemann surfaces at genus $g$, with $n$ punctures, it turns out that in the double--scaling limit this model is actually related to the usual $c=1$ noncritical string theory, its free energy being a Legendre transform of the free energy of the $c=1$ string compactified at self--dual radius \cite{dv90, dv91}. The Penner matrix model is defined by the potential $V_{\mathrm{P}} (z) = z - \log z$ and one may simply compute its partition function again using orthogonal polynomials. Indeed, one may write the Penner measure as
\be
\rmd\mu(z) = z^{\frac{1}{g_s}}\, \rme^{-\frac{z}{g_s}} \frac{\rmd z}{2\pi},
\ee
\noindent
which is, up to normalization, the measure for the generalized, or associated, Laguerre polynomials $L_n^{(\alpha)} (x) = \frac{1}{n!}\, \rme^x x^{-\alpha} \frac{\rmd^n}{\rmd x^n} \left( \rme^{-x} x^{n+\alpha} \right)$. It thus follows for the Penner matrix model
\be
p_n (z) = (-1)^n g_s^n n!\, L^{\left( 1/g_s \right)}_n \left( \frac{z}{g_s} \right),\ \ \ \
h^{\mathrm{P}}_n = \frac{1}{2\pi}\, g_s^{2n+1+\frac{1}{g_s}}\, n!\, \Gamma\left( n+\frac{1}{g_s}+1 \right).
\ee
\noindent
This immediately leads to the calculation of the partition function in this model as
\be
Z_{\mathrm{P}} = \prod_{n=0}^{N-1} h^{\mathrm{P}}_n = \frac{g_s^{N \left( N+\frac{1}{g_s} \right)}}{(2\pi)^N}\, \frac{G_2 \left( N+1 \right) G_2 \left( N+\frac{1}{g_s}+1 \right)}{G_2 \left( \frac{1}{g_s}+1 \right)},
\ee
\noindent
where we made use of the Barnes function, satisfying
\be
\prod_{n=0}^{N-1} \Gamma \left( n+\alpha+1 \right) = \frac{G_2 (N+\alpha+1)}{G_2 (\alpha+1)}.
\ee
\noindent
The normalized Penner free energy $\CF_{\mathrm{P}} = F_{\mathrm{P}} - F_{\mathrm{G}} = \log \frac{Z_{\mathrm{P}}}{Z_{\mathrm{G}}}$ is given by
\be\label{normalizedfp}
\CF_{\mathrm{P}} = \frac{1}{2} N^2 \log g_s + \frac{N}{g_s} \log g_s - \frac{1}{2} N \log 2\pi + \log G_2 \left( N+\frac{1}{g_s}+1 \right) - \log G_2 \left( \frac{1}{g_s}+1 \right),
\ee
\noindent
and it admits the following genus expansion, obtained from the asymptotic expansion of the logarithm of the Barnes functions,
\bea
\CF_0^{\mathrm{P}} (t) &=& \frac{1}{2} \left(t+1\right)^2 \left( \log \left(t+1\right) - \frac{3}{2} \right) + \frac{3}{4}, \\
\CF_1^{\mathrm{P}} (t) &=& - \frac{1}{12} \log \left(t+1\right),\\
\CF_g^{\mathrm{P}} (t) &=& \frac{B_{2g}}{2g(2g-2)} \left( \left(t+1\right)^{2-2g} - 1 \right), \qquad g \ge 2.
\label{penge}
\eea
\noindent
One immediately notices that all free energies with $g \ge 1$ diverge when $t \to -1$. It is then quite obvious to consider the double--scaling limit, approaching the critical point $t_c = -1$, as
\be
t \to -1, \quad g_s \to 0, \qquad \mu = \frac{t-t_c}{g_s} \quad \mathrm{fixed},
\ee
\noindent
in order to obtain the $c=1$ string at self--dual radius \cite{dv90}
\be
g_s^{2g-2}\, \CF_g^{\mathrm{P}} (t) \to \frac{B_{2g}}{2g(2g-2)}\, \mu^{2-2g}, \qquad g \ge 2.
\ee

Next, let us address the large $N$ expansion of the Penner matrix model by making use of saddle--point techniques \cite{cdl91, akm94}. This time around, the \textit{ans\"atz} for the large $N$, genus zero  resolvent is \cite{cdl91}
\be
\omega_0 (z) = \frac{1}{2t} \left( V'(z) - \frac{1}{z \sqrt{ab}} \sqrt{(z-a)(z-b)} \right),
\ee
\noindent
so that its large $z$ asymptotics, $\omega_0 (z) \sim \frac{1}{z} + \cdots$ as $z \to \infty$,  immediately determine the endpoints of the cut $\CC=[a,b]$ to be
\bea
a &=& 1 + 2t - 2 \sqrt{t(t+1)}, \\
b &=& 1 + 2t + 2 \sqrt{t(t+1)}.
\eea
\noindent
It is now simple to obtain the one--form on the spectral curve of the Penner model
\be
y(z)\, \rmd z = \frac{1}{z} \sqrt{z^2 - 2 \left( 2t+1 \right) z + 1}\, \rmd z
\ee
\noindent
as well as the holomorphic effective potential
\bea
V_{\mathrm{h;eff}}^{\mathrm{P}} (z) &=& \sqrt{z^2 - 2 \left( 2t+1 \right) z + 1} + \log z - \log \left( 1 - \left( 2t+1 \right) z + \sqrt{z^2 - 2 \left( 2t+1 \right) z + 1} \right) - \nonumber \\
&&
- \left( 2t+1 \right) \log \left( z - \left( 2t+1 \right) + \sqrt{z^2 - 2 \left( 2t+1 \right) z + 1} \right) + \left( t+1 \right) \log \big( t(t+1) \big) + \nonumber \\
&&
+ \left( t+1 \right) \log 4 + \rmi \pi,
\label{vep}
\eea
\noindent
where we have normalized the result such that $V_{\mathrm{h;eff}}^{\mathrm{P}} (b) = 0$. In Figures \ref{palgcurve} and \ref{pstokescurve} we plot the Penner algebraic curve for different values of $t$, as well as the real value of the holomorphic effective potential in the complex plane. The structure of Stokes lines for this potential is now more complicated (see, \textit{e.g.}, \cite{cdl91, akm94}) than in the familiar polynomial cases (see, \textit{e.g.}, \cite{d91, d92}).

\FIGURE[ht]{
\label{palgcurve}
\centering
\epsfig{file=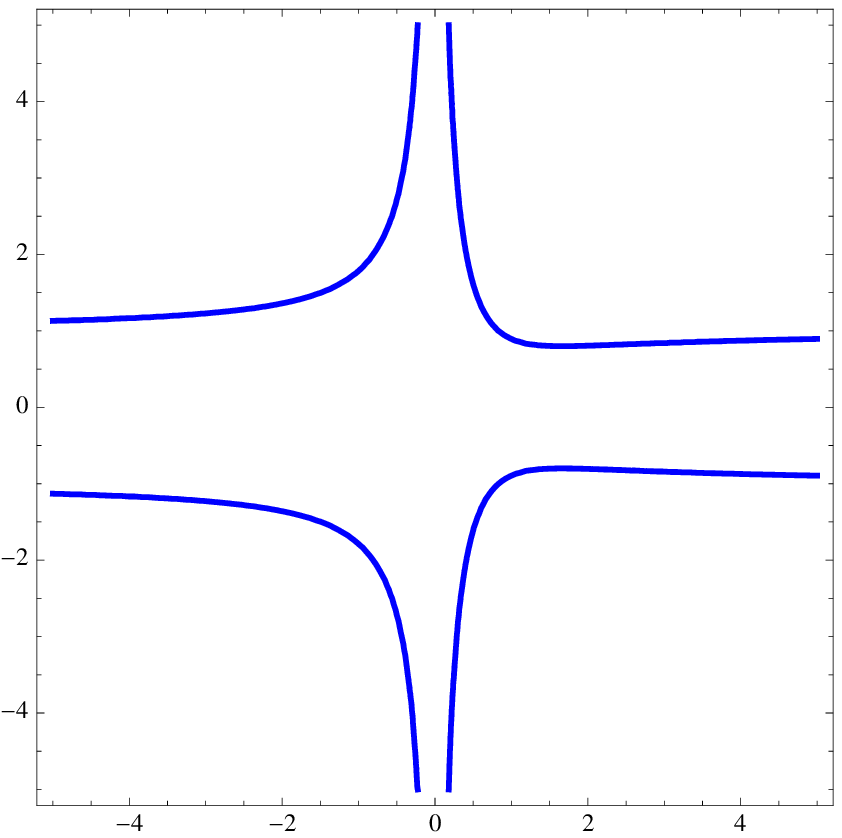, width=3cm, height=3cm}
\epsfig{file=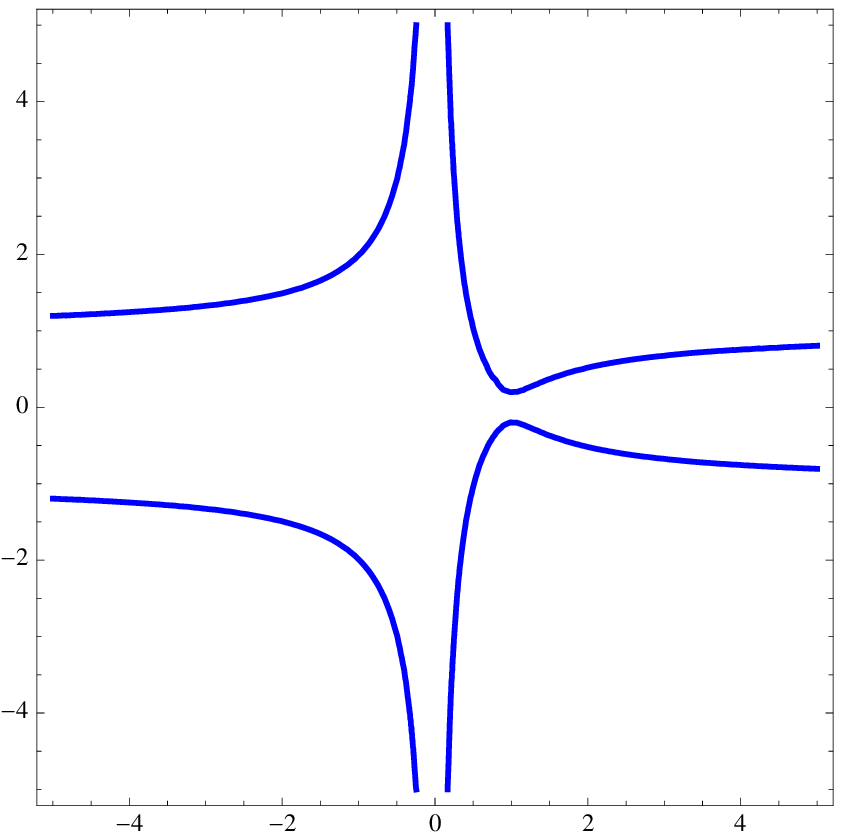, width=3cm, height=3cm}
\epsfig{file=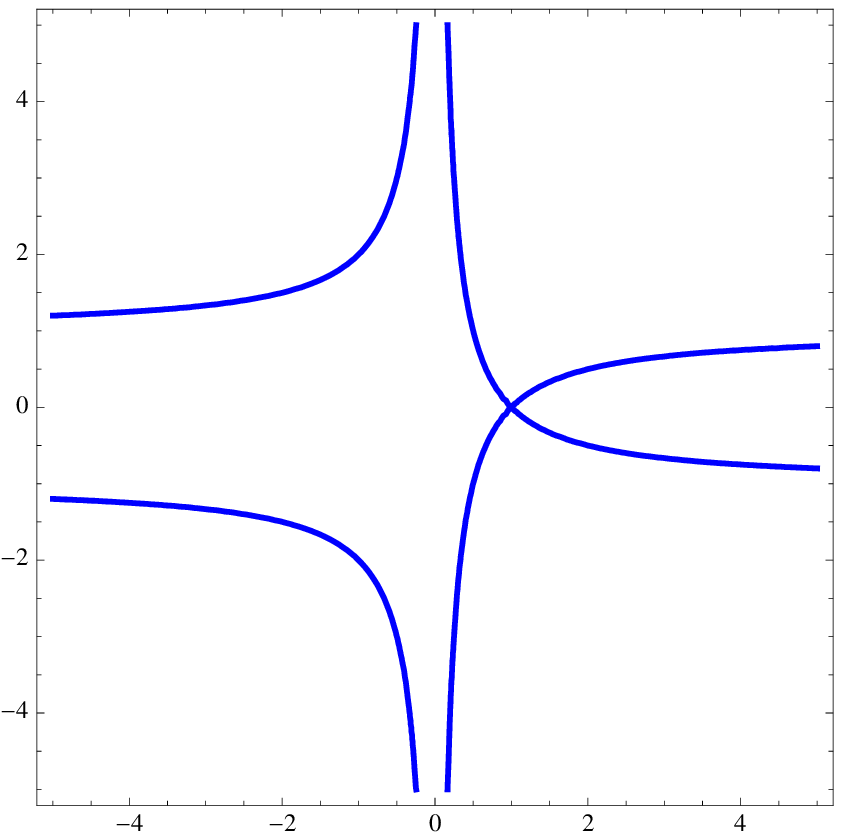, width=3cm, height=3cm}
\epsfig{file=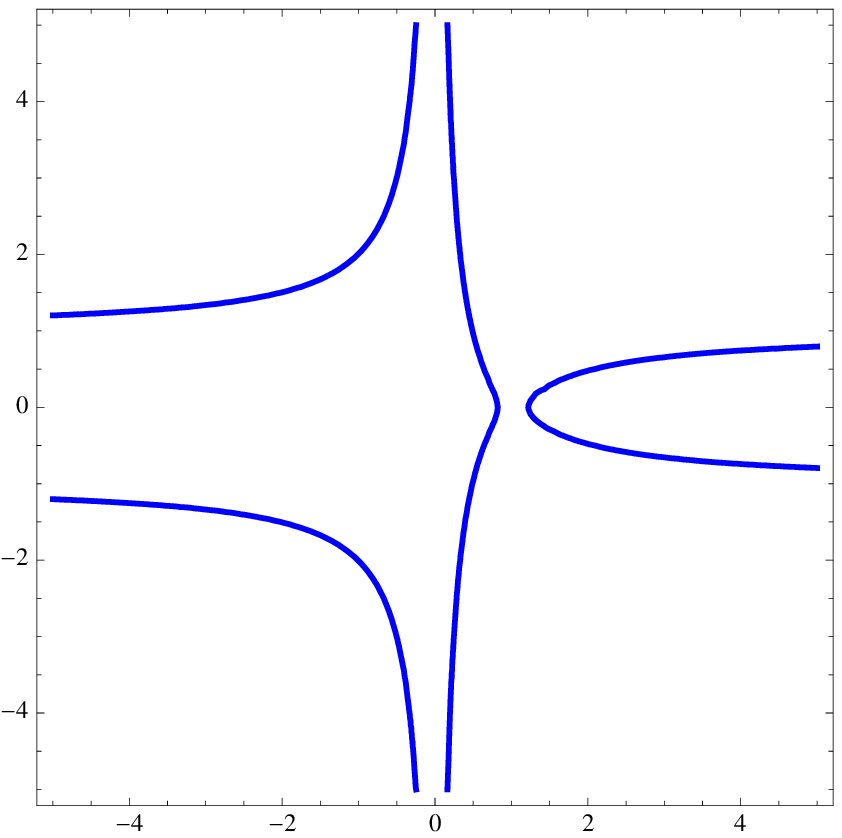, width=3cm, height=3cm}
\epsfig{file=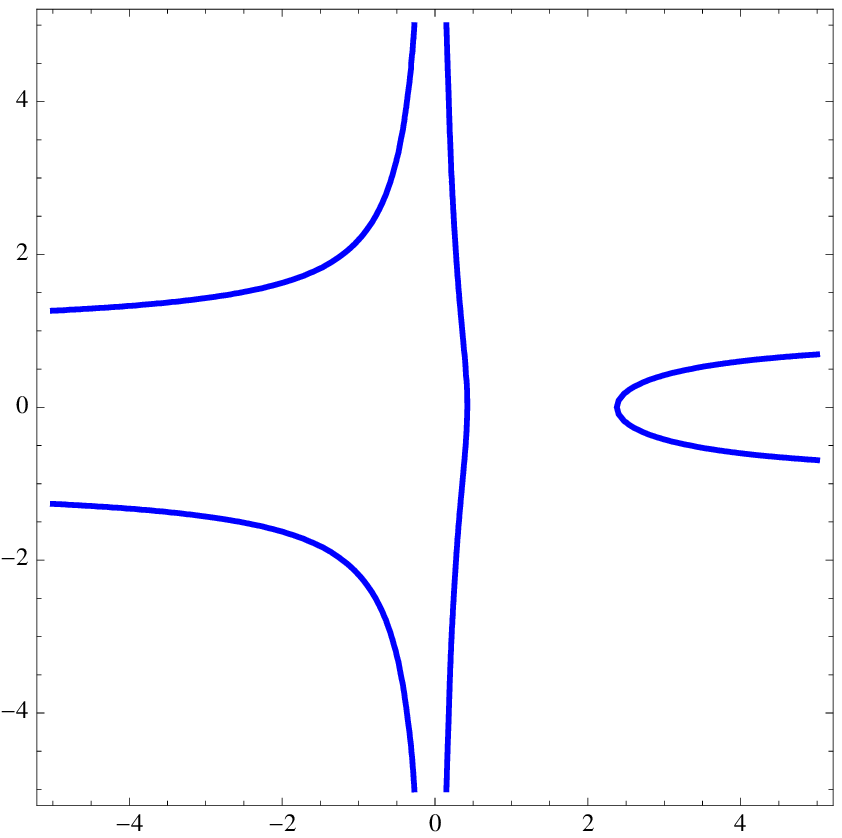, width=3cm, height=3cm}
\caption{The Penner algebraic curve for values of $t = -0.2, -0.01, 0, +0.01, +0.2$, from left to right, respectively. Notice that the algebraic curve is singular for $t=0$.}
}
\FIGURE[ht]{
\label{pstokescurve}
\centering
\epsfig{file=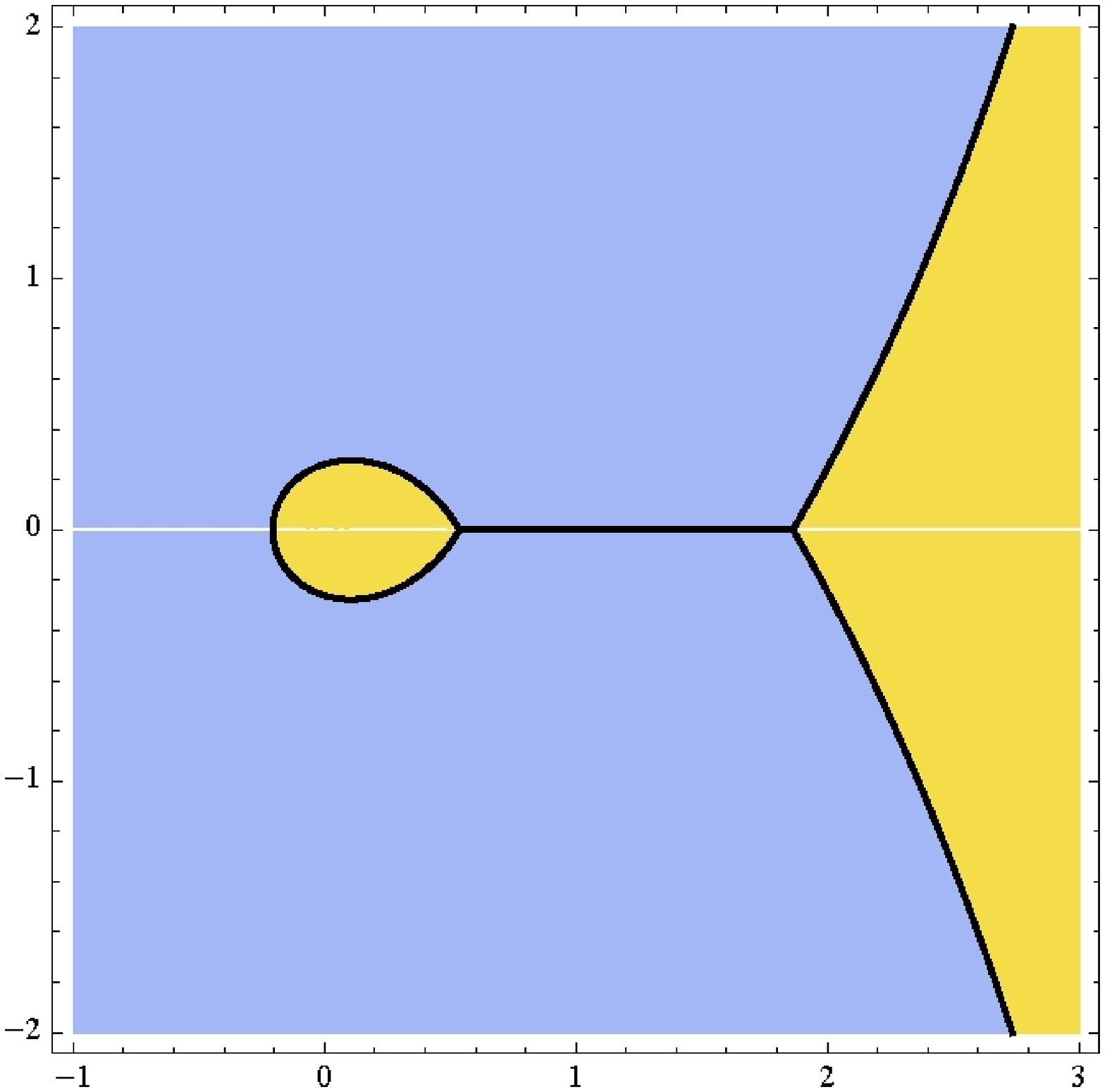, width=9.5cm, height=9.5cm}
\caption{The real part of the Penner holomorphic effective potential in the complex $z$--plane, for $t=0.1$. In blue is the region $\re\, V_{\mathrm{h;eff}}^{\mathrm{P}} (z) < 0$, in yellow the region $\re\, V_{\mathrm{h;eff}}^{\mathrm{P}} (z) > 0$, and the black lines correspond to the Stokes lines $\re\, V_{\mathrm{h;eff}}^{\mathrm{P}} (z) = 0$ (which also include the cut of the spectral curve). The white cuts corresponds to the logarithmic branch cuts.}
}

\subsection{The Chern--Simons Matrix Model}\label{sec:csmm}

We now turn to the Chern--Simons, or Stieltjes--Wigert, matrix model. As we previously stated this model is particularly interesting for its relation, via a large $N$ duality, to topological string theory on the resolved conifold \cite{gv98b}. The SU$(N)$ Chern--Simons gauge theory on a generic three--manifold has been realized as a matrix model in \cite{m02}; see \cite{m04b} for a review.. Here, we shall focus on the resolved conifold case, where the partition function of SU$(N)$ Chern--Simons gauge theory on $\BS^3$ is, up to a factor, given by the Stieltjes--Wigert matrix model \cite{t02} defined by the potential $V_{\mathrm{SW}} (z) = \frac{1}{2} \left( \log z \right)^2$. To be precise, the Chern--Simons partition function relates to the Stieltjes--Wigert partition function by the simple expression $Z_{\mathrm{CS}} = \exp \left( - \frac{t}{12} \left( 7N^2-1 \right) \right) Z_{\mathrm{SW}}$, so that the corresponding free energies equate as
\be
F_{\mathrm{CS}} = - \frac{7}{12}\, \frac{t^3}{g_s^2} + \frac{t}{12} + F_{\mathrm{SW}}.
\ee
\noindent
For a review of the main features of this matrix model, including saddle--point methods and orthogonal polynomial analysis, we refer the reader to, \textit{e.g.}, \cite{m04}.

Let us start by computing the partition function $Z_{\mathrm{CS}}$ using orthogonal polynomials---as we shall see one may regard the Stieltjes--Wigert matrix model as a q--deformation, in the quantum group sense, of the Gaussian matrix model. The logarithmic measure $\rmd \mu(z) = \rme^{- \frac{1}{2g_s} \left( \log z \right)^2} \frac{\rmd z}{2\pi}$ is well--known in the literature precisely because it leads to so--called Stieltjes--Wigert orthogonal polynomials,
\be
p_n (z) = (-1)^n q^{n^2 + \frac{n}{2}} \sum_{k=0}^n {{n}\brack{k}}_q\, q^{\frac{k(k-n)}{2}-k^2} \left( - q^{-\frac{1}{2}} z \right)^k, \qquad h^{\mathrm{SW}}_n = q^{\frac{7}{4}n(n+1)+\frac{1}{2}} \left[ n \right]_{q}! \sqrt{\frac{g_s}{2\pi}},
\ee
\noindent
where we have introduced
\be
q = \rme^{g_s}, \qquad \left[ n \right]_q = q^{\frac{n}{2}} - q^{-\frac{n}{2}}, \qquad {{n}\brack{m}}_q = \frac{\left[ n \right]_q!}{\left[ m \right]_q! \left[ n-m \right]_q!}.
\ee
\noindent
With this information at hand, one may now explicitly compute the Stieltjes--Wigert partition function from definition
\be
Z_{\mathrm{SW}} = \prod_{n=0}^{N-1} h^{\mathrm{SW}}_n = \left( \frac{g_s}{2\pi} \right)^{\frac{N}{2}} q^{\frac{N}{12} (7N^2-1)} \prod_{n=0}^{N-1} \left[ n \right]_{q}!.
\ee
\noindent
A simple glance at (\ref{opgaussian}) immediately shows that, up to normalization, one may indeed regard the Stieltjes--Wigert matrix model as a q--deformation of the Gaussian matrix model, at least at the level of the partition functions. One may further define the q--deformed, or quantum Barnes function as
\be
G_q (N+1) = \prod_{n=0}^{N-1} \left[ n \right]_{q}!,
\ee
\noindent
so that the Stieltjes--Wigert partition function is simply $Z_{\mathrm{SW}} = \left( \frac{g_s}{2\pi} \right)^{\frac{N}{2}} q^{\frac{N}{12} (7N^2-1)} G_q (N+1)$. These expressions may then be used to address the large $N$ topological expansion of the Stieltjes--Wigert matrix model. Standard use of orthogonal polynomial techniques \cite{biz80}, as described, \textit{e.g.}, in \cite{m04}, yield\footnote{Notice that, unlike in the previous example of the Penner model, in here we have \textit{not} normalized the Chern--Simons free energy by the Gaussian free energy.}
\bea
F_0^{\mathrm{CS}} (t) &=& \frac{t^3}{12} - \frac{\pi^2 t}{6} - \mathrm{Li}_3 \left( \rme^{-t} \right) + \zeta(3), \\
F_1^{\mathrm{CS}} (t) &=& - \frac{t}{24} + \frac{1}{12}\, \mathrm{Li}_{1} \left( \rme^{-t} \right) + \zeta'(-1), \\
F_g^{\mathrm{CS}} (t) &=& \frac{B_{2g} B_{2g-2}}{2g \left( 2g-2 \right) \left( 2g-2 \right)!} + \frac{B_{2g}}{2g \left( 2g-2 \right)!}\, \mathrm{Li}_{3-2g} \left( \rme^{-t} \right), \qquad g \ge 2,
\label{csexp}
\eea
\noindent
where $\mathrm{Li}_{p} (z)$ is the polylogarithm of index $p$,
\be
\mathrm{Li}_{p} (z) = \sum_{n=1}^{+\infty} \frac{z^n}{n^p}.
\ee
\noindent
At genus $g \ge 2$ of course the topological expansions of Chern--Simons and Stieltjes--Wigert perturbative free energies coincide, $F_g^{\mathrm{CS}} (t) = F_g^{\mathrm{SW}} (t)$. Furthermore, after analytical continuation $g_s \to \rmi \bar{g}_s$, the free energies $F_g^{\mathrm{CS}} (t)$ coincide with the free energies of topological strings on the resolved conifold (\ref{rc}), once one identifies 't~Hooft coupling and K\"ahler parameter. Finally, notice that all free energies with $g \ge 1$ diverge when $t \to 0$ which corresponds to $\rme^{-t} \to 1$; with this second variable the natural one as the divergences are associated to the singular point of the (negative index) polylogarithm, $\mathrm{Li}_{-p} (1)$. It is then quite natural to consider the double--scaling limit, approaching the critical point $\rme^{-t_c} = 1$, as
\be
\rme^{-t} \to 1, \quad g_s \to 0, \qquad \mu = \frac{\rme^{-t_c} - \rme^{-t}}{g_s} \quad \mathrm{fixed},
\ee
\noindent
in which case one again obtains the $c=1$ string at self--dual radius
\be
g_s^{2g-2}\, F_g^{\mathrm{CS}} (t) \to \frac{B_{2g}}{2g(2g-2)}\, \mu^{2-2g}, \qquad g \ge 2.
\ee

Finally we address the spectral curve and holomorphic effective potential for the Stieltjes--Wigert matrix model by making use of saddle--point techniques. With potential $V_{\mathrm{SW}}(z) = \frac{1}{2} \left( \log z \right)^2$ and $V_{\mathrm{SW}}'(z) = \frac{1}{z} \log z$ one must be a bit careful in applying (\ref{genus0resolvent}) to compute the resolvent: indeed, the deformation of the contour around the cut, $\CC=[a,b]$, must now be done differently due to the logarithmic branch--cut. Instead of capturing the pole at $z$ and the pole at $\infty$, this time around one captures the pole at $z$ and the branch cut along the negative real axis (zero included); we refer the reader to \cite{m04} for further details. The endpoints of the cut are
\be
a,b = 2 \rme^{2t} - \rme^t \pm 2 \rme^{\frac{3t}{2}} \sqrt{\rme^t - 1},
\ee
\noindent
while the one--form on the spectral curve reads
\be\label{spc}
y(z)\, \rmd z = \frac{2}{z}\, \log \frac{1 + \rme^{-t} z + \sqrt{\left( 1 + \rme^{-t} z \right)^2 - 4z}}{2 \sqrt{z}}\, \rmd z,
\ee
which coincides with the one--form $\log Y(Z)\, \frac{\rmd Z}{Z}$ on the mirror curve $\CH(Z,Y)=0$ of the resolved conifold, written in terms of the $\BC^*$ variables $Z=\rme^z$ and $Y=\rme^y$. One further computes the holomorphic effective potential as
\bea\label{vecs}
V_{\mathrm{h;eff}}^{\mathrm{CS}} (z) &=& - \frac{1}{2} \log^2 z + \log^2 \xi - 2\, \Big( \log \xi\, \log (1-\rme^{-t} \xi) + \mathrm{Li}_2 (1-\xi) + \mathrm{Li}_2 (\rme^{-t} \xi) \Big) - V_0 \nonumber \\
&=& - \frac{1}{2} \log^2 z + \log^2 \xi - 2\, \Big( \log \xi\, \log (1-\rme^{-t} \xi) + \mathrm{Li}_2 (\rme^{-t} \xi) - \mathrm{Li}_2 \left( \xi \right) - \nonumber \\
&& - \log \left( 1-\xi\right)  \log \xi+\frac{\pi^2}{6} \Big) - V_0,
\eea
\noindent
where equality holds due to the Euler's reflection formula for dilogarithms
\be
\mathrm{Li}_2 (\xi) + \mathrm{Li}_2 (1-\xi) = \frac{\pi^2}{6} - \log \left( 1-\xi \right) \log\xi.
\ee
\noindent
In here we have set
\be
\xi (z) = \frac{1 + \rme^{-t} z + \sqrt{\left( 1 + \rme^{-t} z \right)^2 - 4z}}{2}
\ee
\noindent
to simplify notation\footnote{Using this variable, the spectral curve is also compactly re--written as $y(z) = \frac{2}{z} \log \frac{\xi}{\sqrt{z}}$.} and we have defined
\bea
V_0 &=& -\frac{1}{2} \log^2 \left( 2\, \rme^{2t} - \rme^t + 2\, \rme^{\frac{3t}{2}} \sqrt{\rme^t-1} \right) + \log^2 \left( \rme^t + \rme^{\frac{t}{2}} \sqrt{\rme^t-1} \right) - \nonumber \\
&&
-2 \log \left(\rme^t+\rme^{\frac{t}{2}} \sqrt{\rme^t-1}\right) \log \left(-\rme^{-\frac{t}{2}} \sqrt{\rme^t-1}\right) - 2\, \mathrm{Li}_2 \left( 1-\rme^t-\rme^{\frac{t}{2}} \sqrt{\rme^t-1} \right) - \nonumber \\
&&
-2\, \mathrm{Li}_2 \left( 1+\rme^{-\frac{t}{2}} \sqrt{\rme^t-1} \right),
\eea
\noindent
to ensure that the result is normalized such that $V_{\mathrm{h;eff}}^{\mathrm{CS}} (b) = 0$. In Figures \ref{csalgcurve} and \ref{csstokescurve} we plot the Stieltjes--Wigert algebraic curve for different values of $t$, as well as the real value of the holomorphic effective potential in the complex plane. Given that $\mathrm{Li}_2 (z)$ is the standard dilogarithm function, with its intricate branch structure, it is not too hard to realize that the structure of Stokes lines of the present effective potential is now much more complicated than usual.

\FIGURE[ht]{
\label{csalgcurve}
\centering
\epsfig{file=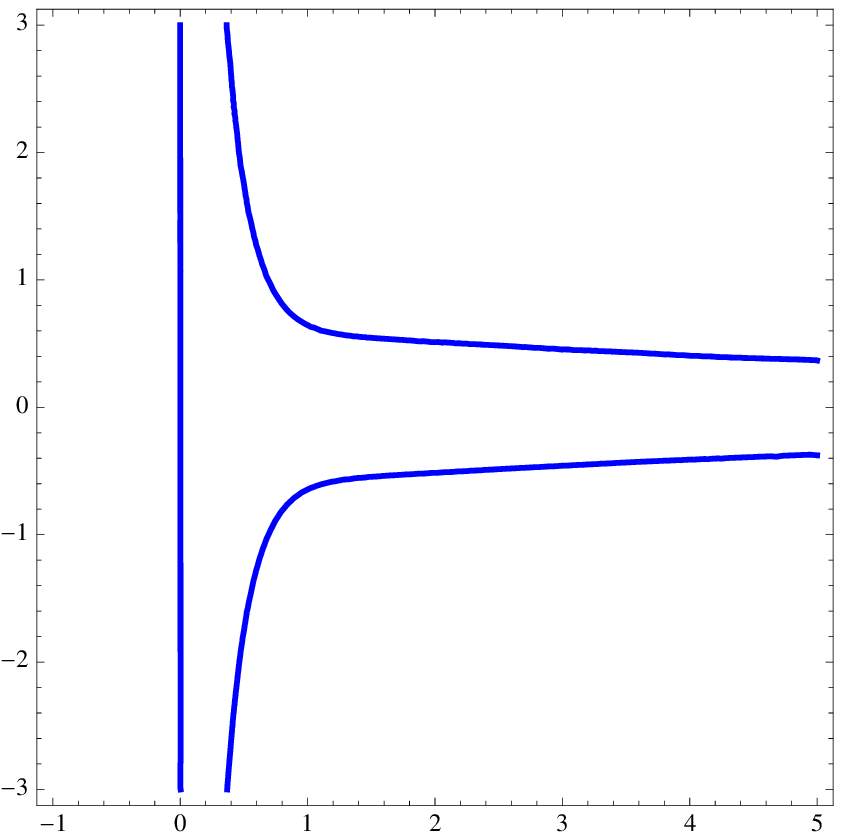, width=3cm, height=3cm}
\epsfig{file=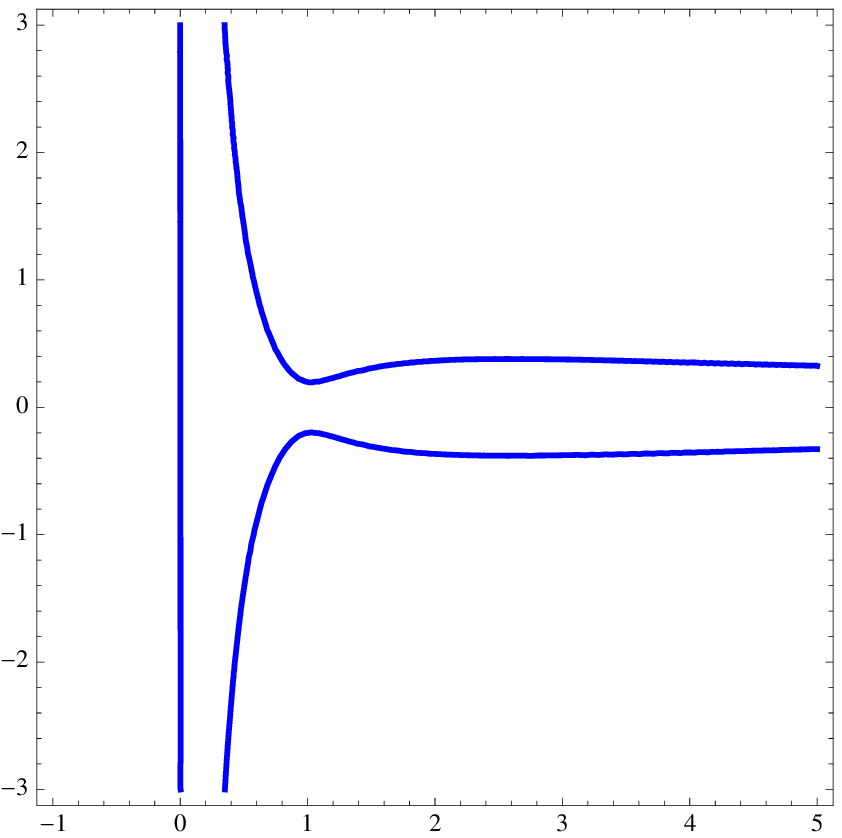, width=3cm, height=3cm}
\epsfig{file=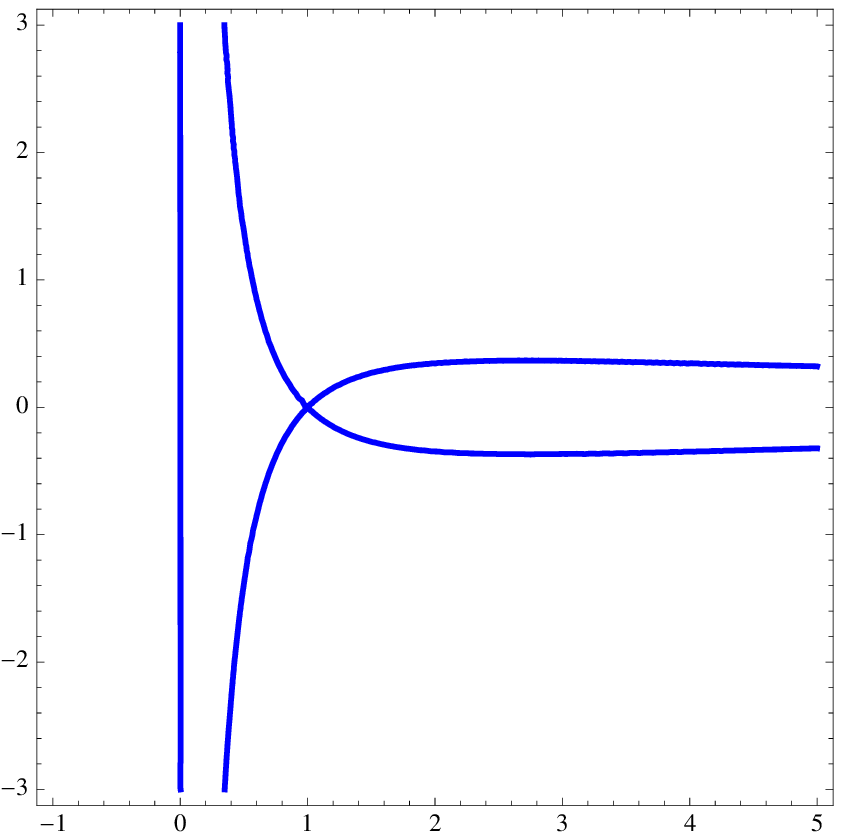, width=3cm, height=3cm}
\epsfig{file=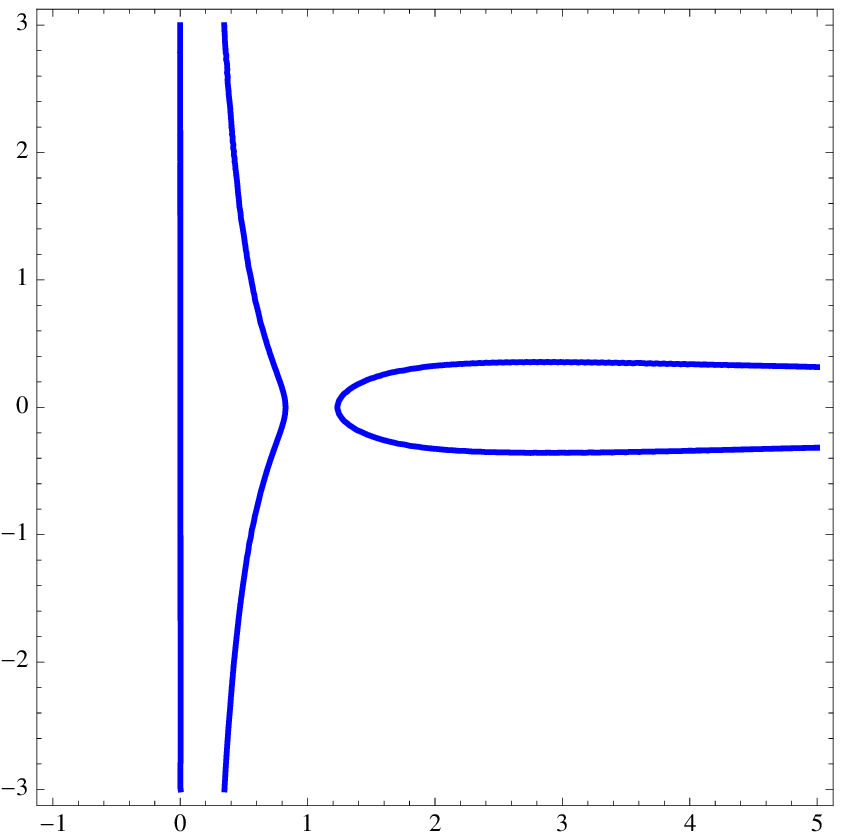, width=3cm, height=3cm}
\epsfig{file=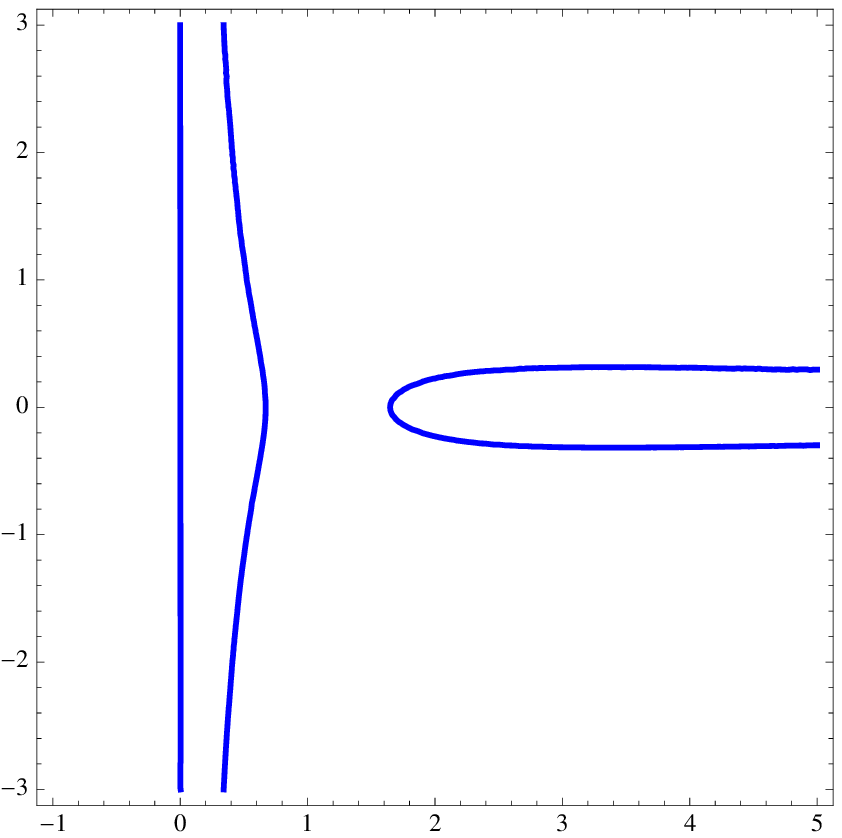, width=3cm, height=3cm}
\caption{The Stieltjes--Wigert algebraic curve for values of $t = -0.1, -0.01, 0, +0.01, +0.05$, from left to right, respectively. Notice that the algebraic curve is singular for $t=0$.}
}
\FIGURE[ht]{
\label{csstokescurve}
\centering
\epsfig{file=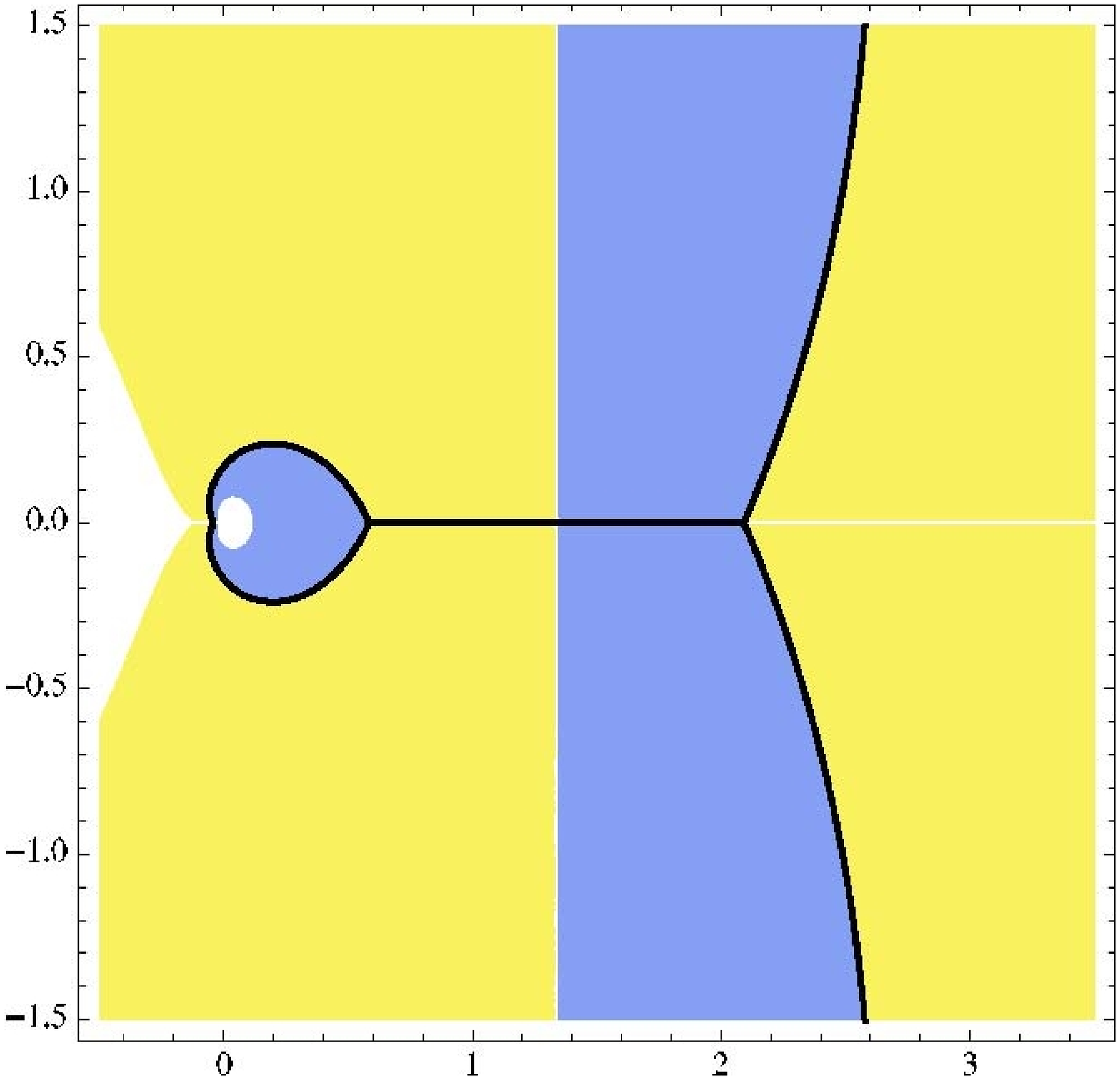, width=9.5cm, height=9.5cm}
\caption{The real part of the Stieltjes--Wigert holomorphic effective potential in the complex $z$--plane, for $t=0.1$. The black lines correspond to the Stokes lines $\re\, V_{\mathrm{h;eff}}^{\mathrm{CS}} (z) = 0$ (which also include the cut of the spectral curve). The white cuts and regions correspond to the logarithmic and dilogarithmic branch cuts. Because of the choice of principal sheets in \textsf{\textit{Mathematica}} the colored regions are now not so clear. Akin to the Penner model, the region in yellow, to the right of the vertical black lines, and the region in blue, inside the ``closed bubble'', have $\re\, V_{\mathrm{h;eff}}^{\mathrm{CS}} (z) > 0$, in the principal sheet. The rest is $\re\, V_{\mathrm{h;eff}}^{\mathrm{CS}} (z) < 0$.}
}

\subsection{Double--Scaling Limit and $c=1$ Behavior}

We have just seen that the Gaussian, Penner and Chern--Simons free energies admit rather simple double--scaling limits to the $c=1$ string at self--dual radius. In the Chern--Simons case this relates to our earlier discussion in section \ref{sec:gwt}, where we pointed out that, at the conifold point of moduli space, the A--model may still be characterized by its critical behavior in the double--scaling limit (\ref{dsl}). Free energies of the topological string reduce, in this situation, to free energies of the $c=1$ string. Let us now briefly discuss, in the example of the Chern--Simons matrix model, how one may also study the destiny of the open string sector, \textit{i.e.}, of the matrix model correlators $W_h (p_1, \ldots, p_h)$ introduced in (\ref{obj}), in this $c=1$ double--scaling limit.

Introducing a parameter $\zeta$ as
\be
\rme^{-t} \equiv 1 - \zeta,
\ee
\noindent
the conifold point of the Chern--Simons model is thus located at  $\zeta=0$. The expansion of the branch points of the spectral curve (\ref{spc}) near the conifold point yields
\be
a,b = 1 \pm 2 \zeta^{\frac{1}{2}} + \cdots, 
\ee
\noindent
in which case it is natural to scale also the $z$ variable in the spectral curve as
\be
z = 1 + \zeta^{\frac{1}{2}}\, s
\ee
\noindent
in order to appropriately zoom into the critical region. In here, $s$ is the double--scaled open coordinate. In these variables the Chern--Simons one--form $y(z)\, \rmd z$, at criticality, scales to
\be\label{concurve}
y(z)\, \rmd z \rightarrow y(s)\, \rmd s = \zeta\, \sqrt{s^2-4}\, \rmd s
\ee
\noindent
which one immediately recognizes as the one--form of the Gaussian matrix model. The interesting point is that, in the same variables, also the two--point correlator $W_0(p,q)$ reduces to the Gaussian one
\be
W_{0,2}(p,q)\, \rmd p\, \rmd q \rightarrow \frac{1}{2}\, \frac{1}{(s-t)^2} \left( \frac{st-4}{\sqrt{\left( s^2-4 \right) \left( t^2-4 \right)}} - 1 \right) \rmd s\, \rmd t.
\ee
\noindent
In fact, there is a property of the topological recursion, proved in \cite{eo07}, which states that one may either first compute matrix model amplitudes and then take their double--scaling limits, or else recursively compute amplitudes directly from the double--scaled curve, the result being the same (\textit{i.e.}, the operations commute). As such, in the $c=1$ double--scaling limit Chern--Simons open correlators will all reduce to Gaussian open correlators
\be\label{opends}
W^{\mathrm{CS}}_{g,h} \left( z_1, \ldots, z_h \right) \rmd z_1 \cdots \rd z_h \rightarrow \zeta^{2-2g-h}\, W^{\mathrm{G}}_{g,h} \left( s_1, \ldots, s_h \right) \rd s_1 \cdots \rd s_h.
\ee
\noindent
Now recall that open topological string amplitudes may be computed from the matrix model correlators $W_{g,h} (z_1, \ldots, z_h)$ as \cite{av00, akv01, m06, bkmp07}
\be
A^{(g)}_h (p_1, \ldots, p_h) = \int^{p_1} \cdots \int^{p_h} \rmd z_1 \cdots \rmd z_h\, W_{g,h} (z_1, \ldots, z_h),
\ee
\noindent
where the $\{ p_i \}$ are the open string parameters which parametrize the moduli space of the brane. As such, (\ref{opends}) shows how, near the conifold point, open amplitudes of topological strings on the resolved conifold reduce to Gaussian amplitudes. This is actually generic for topological string theory near the conifold point \cite{bkmp08}. We shall now relate these Gaussian amplitudes with open amplitudes in the dual $c=1$ model.

We first need to recall some results in non--critical strings, holographically duals to matrix models. We are interested in minimal models obtained by coupling 2d gravity to minimal $(p,q)$ matter models, with central charge $c_{p,q} = 1 - 6(p-q)^2/pq$. The coupling to gravity leads to the appearance of the Liouville field, $\phi$, with world--sheet action
\be
S_{\mathrm{L}} = \int \frac{\rmd^2 z}{4\pi}\, \sqrt{g} \left( \partial\phi^2 + Q R \phi + 4 \pi \mu_{\mathrm{L}} \rme^{2 b \phi} \right),
\ee
\noindent
where $\mu_{\mathrm{L}}$ is the bulk cosmological constant. The central charge of the Liouville sector is $c_{\mathrm{L}} = 1+6 Q^2$, and the parameter $b$ above relates to the background charge $Q$ as $Q = b + 1/b$. The bosonic string requirement that the total central charge of Liouville theory plus minimal matter equals $c=26$ eventually fixes $b=\sqrt{\frac{p}{q}}$. There are two distinct types of boundary conditions in Liouville theory \cite{fzz00, zz01}. There is a one--parameter family of Neumann boundary conditions, the so--called FZZT branes, parameterized by the boundary cosmological constant $\mu_{\mathrm{B}}$, usually expressed in terms of a parameter $s$ as
\be
\mu_{\mathrm{B}} = \sqrt{\frac{\mu_{\mathrm{L}}}{\sin \left( \pi b^2 \right)}} \cosh \left( \pi b s \right).
\ee 
\noindent
Besides FZZT branes, there are also ZZ branes, associated to Dirichlet boundary conditions. These correspond to a two--parameter family, parameterized by the pair of integers $(m,n)$, and are localized at $\phi=\infty$. At the quantum level FZZT and ZZ boundary conditions, or, respectively, the $\langle B_s|$ and $\langle B_{(m,n)}|$ boundary states, are related as \cite{h01, m0305, ss03}
\be\label{rela}
\langle B_{(m,n)}| = \langle B_{s(m,n)}| - \langle B_{s(m,-n)}|, \qquad \mathrm{with} \qquad s(m,n) = \rmi \left( \frac{m}{b} + bn \right).
\ee

Both types of branes have been given a geometrical interpretation in terms of a complex curve, in \cite{ss03}. This is accomplished by introducing the variables
\be\label{xycurve}
x = \mu_{\mathrm{B}} \sim \cosh \left( \pi b s \right), \qquad y = \frac{\p}{\p\mu_{\mathrm{B}}} Z^{\mathrm{FZZT}} \sim \sinh \left( \frac{\pi s}{b} \right).
\ee
\noindent
Considered as complex variables, the coordinates $\{x,y\}$ define an algebraic curve $F(x,y)=0$ embedded into $\BC^2$, which is identified with the spectral curve of the dual matrix model (\textit{i.e.}, a double--scaled hermitian one--matrix model) \cite{ss03}. The FZZT brane disk partition function may be equivalently written as the line integral of the one--form $y\, \rmd x$ as
\be
Z^{\mathrm{FZZT}} (\mu_{\mathrm{B}}) = \int^{\mu_{\mathrm{B}}} \rmd x\, y.
\ee
\noindent
Analogously the $h$--point matrix model correlators (\ref{obj}) are identified with open amplitudes with FZZT boundary conditions, \textit{e.g.}, the two--point function $W_{0,2} \left( p,q \right)$ is identified as the annulus amplitude for FZZT branes and so on. The ZZ brane disk partition function is instead defined as the line integral of $y\, \rmd x$ over a closed contour
\be
Z^{\mathrm{ZZ}}_{(m,n)} = \oint_{\gamma_{m,n}} \rmd x\, y,
\ee
\noindent
where $\gamma_{m,n}$ is a non--contractible contour conjugate to a ``pinched cycle'', starting and ending at the singular point $x_{(m,n)} = x \left( s(m,n) \right)$ and $y_{(m,n)} = y \left( s(m,n) \right)$ \cite{ss03}.

The case of $c=1$ is a bit more subtle since one has to consider the singular $b\to 1$ limit. It is first necessary to introduce the renormalized couplings
\be\label{renorm}
\mu_{c=1} = \lim_{b\to 1} \left( \pi \left( 1-b^2 \right) \mu_{\mathrm{L}} \right), \qquad 
\mu_{\mathrm{B},c=1} = \lim_{b\to 1} \left( \pi \left( 1-b^2 \right) \mu_{\mathrm{B}} \right).
\ee
\noindent
Furthermore, an appropriate subtraction is required in order to define the FZZT disk partition function. This may be expressed in terms of the one--form $w(s)\, \rmd \mu_{\mathrm{B},c=1} (s)$ with
\be\label{renw}
w(s) \equiv \lim_{b \to 1} \left( \frac{\p_{\mu_{\mathrm{B}}} Z^{\mathrm{FZZT}}}{\pi \left( 1-b^2 \right)}  + \frac{4}{\pi}\, Z_{\mathrm{D}}\,\mu_{\mathrm{B}} \right),
\ee
\noindent
where $Z_{\mathrm{D}}$ is the disk partition function in the $c=1$ CFT. The relevant $c=1$ curve then reads
\be\label{rencurve}
x(s) = \mu_{{\mathrm{B}},c=1} (s) = \sqrt{\mu_{c=1}} \cosh \left( \pi s \right), \qquad y(s) = w(s) = - D \sqrt{\mu_{c=1}}\, \pi s \sinh \left( \pi s \right),
\ee
\noindent
where $D$ is some constant. The identification of the curve (\ref{rencurve}), arising from CFT considerations, with the curve of the dual matrix model is another delicate point. Here, the relevant matrix model is a double--scaled version of Matrix Quantum Mechanics (MQM) with a Sine--Liouville perturbation. It is know for quite some time \cite{bipz78} that the singlet sector of this MQM can be reduced to a system of free fermions, in an inverted harmonic oscillator. In the semiclassical limit the ground state of this system is completely determined by the shape of the Fermi sea, which can be parameterized in terms of an uniformization parameter $\tau$ as \cite{k91}
\be\label{mqmc}
x(\tau)=\sqrt{2\mu}\cosh(\tau), \qquad y(\tau)=\sqrt{2\mu} \sinh(\tau),
\ee
\noindent
where $\mu$ denotes the Fermi level. In analogy with the $c<1$ case, one would like to identify the above MQM curve with the CFT curve (\ref{rencurve}). However, these two curves are clearly distinct. A solution to this puzzled has been offered in \cite{ak04, a04b, kk04}, where it was proposed that one should instead identify $w(s)$ with the resolvent, rather than directly with the spectral curve of the dual matrix model. The spectral curve can then be extracted, following a very standard matrix model procedure, from the discontinuity of $w(s)$,
\be
\rho (s) \equiv -\frac{1}{2\pi\rmi} \left( w (s+\rmi\epsilon) - w(s-\rmi\epsilon) \right) = - D \sqrt{\mu_{c=1}}\, \sinh \left( \pi s \right).
\ee
\noindent
Clearly the new CFT curve, defined as
\be
x(s) = \mu_{\mathrm{B},c=1}(s), \qquad y(s) = \rho(s),
\ee
\noindent
agrees with the one--matrix model spectral curve (\ref{mqmc}) after an appropriate identification of parameters. The FZZT brane partition function can then be obtained from the line integral of the one--form $y(\tau)\, \partial_{\tau}x(\tau)\, \rmd\tau$, while the ZZ brane partition function can be defined  
by the following closed integral on the MQM curve \cite{a04b}
\be\label{zz1}
Z^{\mathrm{ZZ}}_{(n,1)} = \rmi \oint_{\gamma_{n}} \rmd x\, y = \rmi \int_{-\rmi\pi n}^{\rmi\pi n} \rmd\tau\, \partial_\tau x(\tau)\, y(\tau) = 2 \pi n \mu, \qquad n \in \BZ,
\ee
\noindent
corresponding to a $(n,1)$ ZZ brane partition function. Indeed, it has been shown that only an one--parameter set of the $c=1$ ZZ branes may be identified in the dual MQM \cite{a04a}.

Let us further notice that the above matrix quantum mechanics spectral curve, (\ref{mqmc}), with the uniformization parameter $\tau$, is just an infinite covering of the hyperboloid
\be
x^2-y^2=2\mu,
\ee
\noindent
which is precisely the spectral curve of the Gaussian matrix model. In particular, this explains how open matrix model correlators of the Gaussian model, $W^{\mathrm{G}}_{g,h}$, get identified with D--brane amplitudes with FZZT boundary conditions in the $c=1$ model at self--dual radius. Indeed, in \cite{bh06} it was checked that the double--scaled Gaussian correlators are related to macroscopic loop operators in the $c=1$ theory. Finally, the limit (\ref{opends}) shows that topological string amplitudes with toric--brane boundary conditions reduce to $c=1$ amplitudes for FZZT branes. Hence, and as already pointed out in a related context in \cite{m06}, toric branes reduce to FZZT branes in the double--scaling limit, at the conifold point.

\section{Nonperturbative Effects, Large Order and the Borel Transform}\label{sec:lomm}

We may now turn to the study of the asymptotic perturbative expansions for the free energies of the matrix models and topological strings we are interested in. In particular, we shall perform a detailed Borel analysis of each case, and thus understand what type of nonperturbative effects control the large--order behavior of the distinct perturbative expansions.

\subsection{The Gaussian Matrix Model and $c=1$ Strings}\label{sec:ga}

Let us begin with the Gaussian matrix model. The genus expansion of its free energy, (\ref{gs}), is clearly an asymptotic expansion with  $F^{\mathrm{G}}_g\sim (2g-3)!$, given the growth of Bernoulli numbers as $B_{2g}\sim (2g)!$. Recalling our discussion in section \ref{sec:ds}, we may then consider the Borel transform of the divergent Bernoulli sum---\textit{i.e.}, restricting to genus $g \ge 2$---and obtain
\be
\CB [F_{\mathrm{G}}] (\xi) = \sum_{g=2}^{+\infty} \frac{F_g^{\mathrm{G}} (t)}{(2g-3)!}\, \xi^{2g-2} = - \frac{1}{12} + \frac{t^2}{\xi^2} - \frac{1}{4}\, \frac{1}{\sinh^2 \left( \frac{\xi}{2t} \right)}.
\ee
\noindent
This function has no poles on the positive real axis, for real argument (the genus expansion (\ref{gs}) is an alternating series). As such, one can define its inverse Borel transform\footnote{Notice that since $F^{\mathrm{G}}_g\sim (2g-3)!$ the inverse of the Borel transform will now have an extra factor of $\frac{1}{s}$ with respect to the definition of section \ref{sec:ds}, which dealt with asymptotic growths of the type $\sim (\beta n)!$.}
\be\label{gsc}
\widetilde{F}_{\mathrm{G}} (g_s) = -\frac{1}{4} \int_0^{+\infty} \frac{\rmd s}{s} \left( \frac{1}{\sinh^2 \left( \frac{g_s}{2 t}\, s \right)} - \left( \frac{2 t}{g_s} \right)^2 \frac{1}{s^2} + \frac{1}{3} \right) \rme^{-s},
\ee
\noindent
providing a nonperturbative completion for the asymptotic expansion of the free energy in the Gaussian matrix model. It is quite interesting to notice that, upon the trivial change of variables $s \to \sigma = \frac{g_s}{2t}\, s$, this expression precisely coincides with the one--loop effective Lagrangian for a charged scalar particle in a constant self--dual electromagnetic field (of magnetic type) introduced in section \ref{sec:se}. Comparing with (\ref{sd}) we see that in here 
\be
\gamma = \frac{2 e \CF}{m^2} = \frac{1}{N}.
\ee

If one instead considers imaginary string coupling, $\bar{g}_s = \rmi g_s$, the asymptotic expansion (\ref{gsc}) will coincide with the one--loop effective Lagrangian corresponding to a self--dual background of electric type, which is exactly the same as that for $c=1$ strings at self--dual radius. This time around the perturbative series is not alternating in sign, and the Borel integral representation
\be
\widetilde{F}_{c=1} (\bar{g}_s) = \frac{1}{4} \int_0^{+\infty} \frac{\rmd \sigma}{\sigma} \left( \frac{1}{\sin^2 \sigma} - \frac{1}{\sigma^2} - \frac{1}{3} \right) \rme^{- \frac{2 t \sigma}{\bar{g}_s}}
\ee
\noindent
has an integrand with poles on the positive real axis, in principle leading to ambiguities in the reconstruction of the function, as discussed in an earlier section. However, we may now use the analogy of this expression to the results in section \ref{sec:se} in order to use the unitarity prescription to perform an unambiguous calculation---which basically yields an $\rmi \epsilon$ prescription which reduces the imaginary part of the integral to a sum over the residues of its integrand. The nonperturbative imaginary contribution to the above free energy is thus simple to compute as\footnote{Notice that the pole at $\sigma = 0$ has vanishing residue.}
\bea
\im\, \widetilde{F}_{c=1} (\bar{g}_s) &=& \frac{\pi}{4}\, \sum_{n=0}^{+\infty} \oint_{n\pi} \frac{\rmd \sigma}{2\pi\rmi}\, \frac{1}{\sigma} \left( \frac{1}{\sin^2 \sigma} - \frac{1}{\sigma^2} - \frac{1}{3} \right) \rme^{- \frac{2 t \sigma}{\bar{g}_s}} = \nonumber \\
&=& - \frac{1}{4\pi \bar{g}_s} \sum_{n=1}^{+\infty} \left( \frac{2\pi t}{n} + \frac{\bar{g}_s}{n^2} \right) \exp \left(- \frac{2\pi t\, n}{\bar{g}_s} \right).
\label{gim}
\eea
\noindent
As expected from the discussion above, this formula precisely matches with the Schwinger result in a self--dual background, expressed in (\ref{imlsdb}). It can also be obtained from the ``alternating'' result (\ref{gsc}) by analytic continuation and contour rotation. Furthermore, as discussed in section \ref{sec:lo}, it follows that the discontinuity of the free energy across its branch cut consists of an instanton expansion given by
\be\label{gdisc}
{\rm Disc}~\widetilde{F}_{c=1} (\bar{g}_s) = - \frac{\rmi}{2\pi \bar{g}_s} \sum_{n=1}^{+\infty} \left( \frac{2\pi t}{n} + \frac{\bar{g}_s}{n^2} \right) \rme^{- \frac{2\pi t\, n}{\bar{g}_s}} = F^{(1)} (\bar{g}_s) + F^{(2)} (\bar{g}_s) + \cdots.
\ee

We may now relate this instanton series to the full $c=1$ perturbative expansion, (\ref{cone}), by means of the Cauchy formula (\ref{disc}). One first observes that the integral over the contour at infinity in (\ref{disc}) has, in here, no contribution, since the Barnes function is regular at infinity (see, \textit{e.g.}, \cite{a03}). As such, the dispersion relation (\ref{disc}) reads\footnote{Recall that for matrix models and strings one uses $z=\bar{g}_s^2$; see section \ref{sec:lo}.}, after power series expansion of the integrand's denominator,
\bea
\widetilde{F}_{c=1} (\bar{g}_s) &=& - \sum_{n=1}^{+\infty} \sum_{k=0}^{+\infty} \frac{\bar{g}_s^{2k}}{\left( 2\pi n \right)^2} \int_{-\infty}^0 \frac{\rmd z}{z^{k+1}} \left( 1 + \frac{2\pi t\, n}{\sqrt{z}} \right) \rme^{- \frac{2\pi t\, n}{\sqrt{z}}} = \nonumber \\
&=& \sum_{n=1}^{+\infty} \sum_{k=1}^{+\infty} \frac{2 \left( 2k+1 \right)}{\left( 2\pi n \right)^{2k+2}} \left( \frac{\bar{g}_s}{t} \right)^{2k} \Gamma (2k) = \sum_{g=2}^{+\infty} \frac{2 \left( 2g-1 \right)}{\left( 2\pi \right)^{2g}}\, \zeta (2g) \Gamma (2g-2) \left( \frac{\bar{g}_s}{t} \right)^{2g-2},
\eea
\noindent
where we used the definition of the Riemann zeta function as
\be
\zeta (z) = \sum_{n=1}^{+\infty} \frac{1}{n^z}.
\ee
\noindent
Notice that, from the first to the second line, we truncated $k=0$ from the $k$ sum. Indeed, for this particular value of $k$ the integral would require regularization. However, this would only contribute to terms at genus zero and one, which we are not considering here in any case. As such we shall simply truncate the $k=0$ contribution from the sum, without the need to regularize the divergence, and focus on the genus $g \ge 2$ contributions. In this way, if in the above formula for $\widetilde{F}_{c=1} (\bar{g}_s) $ we further relate the Riemann zeta function to the Bernoulli numbers via
%
%
%
%
%
\be
\zeta (2n) = (-1)^{-(n+1)}\, \frac{(2\pi)^{2n}}{2 \left( 2n \right)!}\, B_{2n},
\ee
\noindent
it immediately follows
\be\label{pg}
\widetilde{F}_{c=1} (\bar{g}_s) = \sum_{g=2}^{+\infty} \frac{\left| B_{2g} \right|}{ 2g \left( 2g-2 \right)}\, \left( \frac{\bar{g}_s}{t} \right)^{2g-2}
\ee
\noindent
which is indeed the $c=1$ perturbative expansion, for genus $g \ge 2$, with $\mu = \frac{t}{\bar{g}_s}$.

In some sense, (\ref{pg}) takes us back to where we started the discussion, \textit{i.e.}, the Gaussian matrix model perturbative series. Indeed we have seen that the alternating Gaussian perturbative series admits a simple Borel transform, which may be inverted unambiguously to provide a nonperturbative completion of the theory. Upon ``Wick rotation'' of the coupling constant, this completion also describes the non--alternating $c=1$ string theory alongside with its instanton effects (obtained in a fashion very similar to our earlier discussion of the Schwinger effect). Of course that a key aspect of this analysis is the fact that the integral representation of the free energy, provided by the inverse Borel transform, precisely coincides with the nonperturbative integral formulation of the $c=1$ theory put forward in \cite{gm90, gk90}. One may thus consistently pick either starting point and obtain the very same results.

To end our analysis, we shall now address the large--order behavior of perturbation theory and see that it is controlled---as expected---by one--instanton contributions, \textit{i.e.}, by the closest pole to the origin in the complex Borel plane. If one considers the first term in the instanton expansion (\ref{gdisc}) and, following the discussion in section \ref{sec:lo}, one sets
\be
F^{(1)} (\bar{g}_s) = - \frac{\rmi}{\bar{g}_s} \left( t + \frac{\bar{g}_s}{2\pi}  \right) \rme^{- \frac{2\pi t}{\bar{g}_s}},
\ee
\noindent
a comparison with (\ref{oneinstantonlargeorder}) immediately yields
\be\label{ideg}
A = 2 \pi t, \qquad b = -1, \qquad F^{(1)}_1 = t, \qquad F^{(1)}_2 = \frac{1}{2\pi},
\ee
\noindent
thus identifying the instanton action, the characteristic exponent, and the loop expansion around the one--instanton configuration. In fact, it is rather interesting to observe that in this situation the loop expansion around each $\ell$--instanton sector is \textit{finite}. This is quite unusual; typically the $\ell$--instanton loop expansion is itself asymptotic, with its large--order behavior being controlled by the $(\ell+1)$--instanton configuration. What we observe is that in this case only the zero--instanton sector displays non--trivial large--order behavior. Now, with the identifications (\ref{ideg}) the large--order equation (\ref{lomm}) implies
\be\label{loge}
F^{(0)}_g (t) \sim \frac{\left( 2g-1 \right) \left( 2g-3 \right)!}{2\pi^2 \left( 2\pi t \right)^{2g-2}}.
\ee
\noindent
Checking the expected large--order behavior in this case is straightforward, as one simply needs to use the standard relation
\be\label{bern}
|B_{2g}| = \frac{2 \left( 2g \right)!}{\left( 2\pi \right)^{2g}}\, \zeta (2g) = \frac{2 \left( 2g \right)!}{\left( 2\pi \right)^{2g}} \left( 1 + \rme^{-2g \log 2} + \rme^{-2g \log 3} + \cdots \right)
\ee
\noindent
in the asymptotic series (\ref{pg}) and the above (\ref{loge}) immediately follows; exponentially suppressed contributions in (\ref{bern}) are not contributing to the large order of the zero--instanton sector. If one is instead interested in the large--order behavior of the Gaussian matrix model, one essentially just needs to use the analytically continued instanton action $A=2\pi\rmi t$ and everything else follows in a similar fashion.

\subsection{The Penner Matrix Model}\label{sec:pe}

Having worked out the Borel analysis of the free energy in the Gaussian matrix model, which essentially reduces to the Borel analysis of the logarithm of the Barnes function $G_2(z)$, we have all we need in order to write down the nonperturbative part of the free energy of the Penner model (\ref{penge}). In fact the whole procedure is essentially the same as before and we shall leave most calculations to the reader. The Borel transform is now
\be
\CB [\CF_{\mathrm{P}}] (\xi) = \sum_{g=2}^{+\infty} \frac{\CF_g^{\mathrm{P}} (t)}{(2g-3)!}\, \xi^{2g-2} = \frac{t \left( t+2 \right)}{\xi^2} + \frac{1}{4}\, \frac{1}{\sinh^2 \left( \frac{\xi}{2} \right)} - \frac{1}{4}\, \frac{1}{\sinh^2 \left( \frac{\xi}{2 \left( t+1 \right)} \right)}.
\ee
\noindent
It should be simple to spot the similarities to the Gaussian case, as the free energy of the Penner model may be written in terms of Barnes functions as in (\ref{normalizedfp}). As such, the total discontinuity (for $g_s \to \rmi \bar{g}_s$) is now given by the sum of the discontinuities of $\log G_2 \left( \frac{t+1}{g_s}+1 \right)$ and $\log G_2 \left( \frac{1}{g_s}+1 \right)$, which yields
\be\label{pendisc}
{\rm Disc}~\widetilde{\CF}_{\mathrm{P}} (\bar{g}_s) =-\frac{\rmi}{2 \pi\bar{g}_s} \left[ \sum_{n=1}^{+\infty} \left( \frac{2\pi (t+1)}{n} + \frac{\bar{g}_s}{n^2} \right) \rme^{-\frac{2\pi \left( t+1 \right) n}{\bar{g}_s}} - \sum_{m=1}^{+\infty} \left( \frac{2\pi}{m} + \frac{\bar{g}_s}{m^2} \right) \rme^{-\frac{ 2\pi m}{\bar{g}_s}} \right].
\ee
\noindent
This is quite simple to obtain by following a procedure identical to what we used in the Gaussian case, but starting from the above Penner Borel--transform. Notice that in this case we have two sets of nonperturbative contributions, with instanton actions $2\pi \left( t+1 \right)$  and $2\pi$, respectively. The large--order behavior of the theory is controlled, as usual, by the closest pole to the origin in the Borel plane. For $t\geq 0$ the relevant pole is located at $2\pi$; however, close to criticality $t\to -1$, the first instanton tower in (\ref{pendisc}) is the relevant one.

\subsection{The Chern--Simons Matrix Model and the Resolved Conifold}\label{sec:locs}

We may now turn to the Chern--Simons matrix model, holographically describing topological strings on the resolved conifold. The genus expansion of its free energy, (\ref{csexp}), is asymptotic, and in here we wish to analyze this divergent series from the viewpoint of Borel analysis, as we did earlier with both the Gaussian and the Penner matrix models.

Let us start with  the Chern--Simons genus expansion (\ref{csexp}) where, for the moment, we drop the constant map contribution and focus on genus $g \ge 2$. This will allow us to better understand the divergence of the series arising from the term with both a Bernoulli number and a polylogarithm function contributions. One has in this case\footnote{Notice that the $m=0$ contribution, in the sum in the second expression, equals $\sum_{g=2}^{+\infty} \frac{B_{2g}}{2g \left( 2g-2 \right)} \left( \frac{g_s}{t} \right)^{2g-2}$, which is of course the Gaussian free energy at genus $g \ge 2$. This is a consequence of working with a Chern--Simons free energy which is \textit{not} normalized against the Gaussian free energy; see section \ref{sec:csmm}.}
\be
F_{\mathrm{CS}} (g_s) = \sum_{g=2}^{+\infty} g_s^{2g-2}\, \frac{B_{2g}}{2g \left( 2g-2 \right)!}\, {\mathrm{Li}}_{3-2g} \left( \rme^{-t} \right) = \sum_{g=2}^{+\infty} g_s^{2g-2}\, \frac{B_{2g}}{2g \left( 2g-2 \right)} \sum_{m \in \BZ} \frac{1}{\left( t + 2\pi\rmi m \right)^{2g-2}},
\ee
\noindent
where we have used an integral representation of the polylogarithm in terms of a Hankel contour, re--written as a sum over residues \cite{gr07}, to express
\be
{\mathrm{Li}}_{3-2g} \left( \rme^{-t} \right) = \Gamma \left( 2g-2 \right) \sum_{m \in \BZ} \frac{1}{\left( t + 2\pi\rmi m \right)^{2g-2}}, \qquad g \ge 2.
\ee
\noindent
Now, since the Bernoulli numbers grow as $B_{2g} \sim (2g)!$ and the polylogarithm functions behave, in worse growth scenario\footnote{At large $t$ the polylogarithm's growth is not factorial in genus, as one has $\lim_{\re\, t \to \pm\infty} {\mathrm{Li}}_{3-2g} \left( \rme^{-t} \right) = 0$.}, as $\lim_{|t| \to 0} {\mathrm{Li}}_{3-2g} \left( \rme^{-t} \right) \sim \Gamma \left( 2g-2 \right) t^{2-2g}$, this series is asymptotic and, like in Gaussian and Penner models, its coefficients grow factorially as $(2g-3)!$. In this case one is led to the Borel transform
\bea
\CB [F_{\mathrm{CS}}] (\xi) &=& \sum_{g=2}^{+\infty} \frac{F_g^{\mathrm{CS}} (t)}{(2g-3)!}\, \xi^{2g-2} = \sum_{g=2}^{+\infty} \frac{B_{2g}}{2g \left( 2g-2 \right)!} \sum_{m \in \BZ} \frac{\xi^{2g-2}}{\left( t + 2\pi\rmi m \right)^{2g-2}} = \nonumber \\
&=& \sum_{m \in \BZ} \left( - \frac{1}{12} + \frac{\left( t + 2\pi\rmi m \right)^2}{\xi^2} - \frac{1}{4}\, \frac{1}{\sinh^2 \left( \frac{\xi}{2 \left( t + 2\pi\rmi m \right)} \right)} \right).
\eea
\noindent
This function has no poles in the positive real axis, for real argument. This is expected since we started off with an alternating sign expansion and, in this case, we may define the free energy via the inverse Borel transform
\be\label{bc}
\widetilde{F}_{\mathrm{CS}} (g_s) = -\frac{1}{4} \sum_{m \in \BZ} \int_0^{+\infty} \frac{\rmd s}{s} \left( \frac{1}{\sinh^2 \left( \frac{g_s}{2 \left( t + 2\pi\rmi m \right)}\, s \right)} - \left( \frac{2 \left( t + 2\pi\rmi m \right)}{g_s} \right)^2 \frac{1}{s^2} + \frac{1}{3} \right) \rme^{-s}.
\ee
\noindent
This provides an unambiguous nonperturbative completion for the asymptotic expansion of the free energy in the Chern--Simons model. Interestingly enough, and like in previous examples, for each distinct $m$ a trivial change of variables turns the corresponding expression into the one--loop effective Lagrangian for a charged scalar particle in a constant self--dual electromagnetic field (of magnetic type) introduced in section \ref{sec:se}, the sum over all integer $m$ thus corresponding to a sum over an infinite number of Lagrangians of this type. If we further report to the discussion in section \ref{sec:gv}, we see that our Borel resummation is essentially (up to analytic continuation, as we move between alternating and non--alternating perturbative series) equal to the GV integral representation of the free energy of topological strings on the resolved conifold.

Let us thus consider the case of the resolved conifold in greater detail, which is obtained by the simple analytic continuation $g_s \to \rmi \bar{g}_s$, and corresponds to the electric version of the above result. In this case the free energy perturbative series is non--alternating and the Borel transform has poles on the positive real axis, making the reconstruction of the free energy possibly affected by nonperturbative ambiguities, which we may, however, understand in the computation of the imaginary part of the integral,
\be\label{bc1}
\widetilde{F}_{\mathrm{conif}} (\bar{g}_s) = \frac{1}{4} \sum_{m \in \BZ} \int_0^{+\infty} \frac{\rmd\sigma}{\sigma} \left( \frac{1}{\sin^2 \sigma} - \frac{1}{\sigma^2} - \frac{1}{3} \right) \rme^{-\frac{2 \left( t + 2\pi\rmi m \right)}{\bar{g}_s}\, \sigma}.
\ee
\noindent
Moreover, since (\ref{bc1}) agrees, after a simple change of variables, with the GV integral representation (\ref{sgvir}), at least for genus $g \ge 2$, we have a physical interpretation for the nonperturbative terms we find: as observed earlier, in section \ref{sec:gv}, the imaginary part of the integral will compute the BPS pair--production rate in the presence of a constant self--dual graviphoton background.

\FIGURE[ht]{
\label{polesfig}
\centering
\psfrag{c1}{$c=1$}
\psfrag{fpi2gs}{$\frac{4 \pi^2}{g_s}$}
\psfrag{tpii}{$\frac{2\pi\rmi t}{g_s}$}
\epsfig{file=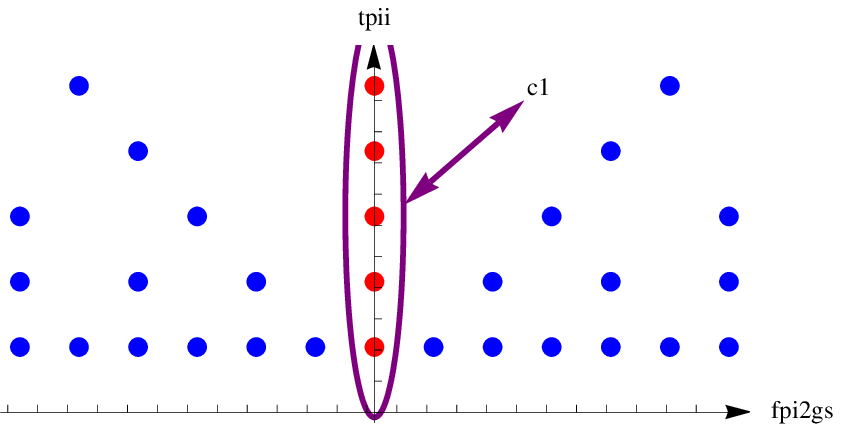, width=12cm, height=6cm}
\caption{Poles in the complex Borel plane, for the case of the resolved conifold free energy. The poles in red coincide with the spurious $c=1$ string contribution, arising from the Gaussian normalization.}
}

The imaginary part of the integral (\ref{bc1}) may be computed by the use of the unitarity $+\rmi\epsilon$ prescription, yielding a sum over residues of the integrand. Equivalently it equals one--half of the integral over the whole real axis (the imaginary part is symmetric), which may be computed by closing the contour on the upper half of the complex plane, thus enclosing the poles of the hyperbolic sine; see Figure \ref{polesfig}. It follows
\bea
\im\, \widetilde{F}_{\mathrm{conif}} (\bar{g}_s) &=& \frac{\pi}{4}\, \sum_{n=1}^{+\infty}
\sum_{m \in \BZ} \oint_{n\pi} \frac{\rmd \sigma}{2\pi\rmi}\, \frac{1}{\sigma} \left( \frac{1}{\sin^2 \sigma} - \frac{1}{\sigma^2} - \frac{1}{3} \right) \rme^{-\frac{2 \left( t + 2\pi\rmi m \right)}{\bar{g}_s}\, \sigma} = \nonumber \\
&=& - \frac{1}{4\pi \bar{g}_s} \sum_{n=1}^{+\infty} \sum_{m \in \BZ} \left( \frac{2\pi \left( t + 2\pi\rmi m \right)}{n} + \frac{\bar{g}_s}{n^2} \right) \exp \left(- \frac{2\pi \left( t + 2\pi\rmi m \right) n}{\bar{g}_s} \right).
\label{csim}
\eea
\noindent
One observes without surprise that this formula matches (an infinite sum of) the Schwinger result in a self--dual background. The discontinuity of the free energy across its branch cut is thus given by the instanton expansion
\be\label{csdisc}
{\rm Disc}~\widetilde{F}_{\mathrm{conif}} (\bar{g}_s) = - \frac{\rmi}{2\pi \bar{g}_s} \sum_{n=1}^{+\infty} \sum_{m \in \BZ} \left( \frac{2\pi \left( t + 2\pi\rmi m \right)}{n} + \frac{\bar{g}_s}{n^2} \right) \rme^{- \frac{2\pi \left( t + 2\pi\rmi m \right) n}{\bar{g}_s}} = F^{(1)} (\bar{g}_s) + F^{(2)} (\bar{g}_s) + \cdots.
\ee
\noindent
Making use of
\be
\sum_{m \in \BZ} \exp \left( - 2\pi\rmi m\, \frac{2 \pi n}{g_s} \right) = \sum_{k \in \BZ} \delta \left( \frac{2 \pi n}{g_s} - k \right)
\ee
\noindent
this discontinuity may also be written as
\be
{\rm Disc}~\widetilde{F}_{\mathrm{conif}} (\bar{g}_s) = - \frac{\rmi}{2\pi \bar{g}_s} \sum_{n=1}^{+\infty} \sum_{k \in \BZ} \left[ \left( \frac{2\pi t}{n} + \frac{\bar{g}_s}{n^2} \right) \delta \left( \frac{2 \pi n}{\bar{g}_s} - k \right) + \frac{\bar{g}_s^2}{n^2}\, \frac{\partial}{\partial \bar{g}_s}\, \delta \left( \frac{2 \pi n}{\bar{g}_s} - k \right) \right] \rme^{- \frac{2\pi t\, n}{\bar{g}_s}}.
\ee

As we did in the previous cases, we may now use the Cauchy formula to relate this instanton series to the perturbative expansion of the resolved conifold's free energies. Once again, the integral over the contour at infinity in (\ref{disc}) has no contribution (see the appendix), and the dispersion relation thus reads, successively,
\bea
\widetilde{F}_{\mathrm{conif}} (\bar{g}_s) &=& - \sum_{n=1}^{+\infty} \sum_{k \in \BZ} \sum_{m=0}^{+\infty} \frac{\bar{g}_s^{2m}}{\left( 2\pi n \right)^2} \int_{-\infty}^0 \frac{\rmd z}{z^{m+1}} \left[ \left( 1 + \frac{2\pi t\, n}{\sqrt{z}} \right) \delta \left( \frac{2 \pi n}{\sqrt{z}} - k \right) + \right. \nonumber \\
&&
\left. + \sqrt{z}\, \frac{\partial}{\partial \sqrt{z}}\, \delta \left( \frac{2 \pi n}{\sqrt{z}} - k \right) \right] \rme^{- \frac{2\pi t\, n}{\sqrt{z}}} = \sum_{n=1}^{+\infty} \sum_{k=1}^{+\infty} \sum_{m=0}^{+\infty} \frac{2 \left( 2m+1 \right) \bar{g}_s^{2m}}{\left( 2\pi n \right)^{2m+2}}\, \frac{\left( \rme^{-t} \right)^k}{k^{1-2m}} = \nonumber \\
&=& \sum_{g=1}^{+\infty} \bar{g}_s^{2g-2}\, \frac{2 \left(2g-1\right)}{\left( 2\pi \right)^{2g}}\, \zeta(2g)\, {\mathrm{Li}}_{3-2g} \left( \rme^{-t} \right) = \sum_{g=1}^{+\infty} \bar{g}_s^{2g-2} \frac{|B_{2g}|}{2g \left( 2g-2 \right)!}\, {\mathrm{Li}}_{3-2g} \left( \rme^{-t} \right).
\label{lg}
\eea
\noindent
where we made use of
\be
\int_{-\infty}^{+\infty} \rmd x\, f(x)\, \delta \left( g(x) \right) = \sum_{i} \frac{f(x_i)}{|g'(x_i)|},
\ee
\noindent
with $x_i$ the real simple roots of $g(x)$; of the definition of the polylogarithm of index $p$
\be
{\mathrm{Li}}_p (z) = \sum_{n=1}^{+\infty} \frac{z^n}{n^p};
\ee
\noindent
and the definition of the Riemann zeta function alongside with its relation to the Bernoulli numbers. This result shows that the instanton expansion (\ref{csdisc}) indeed has enough information to completely rebuild the full free energy perturbative expansion for topological strings on the resolved conifold, at genus $g \ge 2$ (at genus $0$ and $1$ one still needs to take into consideration the relation between Chern--Simons and Stieltjes--Wigert perturbative free energies, and additional regularizations may be needed as, \textit{e.g.}, in the Gaussian case).

One may further show that, in particular, the closest pole to the origin in the Borel complex plane controls the large--order behavior of perturbation theory, corresponding to the familiar one--instanton contribution. A glance at Figure \ref{polesfig} makes it clear that the closest pole to the origin corresponds to the one--instanton contribution of the $c=1$ string. This, of course, is due to the fact that the Chern--Simons free energy in (\ref{csexp}) is not normalized by the Gaussian free energy. This is easily corrected by considering, in the following, the normalized free energy
\be
\mathcal{F}_{\mathrm{conif}} (\bar{g}_s) = F_{\mathrm{conif}} (\bar{g}_s) - F_{c=1} (\bar{g}_s).
\ee
\noindent
This will guarantee that our large--order tests will precisely look at the true ``resolved conifold contribution'', without being plagued by ghost effects due to the Gaussian measure. In this normalized case, and as is simple to check by looking at Figure \ref{polesfig} again, we have two complex conjugate poles equally distant from the origin, at $\xi = 2\pi \left( t \pm 2\pi\rmi \right)/\bar{g}_s$, and we have to consider the contributions from them both. Let us consider the first terms in the instanton expansion (\ref{csdisc})
\be
F^{(1)} (\bar{g}_s) = - \frac{\rmi}{\bar{g}_s} \sum_{m \in \{\pm1\}} \left( t + 2\pi\rmi m + \frac{\bar{g}_s}{2\pi} \right) \rme^{- \frac{2\pi \left( t + 2\pi\rmi m \right)}{\bar{g}_s}},
\ee
\noindent
and thus define, following section \ref{sec:lo},
\bea
A_{\pm} &=& 2\pi \left( t \pm 2\pi\rmi \right) \qquad \Rightarrow \qquad A_\pm = |A|\, \rme^{\pm \rmi\theta_A} = 2\pi \sqrt{t^2+4\pi^2}\, \exp \left( \pm \rmi \arctan \frac{2\pi}{t}\right), \\
F^{(1)}_{1\pm} &=& t \pm 2\pi\rmi \qquad \Rightarrow \qquad F^{(1)}_{1\pm} = |F^{(1)}_1|\, \rme^{\pm\rmi\theta_F} = \sqrt{t^2+4\pi^2}\, \exp \left( \pm \rmi \arctan \frac{2\pi}{t} \right), \\
F^{(1)}_2 &=& \frac{1}{2\pi}, \qquad b=-1.
\label{thetas}
\eea
\noindent
Again, as in the previous examples, the loop expansion around each $\ell$--instanton sector is finite. Thus, also for the resolved conifold only the zero--instanton sector displays non--trivial large--order behavior. In particular, the large--order equation (\ref{lomm}) implies
\be\label{expe}
\CF^{(0)}_g (t) \sim \frac{\Gamma(2g-1)}{\pi^2 |A|^{2g-2}} \left( 1 + \frac{1}{2g-2} \right) \cos \left( \left( 2g-2 \right) \theta_A \right).
\ee
\noindent
The check of this behavior is now harder than before, and we shall need to perform numerical tests to confirm its validity. In this sense, one constructs the test ratio
\be
R_g \equiv \frac{\pi\, \mathcal{F}_g^{(0)}\, |A|^{2g-1}}{2\, |F^{(1)}_1|\, \Gamma(2g-1)} = \cos \left( \left( 2g-2 \right) \arctan \frac{2\pi}{t} \right) \left( 1 + \CO \left( \frac{1}{g} \right) \right),
\ee
\noindent
where equality holds at large $g$ due to (\ref{expe}). In Figure \ref{rcnumtests} we can see that numerical analysis undoubtedly confirms our prediction.

\FIGURE[ht]{
\label{rcnumtests}
\centering
\psfrag{gg}{$g$}
\epsfig{file=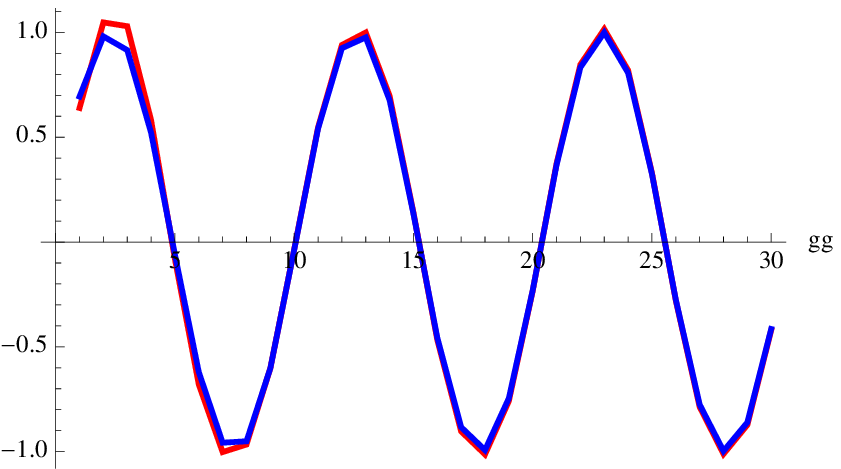, width=7.5cm, height=5cm}
$\quad$
\epsfig{file=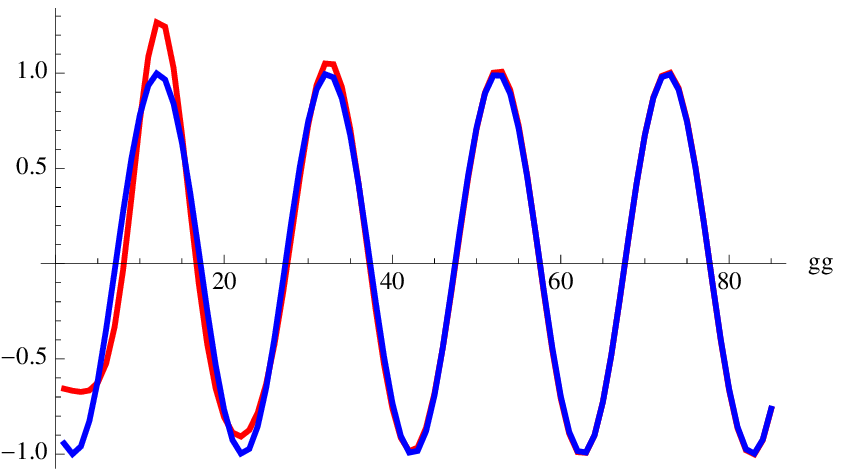, width=7.5cm, height=5cm}
\epsfig{file=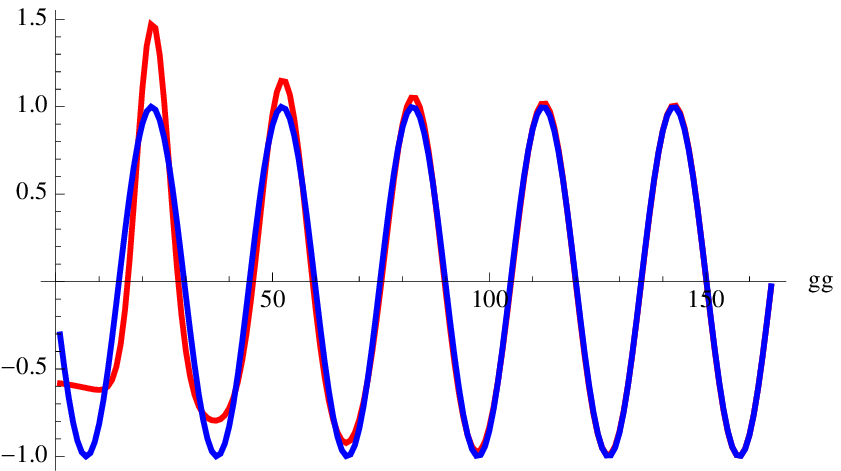, width=7.5cm, height=5cm}
$\quad$
\epsfig{file=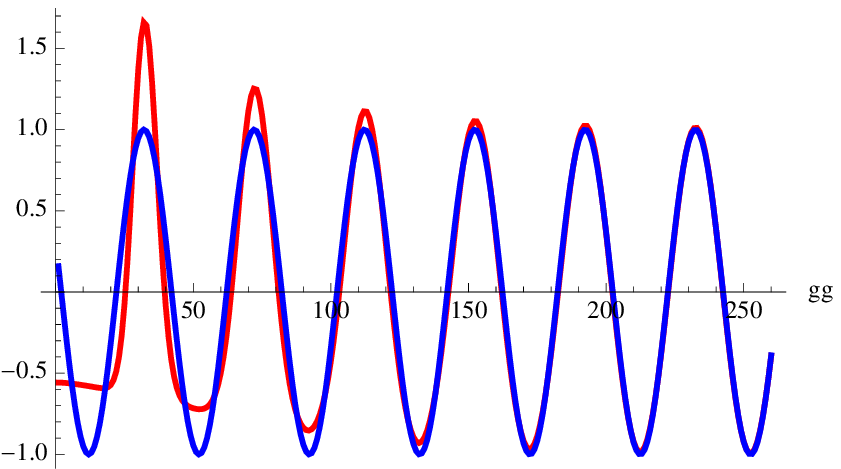, width=7.5cm, height=5cm}
\caption{Test of large--order behavior for the resolved conifold. The plots represent the test ratio $R_g(t)$, in red, versus the expected behavior $\cos \left( \left( 2g-2 \right) \arctan \frac{2\pi}{t} \right)$, in blue, as a function of genus $g$, and for $t=20,40,60,80$, from left top to right bottom, respectively. The matching at high genera is evident.}
}

The analysis so far has focused only on the contribution to the resolved conifold free energy arising from the D2D0 bound states of branes. As is clear in (\ref{csexp}), to this term one must still add the contribution from bound states of D0--branes, \textit{i.e.}, the contribution of constant maps. As it turns out, in this case the discontinuity is obtained from (\ref{csdisc}) by simply setting $t=0$ in that expression, \textit{i.e.},
\be
{\rm Disc}~\widetilde{F}_{\mathrm{K}} (\bar{g}_s) = - \frac{\rmi}{\bar{g}_s} \sum_{n=1}^{+\infty} \sum_{m \in \BZ} \left( \frac{2\pi\rmi m}{n} + \frac{\bar{g}_s}{2\pi n^2} \right) \rme^{- \frac{4\pi^2\rmi\, m\, n}{\bar{g}_s}} = F^{(1)}_{\mathrm{K}} (\bar{g}_s) + F^{(2)}_{\mathrm{K}} (\bar{g}_s) + \cdots.
\ee
\noindent
A procedure that should be familiar by now yields back the perturbative expansion (\ref{fpK}) or (\ref{csexp}) via the Cauchy formula and an integration of the above discontinuity over the free energy branch cut:
\bea
\widetilde{F}_{\mathrm{K}} (\bar{g}_s) &=& - \sum_{n=1}^{+\infty} \sum_{k \in \BZ} \sum_{m=0}^{+\infty} \frac{\bar{g}_s^{2m}}{\left( 2\pi n \right)^2} \int_{-\infty}^0 \frac{\rmd z}{z^{m+1}} \left[ \delta \left( \frac{2 \pi n}{\sqrt{z}} - k \right) + \sqrt{z}\, \frac{\partial}{\partial \sqrt{z}}\, \delta \left( \frac{2 \pi n}{\sqrt{z}} - k \right) \right] = \nonumber \\
&=& \sum_{n=1}^{+\infty} \sum_{k=1}^{+\infty} \sum_{m=0}^{+\infty} \frac{2 \left( 2m+1 \right) \bar{g}_s^{2m}}{\left( 2\pi n \right)^{2m+2}}\, \frac{1}{k^{1-2m}} = \sum_{g=1}^{+\infty} \bar{g}_s^{2g-2} \frac{2 \left( 2g-1 \right)}{\left( 2\pi \right)^{2g}}\, \zeta \left( 2g \right) \zeta \left( 3-2g \right) = \nonumber \\
&=& \sum_{g=1}^{+\infty} \bar{g}_s^{2g-2} \frac{(-1)^g\, B_{2g}\, B_{2g-2}}{2g \left( 2g-2 \right) \left( 2g-2 \right)!},
\label{lgk}
\eea
\noindent
where we have used familiar properties of the zeta function and Bernoulli numbers, including
\be
\zeta(-n) = \frac{(-1)^n}{n+1}\, B_{n+1}.
\ee

The last thing we want to show is that, as expected, it is the closest pole to the origin in the Borel complex plane that controls the large--order behavior of the theory. Repeating our earlier discussion we find the one--instanton contribution
\be
F^{(1)} (\bar{g}_s) = - \frac{\rmi}{\bar{g}_s} \sum_{m \in \{\pm1\}} \left( 2\pi\rmi m + \frac{\bar{g}_s}{2\pi} \right) \rme^{- \frac{4\pi^2 \rmi m}{\bar{g}_s}},
\ee
\noindent
leading to
\bea
&&
A_\pm = \pm 4\pi^2 \rmi, \qquad b=-1, \nonumber\\
&&
F^{(1)}_{1\pm} = \pm 2\pi\rmi, \qquad F^{(1)}_2 = \frac{1}{2\pi}.
\label{thetasK}
\eea
\noindent
In this case the large--order equation (\ref{lomm}) yields
\be
F^{(0)}_g (t) = \frac{\Gamma(2g-1)}{\pi^2 \left( 4\pi^2\rmi \right)^{2g-2}} \left( 1 + \frac{1}{2g-2} \right) = \left(-1\right)^{g-1} \frac{16\pi^2}{\left( 2\pi \right)^{4g}} \left( 2g-3 \right)! \left( 2g-1 \right).
\ee
\noindent
To check that this is the right answer all one has to do is to use
\bea
(-1)^g\, B_{2g}\, B_{2g-2} &=& (-1)^{g+1} \frac{16\pi^2}{\left( 2\pi \right)^{4g}} \left( 2g \right)! \left( 2g-2 \right)!\, \zeta (2g)\, \zeta (2g-2) = \nonumber \\
&=& (-1)^{g-1} \frac{16\pi^2}{\left( 2\pi \right)^{4g}} \left( 2g \right)! \left( 2g-2 \right)!  \left( 1 + 20\, \rme^{-6g \log 2} + 4\, \rme^{-4g \log 2} + \cdots \right).
\eea
\noindent
Notice that the above exponentially suppressed contributions do not contribute to the large order of the zero--instanton sector.

\section{Stokes Phenomena and Instantons from Hyperasymptotics}

Having understood the Borel analysis of topological strings and $c=1$ matrix models, benefiting in this course of the identification of instanton effects in these models, we shall now make a brief detour into the realm of hyperasymptotics, as first introduced in \cite{bh90} (see, \textit{e.g.}, \cite{b99} for a review), \textit{i.e.}, a series of techniques to refine optimally truncated asymptotic expansions by the inclusion of exponentially small contributions. In particular, our focus will concern hyperasymptotic approximations for integrals with saddles \cite{bh91, b93}, the prototypical example for problems dealing with the calculation of partition functions or free energies. In this case, the exponentially suppressed contributions arise from saddles \textit{other} than the one chosen in the steepest--descent asymptotic approximation. The main interest of this analysis for the present work is that the Stokes phenomenon---certain ``discontinuities'' which we shall explain below and later relate to instanton effects---is automatically incorporated into the hyperasymptotic scheme.

Suppose one wants to use the method of steepest descents in order to find an asymptotic expansion, as $|\kappa| \to \infty$ with  $\kappa = |\kappa|\, \rme^{\rmi \theta}$, of the one--dimensional ``partition function''
\be\label{1dpart}
Z(\kappa) = \int_{\CC} \rmd z\, \rme^{-\kappa W(z)},
\ee
\noindent
where $\CC$ is a contour we specify below. A typical calculation goes as follows: one begins with the calculation of the saddle points of the ``potential function'' $W(z)$, computed as the set of points $\{ z_k \}_{k=1,2,\cdots}$ such that $W'(z_k)=0$; then, chosen a reference saddle--point $z_{n}$, the contour of integration $\CC$ is deformed to the infinite oriented path of steepest descent through $z_{n}$, which we shall denote by $\CC_{n} (\theta)$. This contour is defined as
\be
\im \left[ \kappa \left( W(z) - W(z_{n}) \right) \right] = 0
\ee
\noindent
and with $\kappa \left( W(z) - W(z_{n}) \right)$ increasing away from $z_{n}$. This immediately implies that the phase of $W(z) - W(z_{n})$ must equal $- \theta + 2\pi m$, $m \in \BZ$. We then introduce the ``partition  function'' $Z_n(\kappa)$ evaluated on the $n$--th saddle
\bea
Z_{n} (\kappa) &\equiv& \frac{1}{\sqrt{\kappa}}\, \rme^{- \kappa W(z_{n})}\, \CZ_{n}(\kappa), \\
\CZ_{n} (\kappa) &=& \sqrt{\kappa} \int_{\CC_{n} (\theta)} \rmd z\, \rme^{-\kappa \left( W(z) - W(z_{n}) \right)}.
\label{partsp}
\eea
\noindent
As we shall see, $\CZ_n (\kappa)$ will display Stokes phenomena in the form of a discontinuity associated to a jump in the steepest--descent path whenever it passes through one of the other saddles, $k \not = n$. The integral (\ref{partsp}) can be evaluated via the steepest--descent method and one obtains a function of $\kappa$ for each saddle, $n$, given by a series in negative powers of $\kappa$
\be\label{Zasympt}
\CZ_{n} (\kappa) \sim \sum_{g=0}^{+\infty} \frac{\zeta_g (n)}{\kappa^g},
\ee
\noindent
with (see \cite{bh91, b93} for details)
\be
\zeta_g (n) = \Gamma \Big( g+\frac{1}{2} \Big) \oint_{z_{n}} \frac{\rmd z}{2\pi\rmi}\, \frac{1}{\left( W(z) - W(z_{n}) \right)^{g+\frac{1}{2}}}.
\ee
\noindent
The series in (\ref{Zasympt}) is asymptotic. The point of view of \cite{bh91} is to understand this divergence as a consequence of the existence of \textit{other} saddles $\{z_{k \not = n}\}$, through which $\CC_{n}$ does not pass. Because one is free to choose the reference saddle $n$ at will, all possible asymptotic series are thus related by a requirement of mutual consistency, also known as the principle of resurgence: each divergent series will contain, in its late terms, and albeit in coded form due to their divergent nature, all the terms associated to the asymptotic series from all other saddles.

Another important point concerning the asymptotic series (\ref{Zasympt}) dwells with the fact that this expression only holds in a wedge of the complex $\kappa$--plane, \textit{i.e.}, for a restricted range of $\theta$, a property which is associated to Stokes phenomena. Suppose that in the above set--up, and once (\ref{Zasympt}) has been computed, we start varying $\theta$ in such a way that we always choose the contour of integration to be the steepest--descent through the saddle $z_{n}$. As it turns out, this is a continuous process only for a finite range of $\theta$: indeed, one faces a discontinuity if $\theta$ reaches a value such that the contour of integration passes through a second saddle, $z_{m}$. This will happen when $\theta$ reaches the value $-\sigma_{nm}$, with
\be
-\sigma_{nm} = - \arg \left( W(z_{m}) - W(z_{n}) \right).
\ee
\noindent
For this value of $\theta$ the steepest--descent contour will change discontinuously and  exponentially suppressed contributions to (\ref{partsp}) will ``suddenly'' become of order one. In this case, and in order for the steepest--descent contour to go through a single saddle, one must restrict $\theta$ to an interval $-\sigma_{nm_1} < \theta < -\sigma_{nm_2}$, where $z_{m_1}$ and $z_{m_2}$ are saddles adjacent to $z_{n}$, \textit{i.e.}, saddles which may be reached from $z_{n}$ through steepest--descent paths\footnote{A saddle $z_m$ is said to be adjacent to the saddle $z_n$ iff there is a path of steepest descent from $z_n$ to $z_m$, \textit{i.e.}, $z_m$ will be adjacent to $z_n$ whenever $\theta = - \sigma_{nm}$ and thus $\arg \left( W(z) - W(z_{n}) \right) = \sigma_{nm}$ (naturally this is also the condition that defines the Stokes lines for $\CZ_n (\kappa)$). One similarly defines the adjacent contour through the adjacent saddle as the steepest--descent contour $\CC_m (-\sigma_{nm})$, through $z_m$.}. This is illustrated in Figure \ref{kplnsdl}. One of the goals of hyperasymptotics \cite{bh90} is to deploy resurgence in order to better understand Stokes phenomena, and this is the aspect we shall be mostly interested in.

\FIGURE[ht]{
\label{kplnsdl}
\centering
\psfrag{sn1}{$-\sigma_{nm_1}$}
\psfrag{sn2}{$-\sigma_{nm_2}$}
\psfrag{zn}{$n$}
\psfrag{m1}{$m_1$}
\psfrag{m2}{$m_2$}
\psfrag{cn1}{$\CC_n (-\sigma_{nm_1})$}
\psfrag{cn2}{$\CC_n (-\sigma_{nm_2})$}
\psfrag{cm1}{$\CC_{m_1} (-\sigma_{nm_1})$}
\psfrag{cm2}{$\CC_{m_2} (-\sigma_{nm_2})$}
\epsfig{file=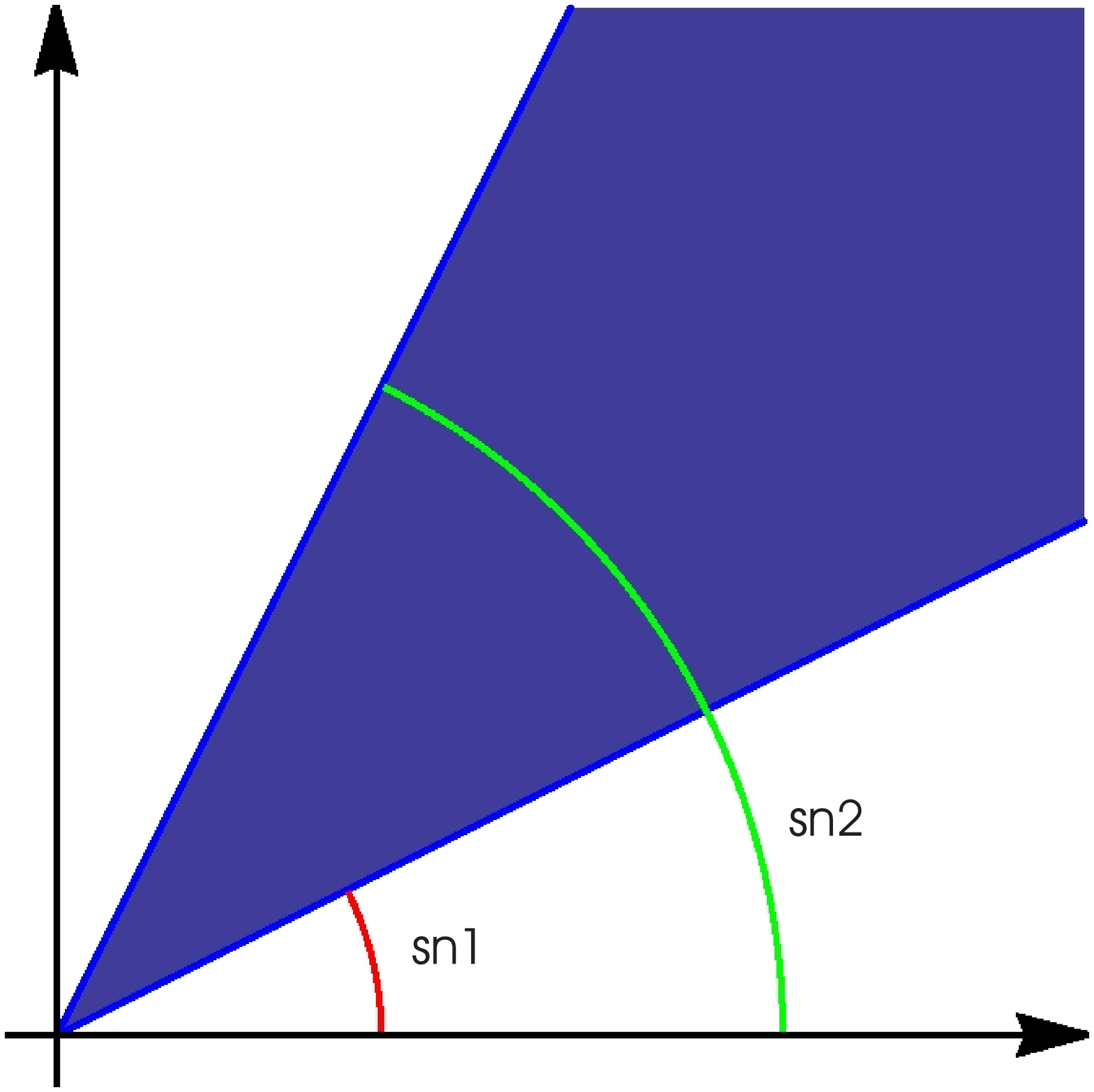, width=6.5cm, height=6.5cm}
$\qquad$
\epsfig{file=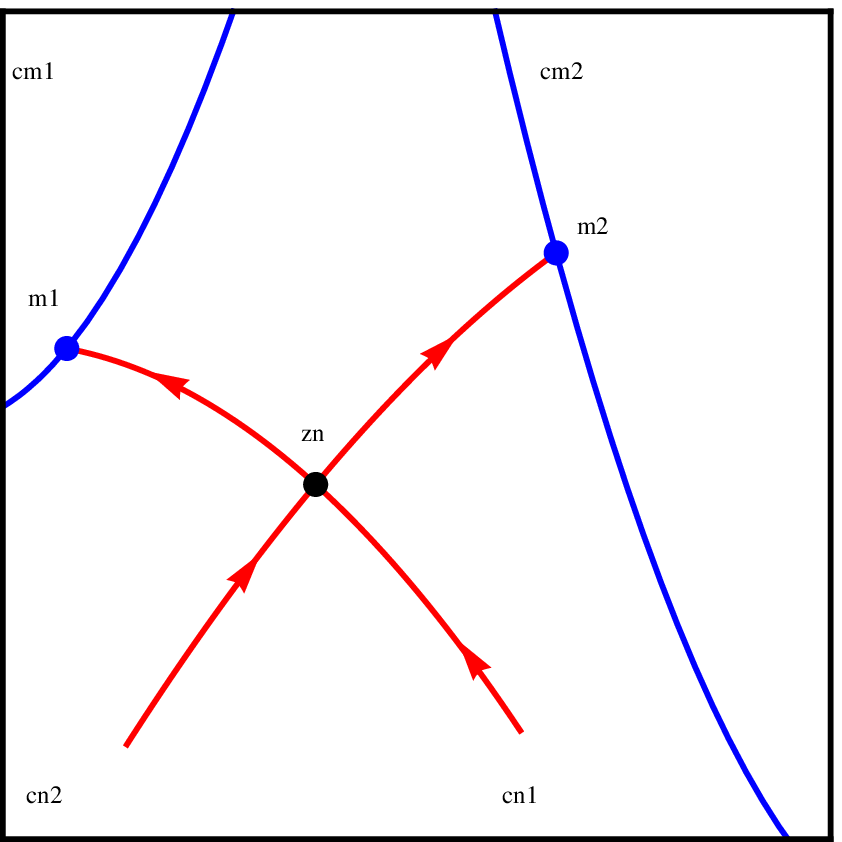, width=7.5cm, height=6.5cm}
\caption{On the left, the complex $\kappa$ plane, showing a region $-\sigma_{nm_1} < \theta < -\sigma_{nm_2}$. On the right, the equivalent region in $z$ space, with saddle $z_n$, adjacent saddles $z_{m_1}$ and $z_{m_2}$, and respective steepest--descent contours through $z_n$ (in red) hitting the adjacent saddles. In blue we plot the adjacent contours.}
}

Let us make these ideas more precise. In hyperasymptotics one begins with (optimal) truncation of the asymptotic series (\ref{Zasympt}). In this case,
\be\label{Zasymptrunc}
\CZ_{n} (\kappa) = \sum_{g=0}^{N-1} \frac{\zeta_g (n)}{\kappa^g} + \CR_{n}^{(N)} (\kappa),
\ee
\noindent
where $\CR_{n}^{(N)} (\kappa)$ is the remainder associated to the finite truncation. The main contribution of the hyperasymptotic calculation in \cite{bh91} was to produce an expression for the reminder which led to an exact resurgence formula for (\ref{Zasympt}), and we shall now present these results. Let us first define the singulant, for every adjacent saddle $z_m$, as
\be
\CW_{nm} \equiv W(z_{m}) - W(z_{n}) \equiv \left|\CW_{nm}\right|\, \rme^{\rmi \sigma_{nm}}.
\ee
\noindent
In this case the remainder term (see \cite{bh91, b93} for details) can be expressed as a sum over integrals through all adjacent saddles to $z_n$, $\{ z_m \}$:
\be
\CR_{n}^{(N)} (\kappa) = \frac{1}{2\pi\rmi}\, \frac{1}{\kappa^N} \sum_{m} \int_0^{\infty\cdot \rme^{-\rmi \sigma_{nm}}} \rmd\eta\, \frac{\eta^{N-1}}{1-\frac{\eta}{\kappa}}\, \rme^{-\eta \CW_{nm}}\, \CZ_m (\eta),
\ee
\noindent
where the $\CZ_m (\eta)$ are defined on the adjacent contours $\CC_m (-\sigma_{nm})$. There are several interesting points to this formula. First, it provides an exact and explicit expression for the reminder and one now explicitly sees that the divergence of 
the asymptotic series (\ref{Zasympt}) is directly related to the existence of adjacent saddles. Second, inserting the above expression back in (\ref{Zasymptrunc}), one obtains the exact resurgence formula
\be\label{resurgentZ}
\CZ_{n} (\kappa) = \sum_{g=0}^{N-1} \frac{\zeta_g (n)}{\kappa^g} + \frac{1}{2\pi\rmi}\, \frac{1}{\kappa^N} \sum_{m} \int_0^{\infty\cdot \rme^{-\rmi \sigma_{nm}}} \rmd\eta\, \frac{\eta^{N-1}}{1-\frac{\eta}{\kappa}}\, \rme^{-\eta \CW_{nm}}\, \CZ_m (\eta),
\ee
\noindent
which is the basis for the hyperasymptotic analysis of \cite{bh91}: indeed, each $\CZ_m (\eta)$ in the above integrands may itself be expanded as an asymptotic series leading, via iterations of the above formula, to exponentially improved asymptotic results for the original $\CZ_{n} (\kappa)$ (in the sense that the error associated to the approximation is reduced from polynomially small to exponentially small\footnote{For a recent discussion in the field theoretic context see \cite{fg09}.}). Third, as the resurgent formula holds for any $N$, one may write it down for $N=0$, obtaining either the resurgent expression \cite{b93}
\be\label{resurgentN=0}
\CZ_{n} (\kappa) = \frac{1}{2\pi\rmi} \sum_{m} \int_0^{\infty \cdot \rme^{-\rmi \sigma_{nm}}} \rmd\eta\, \frac{\rme^{-\eta \CW_{nm}}}{\eta-\frac{\eta^2}{\kappa}}\, \CZ_m (\eta),\ee
\noindent
which leads to interesting functional relations in selected examples; or the (formal) resurgent relation \cite{bh91}
\be
\zeta_g (n) = \frac{1}{2\pi\rmi} \sum_{m} \sum_{h=0}^{+\infty} \frac{(g-h-1)!}{\CW_{nm}^{g-h}}\, \zeta_h (m),
\ee
\noindent
which expresses the late terms ($g \gg 1$) of the asymptotic series at a given saddle as a sum over the early terms of the corresponding asymptotic series at the adjacent saddles. In particular, the leading contribution arises from the adjacent saddle $m^*$ with smallest singulant, \textit{i.e.}, to leading order one obtains
\be
\zeta_g (n) \sim \frac{(g-1)!}{\CW_{nm^*}^{g}}\, \zeta_0 (m^*),
\ee
\noindent
which makes manifest the characteristic factorial behavior of asymptotic series. Fourth, and finally, the resurgence formula (\ref{resurgentZ}) precisely incorporates Stokes phenomenon \cite{bh91}; the appearance of suppressed exponential terms as the steepest--descent contour $\CC_n (\theta)$ sweeps through one of the adjacent saddles, $m$ (see Figure \ref{stksddl}). This, as we mentioned, will happen as $\theta$ crosses the Stokes line $\CC_n(-\sigma_{nm})$, in which case the asymptotic expansion (\ref{Zasympt}) will have a discontinuity
\be
\Delta\, \CZ_{n} (\kappa) \Big|_{\theta = - \sigma_{nm}} \equiv \CZ_{n} \left( \left| \kappa \right| \rme^{\rmi \left( -\sigma_{nm} + 0^+\right)} \right) - \CZ_{n} \left( \left| \kappa \right| \rme^{\rmi \left( -\sigma_{nm} + 0^-\right)} \right) \not = 0.
\ee
\noindent
It is not too hard to compute the precise value of this discontinuity straight from the resurgence formula for the remainder, (\ref{resurgentZ}). One obtains, without surprise,
\bea
\Delta\, \CZ_{n} (\kappa) \Big|_{\theta = - \sigma_{nm}} &=& \frac{1}{\kappa^{N-1}} \oint_{\kappa} \frac{\rmd\eta}{2\pi\rmi}\, \frac{1}{\eta-\kappa}\, \eta^{N-1}\, \rme^{-\eta \CW_{nm}}\, \CZ_m (\eta) \nonumber \\
&=& \rme^{-\kappa \CW_{nm}}\, \CZ_m (\kappa),
\eea
\noindent
where, naturally, any further asymptotic expansion on the adjacent saddle, \textit{i.e.}, for $\CZ_m (\kappa)$, is to be evaluated precisely along the Stokes line $\theta = - \sigma_{nm}$. This discontinuity is exponentially small as, on the Stokes line, $\kappa \CW_{nm}$ is real and positive.

\FIGURE[ht]{
\label{stksddl}
\centering
\psfrag{zn}{$n$}
\psfrag{zm}{$m$}
\psfrag{m2}{$m_2$}
\psfrag{cm}{$\CC_n (-\sigma_{nm} + 0^-)$}
\psfrag{cp}{$\CC_n (-\sigma_{nm} + 0^+)$}
\epsfig{file=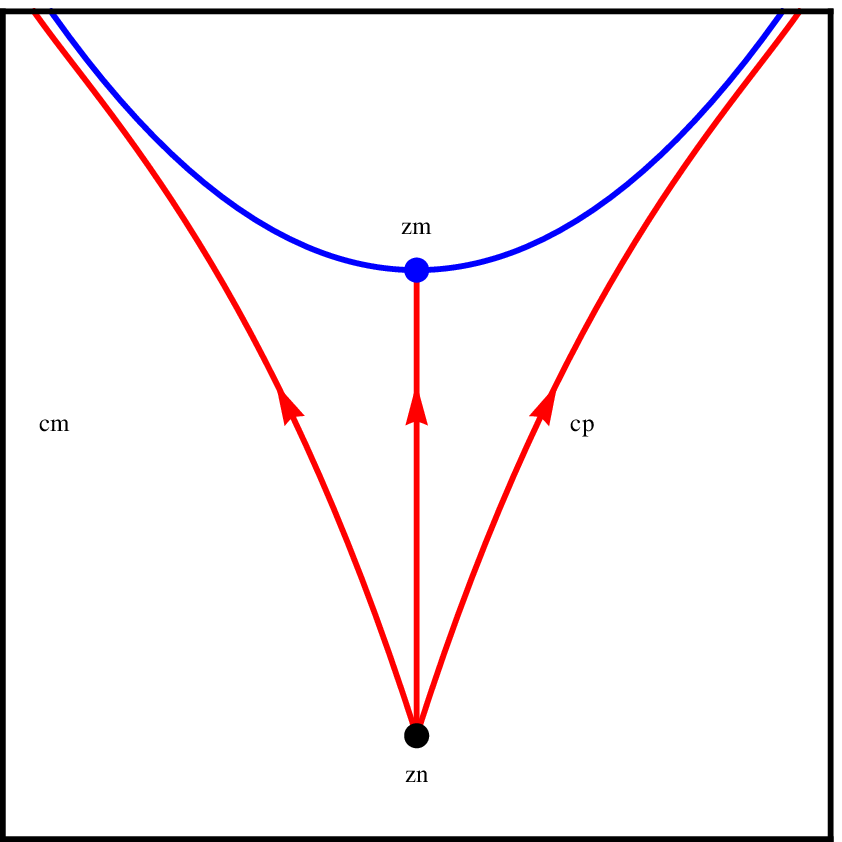, width=9.5cm, height=8cm}
\caption{Crossing the Stokes line, between the saddle $z_n$ and the adjacent saddle $z_{m}$.}
}

The Stokes discontinuity is particularly relevant to us as we later wish to identify it with instanton effects and, as such, we shall dwell upon it later in this section. An important thing to notice is that this is actually not a discontinuity of the function $Z(\kappa)$ but rather a discontinuity of the \textit{asymptotic approximation} to $Z(\kappa)$. A trivial example is the function $\sinh \frac{1}{z}$. In the right half--plane $\re\, z >0$ its asymptotic behavior as $z \to 0^+$ is well described by $\frac{1}{2} \exp \left( \frac{1}{z} \right)$, with the error being exponentially suppressed. However, should we try to rotate this asymptotic approximation to the left half--plane $\re\, z <0$, it would no longer be valid as the subdominant error $-\frac{1}{2} \exp \left( -\frac{1}{z} \right)$ will no longer be small! This is clearly a discontinuity of the asymptotic approximation we chose, and not of the function itself.

We shall next see how to apply this formalism within the case of the Gamma function, which will later lead to a hyperasymptotic understanding of the free energies in the Gaussian and Penner matrix models. In particular, we shall build up our analysis in order to see how to obtain the multiple instanton sectors of these models straight out of the above resurgence formulae.

\subsection{Stokes Phenomena in the Gamma Function}\label{stogamma}

Consider the ``Gamma partition--function''
\be
Z_\Gamma (\kappa) \equiv \Gamma (\kappa).
\ee
\noindent
One usually defines the Gamma function via Euler's integral \cite{pk01}
\be\label{intgamma}
\Gamma(\kappa) = \int_0^{+\infty} \rmd w\, w^{\kappa-1} \rme^{-w}, \qquad \re\, (\kappa) > 0,
\ee
\noindent
where the contour of integration is the positive real axis. The logarithm of the Gamma function, the ``Gamma free--energy'', has a well--known representation \cite{pk01}
\be\label{stand}
F_\Gamma (\kappa) \equiv \log \Gamma(\kappa) = \left( \kappa-\frac{1}{2} \right) \log \kappa - \kappa +\frac{1}{2} \log 2\pi + \Omega (\kappa),
\ee
\noindent
where $\Omega(\kappa)$ is meromorphic with simple poles at $\kappa=-n$, $n \in \BN_0$. One then obtains asymptotic expansions for the Gamma function by first obtaining asymptotic expansions for the function $\Omega(\kappa)$. One such familiar case is the Stirling series, which is the Poincar\'e asymptotic expansion
\be\label{omegaasympexp}
\Omega (\kappa) \sim \sum_{g=1}^{+\infty} \frac{B_{2g}}{2g \left( 2g-1 \right)}\, \frac{1}{\kappa^{2g-1}},
\ee
\noindent
valid as $|\kappa| \to +\infty$, in the sector $|\arg (\kappa)|<\pi$. Our goal in the following is to obtain an exponentially improved version of this asymptotic expansion, in the spirit of our previous discussion on hyperasymptotics and Stokes phenomena, along the guidelines in \cite{b94, pk01}.

Let us begin with the ``Gamma partition--function'', applying the hyperasymptotic analysis in the preceding section as in \cite{b94}. One first changes variables as $w = \kappa\, \rme^z$ and then re--writes Euler's integral (\ref{intgamma}) as
\be
\Gamma (\kappa) = \kappa^\kappa \int_{-\infty}^{+\infty} \rmd z\, \rme^{-\kappa W(z)},
\ee
\noindent
where $W(z) = \rme^z-z$. The saddle points of this function are $z_k = 2\pi\rmi k$, for $k \in \BZ$, with $W(z_k) = 1 - 2\pi\rmi k$. We may now apply the machinery we previously described, with the nuance that one now has an infinite number of saddles. We first select the reference saddle $z_0 = 0$ and define
\bea
\Gamma_0 (\kappa) &\equiv& \sqrt{2\pi}\, \kappa^{\kappa-\frac{1}{2}}\, \rme^{-\kappa}\, \CG_0 (\kappa), \\
\CG_0 (\kappa) &=& \sqrt{\frac{\kappa}{2\pi}} \int_{C_0 (\theta)} \rmd z\, \rme^{-\kappa \left( W(z) - 1 \right)}, \qquad \re\, (\kappa)>0,
\eea
\noindent
with $\log \CG_0 (\kappa) = \Omega (\kappa)$. It is not too hard \cite{b94} to identify \textit{all} saddles $\{ z_m \}_{m \not = 0}$ as adjacent saddles to $z_0$: the singulants are now
\be
\CW_{0m} = - 2\pi\rmi m = 2\pi \left| m \right| \exp \left( \rmi (-1)^{\frac{1}{2}\left(1+\mathrm{sgn} (m)\right)} \frac{\pi}{2} \right)
\ee
\noindent
and the steepest--descent contour from $z_0$ to each $z_m$ is that for which $\theta = - \sigma_{0m}$, \textit{i.e.},
\be
\arg \left( \rme^z - z - 1 \right) = (-1)^{\frac{1}{2}\left(1+\mathrm{sgn} (m)\right)} \frac{\pi}{2}.
\ee
\noindent
As such, for $m>0$ the Stokes line is at $\theta = \frac{\pi}{2}$ and for $m<0$ the Stokes line is at $\theta = -\frac{\pi}{2}$, so that the imaginary axis is a Stokes line for the Gamma function. Parametrizing $z=x+\rmi y$, this contour may also be written as $\cos y = \left( 1+x \right) \rme^{-x}$ with $\rme^{-x} < \frac{\sin y}{y}$ if $m<0$ and greater than if $m>0$ (see Figure \ref{gammacntr}). As such, the remainder associated to the finite truncation of the asymptotic series for $\CG_0 (\kappa)$ follows from (\ref{resurgentZ}) as
\be
\CR_0^{(N)} (\kappa) = \frac{1}{2\pi\rmi \kappa^N} \sum_{m=1}^{+\infty} \left\{ \int_0^{+\rmi\infty} \rmd\eta\, \frac{\eta^{N-1}}{1-\frac{\eta}{\kappa}}\, \rme^{2\pi\rmi \eta m}\, \CG_m (\eta) - \int_{-\rmi\infty}^0 \rmd\eta\, \frac{\eta^{N-1}}{1-\frac{\eta}{\kappa}}\, \rme^{-2\pi\rmi \eta m}\, \CG_{-m} (\eta) \right\},
\ee
\noindent
where one should recall that the $\CG_{m} (\eta)$ are to be evaluated over the adjacent steepest--descent contours $\CC_{m} (-\sigma_{0m})$, \textit{e.g.}, for $m>0$,
\be
\CG_{m} (\eta) = \sqrt{\frac{\eta}{2\pi}} \int_{\CC_m \left( \frac{\pi}{2} \right)} \rmd z\, \rme^{-\eta \left( W(z) - W(z_m) \right)}.
\ee
\noindent
In this integral, consider the shift $w = z - 2\pi\rmi m$, where we move the contour downwards in the complex $z$--plane by $2\pi\rmi m$. It is simple to see that the shifted contour $\bar{\CC}_m \left( \frac{\pi}{2} \right)$ will now go through $z_0$ rather than $z_m$ and the integral becomes
\be
\CG_{m} (\eta) = \sqrt{\frac{\eta}{2\pi}} \int_{\bar{\CC}_m \left( \frac{\pi}{2} \right)} \rmd w\, \rme^{-\eta \left( W(w) - 1 \right)} = \CG_0 (\eta).
\ee
\noindent
The exact same reasoning applies if $m<0$, in which case one just shifts the adjacent contours upwards. Thus
\be
\CR_0^{(N)} (\kappa) = \frac{1}{2\pi\rmi \kappa^N} \sum_{m=1}^{+\infty} \left\{ \int_0^{+\rmi\infty} \rmd\eta\, \frac{\eta^{N-1}}{1-\frac{\eta}{\kappa}}\, \rme^{2\pi\rmi \eta m}\, \CG_0 (\eta) - \int_{-\rmi\infty}^0 \rmd\eta\, \frac{\eta^{N-1}}{1-\frac{\eta}{\kappa}}\, \rme^{-2\pi\rmi \eta m}\, \CG_0 (\eta) \right\}.
\ee

\FIGURE[ht]{
\label{gammacntr}
\centering
\psfrag{x}{$x$}
\psfrag{y}{$y$}
\epsfig{file=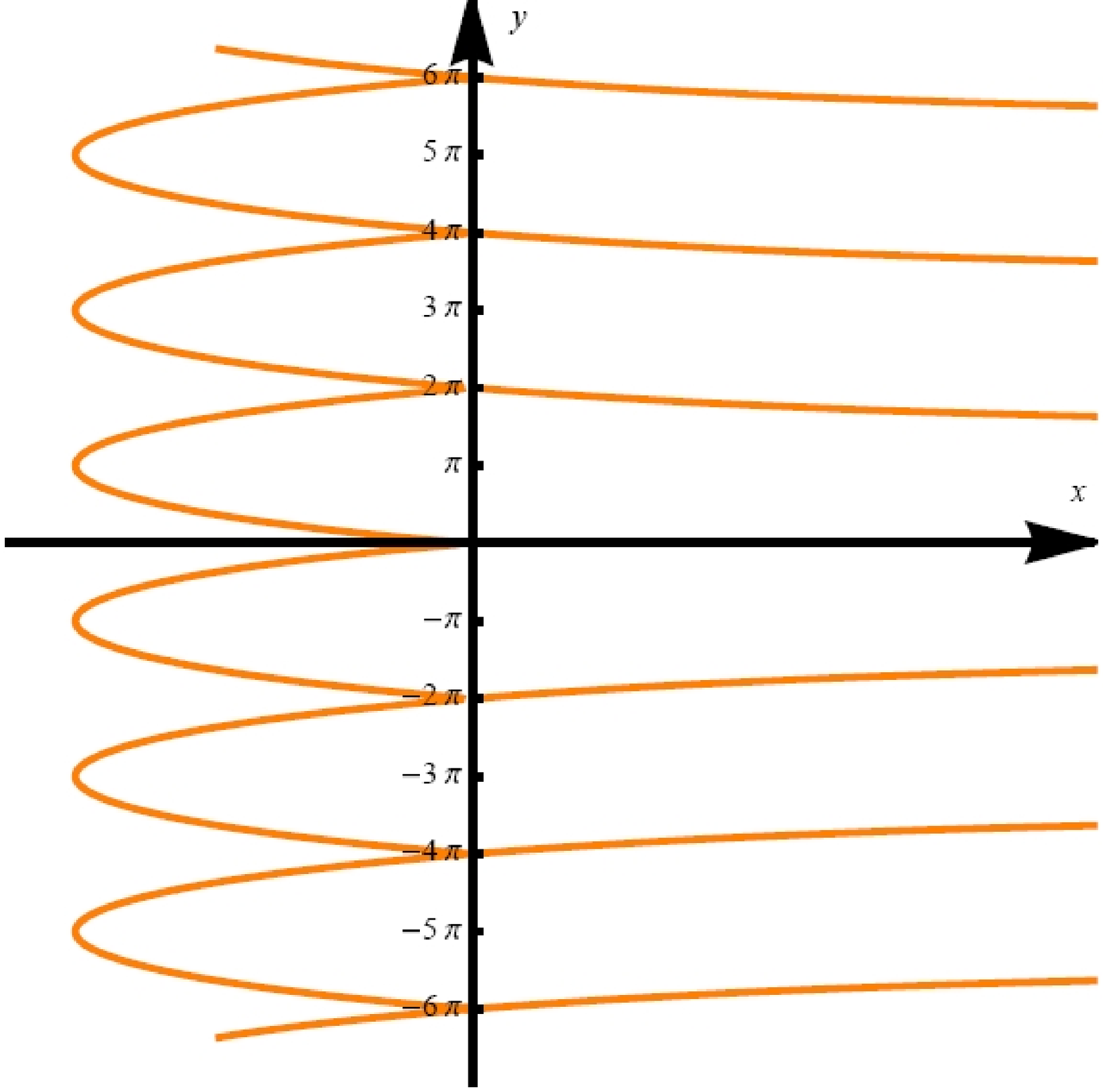, width=9.5cm, height=8.5cm}
\caption{Paths of steepest descent for the Gamma function, in the $(x,y)$ plane, starting at $z_0=0$.}
}

One may now address Stokes phenomena for the ``Gamma partition--function'' and simply confirm that indeed the imaginary axis in the complex $\kappa$--plane is a Stokes line for $\CG_0 (\kappa)$: as $\kappa$ becomes purely imaginary, one of the two integrals above will have a pole. As we have seen before, this leads to the discontinuities
\be\label{gammastokesdisc}
\Delta\, \CG_0 (\kappa) \Big|_{\theta = \pm \frac{\pi}{2}} = \pm \sum_{m=1}^{+\infty} \rme^{\pm 2\pi\rmi \kappa m}\, \CG_0 (\kappa) = \mp \frac{1}{1-\rme^{\mp 2\pi\rmi \kappa}}\, \CG_0 (\kappa).
\ee
\noindent
Notice that the discontinuities are evaluated on the Stokes lines, $\kappa = \pm \rmi |\kappa|$, and are thus always exponentially suppressed. We will interpret these terms as nonperturbative ``instanton'' contributions to the ``Gamma partition--function''.

As we turn to the ``Gamma free--energy'', the same nonperturbative corrections can be obtained very easily from the reflection formula
\be\label{refle}
\Gamma \left( \kappa \right) \Gamma \left( -\kappa \right) = - \frac{\pi}{\kappa \sin \pi \kappa}.
\ee
\noindent
Indeed, one may now obtain the final result from a two line calculation \cite{b91}. The idea is to directly find the (exponentially) improved version of the asymptotic expansion (\ref{omegaasympexp}), in the sector $\frac{\pi}{2} < \theta < \pi$, and which we name $\Omega_+ (\kappa)$. Upon the analytic continuation $-\kappa = \rme^{-\rmi\pi} \kappa$ to the sector $\frac{\pi}{2} < \theta < \pi$, where $- \frac{\pi}{2} < \arg \left( -\kappa \right) < 0$, one may use the expansion (\ref{stand}) for  $\Gamma(-\kappa)$ in (\ref{refle}) to obtain, upon taking the logarithm of (\ref{refle}),
\be\label{sec1}
\Omega_+ (\kappa) = \Omega (\kappa)-\log\left(1-e^{2 \pi i \kappa}\right),
\ee
\noindent
where the plus subscript refers to the sector past $+\frac{\pi}{2}$. In the sector $-\pi < \theta < - \frac{\pi}{2}$ one still obtains the above expression (\ref{sec1}), but with $\rme^{2\pi\rmi\kappa}$ replaced by $\rme^{-2\pi\rmi\kappa}$ instead (and the plus subscript naturally gets replaced by a minus subscript). Thus, in both sectors the corrections to the expansion (\ref{stand}) are always exponentially suppressed, with the discontinuity across the Stokes lines $\theta = \pm \frac{\pi}{2}$ given by
\be\label{loggammastokesdisc}
\Delta\, \Omega(\kappa) \Big|_{\theta = \pm \frac{\pi}{2}} = - \log \left( 1 - \rme^{\pm2\pi\rmi\kappa} \right) = \sum_{m=1}^{+\infty} \frac{\rme^{\pm2\pi\rmi\kappa m}}{m},
\ee
\noindent
which is, as expected, essentially related to the logarithm of the Stokes discontinuity for the ``Gamma partition--function'', (\ref{gammastokesdisc}). It is rather tempting to understand these terms as instanton contributions, with ``instanton action'' $S_{\mathrm{inst}}^{(m)} = W(z_m) - W(z_0) = \CW_{0m} = - 2\pi\rmi m$. For this identification to be valid, one expects that the instanton(s) with least action, $S^{(-1)} = 2\pi\rmi$ and $S^{(1)} = -2\pi\rmi$, will yield the leading contributions controlling the large--order behavior of the perturbative expansion (\ref{omegaasympexp}), as we have discussed before in section \ref{sec:lo}. Setting as usual
\be
\frac{1}{\kappa} \Omega^{(0)} (\kappa) \sim \sum_{g=1}^{+\infty} \kappa^{-2g}\, \Omega_g^{(0)}, \qquad \Omega_g^{(0)} \equiv \frac{B_{2g}}{2g \left( 2g-1 \right)},
\ee
\noindent
and
\be
\frac{1}{\kappa} \Omega ^{(1)} (\kappa) \sim \rmi \kappa^b\, \rme^{- \kappa A} \sum_{g=0}^{+\infty} \kappa^{-g}\, \Omega_{g+1}^{(1)},
\ee
\noindent
the standard large--order analysis yields
\be
\Omega^{(0)}_g \sim \frac{\Gamma \left( 2g+b \right)}{\pi A^{2g+b}} \left( \Omega^{(1)}_1 + \frac{\Omega^{(1)}_2 A}{2g+b-1} + \cdots \right)
\ee
\noindent
and, consequently,
\be
\frac{\Omega_{g+1}^{(0)}}{4 g^2\, \Omega_{g}^{(0)}} = \frac{1}{A^2} + \CO \left( \frac{1}{g} \right).
\ee
\noindent
From the exact expression for the $\Omega_{g}^{(0)}$ coefficients it immediately follows
\be
\frac{\Omega_{g+1}^{(0)}}{4 g^2\, \Omega_{g}^{(0)}} = \frac{B_{2g+2}}{4 g^2\, B_{2g}} \left( 1 - \frac{2}{g} + \CO \left( \frac{1}{g^2} \right) \right) = - \frac{1}{\left( 2\pi \right)^2} + \CO \left( \frac{1}{g} \right),
\ee
\noindent
so that $A = \pm 2\pi\rmi$ just as expected from the instanton analysis. Thus, we see that indeed the one--instanton contributions control the large--order behavior of the perturbation theory, supporting the important identification of Stokes discontinuities with instanton contributions.

An interesting property of our instanton actions is that $S^{(m)} = m S^{(1)}$ so that the action of the $m$--th one--instanton equals the $m$--th multi--instanton action. In this case, one could interpret the result (\ref{loggammastokesdisc}) as being exact, in the sense of including all multi--instanton corrections, and to all loop orders. For this identification to be valid, one expects to fully reconstruct the perturbative coefficients of the ``Gamma free--energy'' out of its complete multi--instanton series, as in (see, \textit{e.g.}, \cite{msw07})
\bea
\Omega_g^{(0)} &=& \frac{1}{2\pi\rmi} \int_{-\rmi \infty}^{+\rmi \infty} \rmd\xi\, \xi^{2g-2}\, \Delta\, \Omega (\xi) = \nonumber \\
&=& \frac{(-1)^{g+1}}{2\pi} \sum_{m=1}^{+\infty} \frac{1}{m} \left( \int_0^{+\infty} \rmd x\, x^{2g-2}\, \rme^{- 2\pi m x} + \int_{-\infty}^0 \rmd x\, x^{2g-2}\, \rme^{2\pi m x} \right).
\eea
\noindent
This is actually just one of the few examples on resurgent relations we have obtained in our earlier hyperasymptotic discussion. It is simple to see that the precise result is obtained, in full accordance with our multi--instanton expectations:
\be
\Omega_g^{(0)} = \frac{2(-1)^{g+1}(2g-2)!}{\left( 2\pi \right)^{2g}} \sum_{m=1}^{+\infty} \frac{1}{m^{2g}},
\ee
\noindent
where one just needs to use the representation of even Bernoulli numbers in terms of the zeta function, that we have used several times before, in order to check the result.

We have thus seen very clearly, at the familiar free--energy level, that the contributions arising from the Stokes line discontinuities are precisely the usual nonperturbative instanton contributions. The exact matching within the matrix model examples we consider in this work will be made complete in the next section.

\subsection{Instantons as Stokes Phenomena in Matrix Models}

In the previous section we have identified the Stokes discontinuities of the Gamma function with instanton effects of the corresponding one--dimensional integral, defining either a partition function or a free energy. We shall now see how this analysis carries through to the matrix models we are interested in. Let us start by considering the Gaussian and Penner matrix models, and how their instantonic sectors follow from a Stokes analysis of the corresponding free energies.

As computed at an earlier stage, the exact free energy for the Gaussian matrix model is given by
\be\label{fg}
F_{\mathrm{G}} = \frac{1}{2} N^2 \log g_s - \frac{1}{2} N \log 2\pi + \log G_2 (N+1),
\ee
\noindent
and the exact free energy for the Penner matrix model (normalized against the Gaussian measure) is
\be\label{fp}
\CF_{\mathrm{P}} = \frac{N}{g_s} \log g_s + \frac{1}{2} N^2 \log g_s - \frac{1}{2} N \log 2\pi + \log G_2 \left( N+\frac{1}{g_s}+1 \right) - \log G_2 \left( \frac{1}{g_s}+1 \right).
\ee
\noindent
It is rather evident from these expressions that both cases have their asymptotic expansions associated to Poincar\'e asymptotic expansions of the Barnes function. As such, the instanton sectors of these two matrix models will be dictated by the Stokes structure of the logarithm of the Barnes function. But due to the integral representation \cite{a03}
\be
\log G_2 (N+1) = \frac{1}{2} N \log 2\pi -\frac{1}{2} N (N-1) + N \log \Gamma (N) - \int_0^N \rmd n\, \log \Gamma (n)
\ee
\noindent
the Stokes structure of the Barnes free--energy is given in terms of the Stokes structure of the Gamma free--energy, which we have previously analyzed in greater detail. In particular
\bea
\Delta\, \log G_2 (N+1) \Big|_{\theta = \pm \frac{\pi}{2}} &=& N\, \Delta\, \Omega (N) \Big|_{\theta = \pm \frac{\pi}{2}} - \int_0^N \rmd n\, \Delta\, \Omega(n) \Big|_{\theta = \pm \frac{\pi}{2}} = \nonumber \\
&=& \sum_{m=1}^{+\infty} \left( \frac{N}{m} \mp \frac{1}{2\pi\rmi m^2} \right) \rme^{\pm2\pi\rmi N m} \mp \frac{\rmi \pi}{12}.
\eea
\noindent
At the Stokes lines $N = \pm \rmi |N|$ and the discontinuities are exponentially suppressed as expected. It immediately follows\footnote{Notice that the factor $\frac{\rmi\pi}{12}$ is a genus one artifact and we drop it in the following. Indeed, when computing the Borel transform of the Gaussian free energy (equivalently, of the logarithm of the Barnes function), we start the sum at genus $g=2$ in order to avoid the problematic logarithmic terms at genus zero and one, which do not contribute to the large--order behavior in any case. If one were to---\textit{incorrectly}---start the sum at genus $g=1$, while still making use of the Bernoulli expression (\ref{gs}) in order to compute this genus $g=1$ term in the sum, one would precisely find this spurious $\frac{1}{12}$ contribution.}
\be
\Delta\, F_{\mathrm{G}} = \frac{\rmi}{2\pi\bar{g}_s}\, \sum_{m=1}^{+\infty} \left( \frac{2\pi t}{m} + \frac{\bar{g}_s}{m^2} \right) \rme^{- \frac{2\pi t\, m}{\bar{g}_s}}
\ee
\noindent
and
\be
\Delta\, \CF_{\mathrm{P}} = \frac{\rmi}{2\pi\bar{g}_s}\, \sum_{m=1}^{+\infty} \left( \frac{2\pi \left( t+1 \right)}{m} + \frac{\bar{g}_s}{m^2} \right) \rme^{- \frac{2\pi \left( t+1 \right) m}{\bar{g}_s}} - \frac{\rmi}{2\pi\bar{g}_s}\, \sum_{m=1}^{+\infty} \left( \frac{2\pi}{m} + \frac{\bar{g}_s}{m^2} \right) \rme^{- \frac{2\pi m}{\bar{g}_s}}.
\ee
\noindent
These results, where we have used $t=g_s N$ and restricted to the Stokes line at $\theta = + \frac{\pi}{2}$, and upon the identification of the $\Delta$ discontinuity with $-{\mathrm{Disc}}$, precisely match our results for the full instanton sector of both Gaussian and Penner matrix models, obtained earlier in section \ref{sec:lomm} via Borel summation (or to be later analyzed via the use of trans--series methods). We have not addressed the case of the Chern--Simons matrix model, as its free energy is given by the logarithm of the quantum Barnes function, for which we do not know of any appropriate hyperasymptotic framework which would allow for a derivation of its Stokes discontinuities. However, we believe it should be possible to study the hyperasymptotics of the quantum Barnes function in a similar fashion to the one above (see also \cite{k07}).

Some interesting lessons may be drawn from our Stokes analysis. Because the appearance of the nonperturbative ambiguity of the matrix models' free energies is related to Stokes phenomena, intimately associated to discontinuities of the \textit{asymptotic approximation}, we see that this nonperturbative ambiguity is in fact an artifact of the \textit{semiclassical}, large $N$ analysis. Clearly, the exact free energies, (\ref{fg}) and (\ref{fp}), are given by the logarithm of entire functions in the complex plane, with no discontinuities. The Stokes lines, and thus the instanton corrections, appear only at the very moment we select a particular saddle and semiclassically evaluate the partition functions or free energies. On the other hand, it is also clear that it is just in this semiclassical limit that the notion of target space in the holographically dual theory emerges. For instance, we have seen in an earlier section that, in the $c=1$ case, the geometry of the target space arises from the matrix model spectral curve, which gets identified with the derivative of the FZZT disk partition function. Even more manifestly, in the case of the Chern--Simons matrix model the spectral curve of the matrix model coincides with the mirror curve of the mirror CY to the resolved conifold. One is led to conclude that if one is to consider the exact free energies, (\ref{fg}) and (\ref{fp}), as the nonperturbative definitions of the holographically dual models, then it appears the ``exact quantum'' target spaces are very different from the semiclassical ones; basically at the nonperturbative level the notion of target space as a smooth geometry is lost. Interestingly enough, this discrepancy between ``semiclassical'' and ``exact quantum'' target spaces has also been advocated in \cite{mmss04}, focusing on the example of non--critical strings with $c<1$. In particular, Stokes phenomena was also identified therein as a source of instanton corrections.

As we shall review in the upcoming section \ref{sec:it}, in the context of matrix models 
the nonperturbative partition function is obtained by summing over all saddles of the matrix integral. In particular, it is the averaging over all possible semiclassical  geometries that leads to the background independence of the nonperturbative partition function, as described in \cite{e08, em08}.

\subsection{Smoothing the Nonperturbative Ambiguity}

As we have explained, the Stokes discontinuities are discontinuities of the asymptotic expansions and not of the functions under approximation, and we have made this clear in some examples. It so happens that in most cases one does not know the function we wish to approximate and, as such, one needs to devise methods to smooth the Stokes discontinuity \cite{b89}. The need for the smoothing should be clear: it maintains the validity of the asymptotic expansions even as we cross the Stokes lines. This is what we shall describe now: the universal smoothing proposed in \cite{b89}, given by an uniform approximation involving the error function, and universally describing Stokes phenomena, which naturally makes the strength of the subdominant contribution grow smoothly from $0$ to $1$ across the Stokes line (upon where it equals $\frac{1}{2}$).

The idea goes as follows. Hoping to maintain the validity of our asymptotic expansions as we cross Stokes lines, let us write these as
\be
\CZ_{n} (\kappa) \sim \sum_{g=0}^{+\infty} \frac{\zeta_g (n)}{\kappa^g} + \rmi \sum_{m} S_{m} (\kappa)\, \Delta\, \CZ_{n} (\kappa) \Big|_{\theta = - \sigma_{nm}},
\ee
\noindent
where we have implicitly included the remainder associated to optimal truncation within the infinite sum, and where we have introduced the Stokes multiplier function $S_{m} (\kappa)$ weighting the subdominant exponentials and which will smooth the transitions across the Stokes lines. The true power of the Stokes multiplier function arises from the fact that, if the appropriate variables are chosen to cross the Stokes lines, then this function is \textit{universal}, within a wide class of problems\footnote{But see \cite{c96} for a counter example.}. This appropriate ``universal'' variable involves the singulant $\CW_{nm}$, specifying contours through adjacent saddles, \textit{i.e.}, specifying the location of the Stokes lines. The Stokes multiplier function may thus be written as \cite{b89}
\be
S_{m} (\kappa) = \frac{1}{2} \big( 1 + \mathrm{erf} \left( s_{nm} (\kappa) \right) \big),
\ee
\noindent
with $\mathrm{erf} (x)$ the error function and where we defined the Stokes variable
\be
s_{nm} (\kappa) \equiv \frac{\im \left( \kappa \CW_{nm} \right)}{\sqrt{2\, \re \left( \kappa \CW_{nm} \right)}}.
\ee
Observe that this is not a solution to the nonperturbative ambiguity problem as there is a choice of a real constant implicit in this result: the choice that before the Stokes line the exponentially suppressed contributions actually vanish. This is of course related to a choice of integration contour in the inverse Borel transform used in the calculation of the Stokes multiplier \cite{b89}. What the Stokes multiplier function does is to reduce the nonperturbative ambiguity to the choice of the real constant describing the intensity of the exponentially suppressed terms before the Stokes line, describing the crossing in an universal fashion. For a recent discussion of the nonperturbative ambiguity and choice of inverse Borel transform, within the matrix model context, see \cite{m08}.

For the ``Gamma partition--function'' with singulant $\kappa \CW_{0m} = - 2\pi\rmi\kappa m$, the Stokes variable becomes
\be
s_{0m} (\kappa) = \frac{\im\, \left( - 2\pi\rmi\kappa m \right)}{\sqrt{2\, \re\, \left( - 2\pi\rmi\kappa m \right)}} = - \sqrt{\pi |\kappa| |m|}\, \frac{\cos \theta}{\sqrt{\sin \theta}}\, \rme^{\rmi \frac{\pi}{4}\left(1 - \mathrm{sgn} (m)\right)}
\ee
\noindent
and the smoothing becomes implemented by the Stokes multiplier
\be\label{smooth}
S_{m>0} (\kappa) = \frac{1}{2} \left( 1 + \mathrm{erf} \left( - \sqrt{\pi |\kappa| m}\, \frac{\cos \theta}{\sqrt{\sin \theta}} \right) \right),
\ee
\noindent
where we have explicitly written the case where one crosses the upper imaginary axis. Notice that in the immediate vicinity of the Stokes line one has $- \frac{\cos \theta}{\sqrt{\sin \theta}} = \left( \theta - \frac{\pi}{2} \right) + \cdots$, simplifying the argument of the error function in that region. At the level of the ``Gamma free--energy'', completely analogous to the partition function analysis as the singulants are precisely the same, this result was interpreted in \cite{b91} as a distinct---but universal---Stokes smoothing for each $m$--th small exponential. One may, in this light, separately understand the appearance of each exponential.

For both Gaussian and Penner models, the discontinuity's singulants are the same as for the Gamma function and, akin to the previous discussion, one is in the presence of infinitely many smoothings \cite{b91} with Stokes multipliers
\be
S_{m>0} (N) = \frac{1}{2} \left( 1 + \mathrm{erf} \left( \left( \theta - \frac{\pi}{2} \right) \sqrt{\pi |N| m} \right) \right),
\ee
\noindent
near the Stokes line $\theta = \frac{\pi}{2}$. This explains the appearance of each suppressed exponential, or each distinct instanton contribution, in a separate but universal and smooth manner. In this way, one may readily obtain formal expressions for the Stokes smoothing, also yielding formal expressions for the free energies of these models which allow, for instance, one to cross the Stokes lines and reach any nonperturbative point in the complex $N$--plane.

\section{Semiclassical Interpretation of Instantons}

At this stage we have a very good understanding of nonperturbative phenomena in $c=1$ matrix models and topological strings, with all the information we have gathered both from Borel and Stokes analysis. However, and because at the end of the day we are analyzing large $N$ matrix models in a saddle--point approximation, we would like to understand these nonperturbative instanton corrections directly from a semiclassical large $N$ point of view, \textit{i.e.}, directly in the matrix model language. This is what we shall do in this section, as we provide a semiclassical interpretation of instantons in terms of eigenvalue tunneling, across a multi--sheeted effective potential, and we also suggest a spacetime interpretation for the nonperturbative effects we have just obtained in the preceding sections.

\subsection{Instantons as Eigenvalue Tunneling}\label{sec:it}

We shall begin by recalling how instanton effects, which control the large--order behavior of the $1/N$ expansion, are interpreted as an eigenvalue tunneling effect, within the matrix model context. Recalling our discussion on matrix models, in diagonal gauge the one--matrix model partition functions is
\be
Z = \frac{1}{N!} \int \prod_{i=1}^N \left( \frac{\rmd \lambda_i}{2\pi} \right) \Delta^2 (\lambda)\, \rme^{- \frac{1}{g_s} \sum_{i=1}^N V(\lambda_i)}.
\ee
\noindent
As it stands this definition cannot be complete, as the integration contours for each eigenvalue $\lambda_i$ still need to be specified. It may happen that the above integral is not well--defined as a convergent, real integral, in which case the model needs to be properly defined by analytic continuation, \textit{i.e.}, by the choice of an appropriate contour in the complex plane such that the integral becomes convergent. Indeed, in general, the various phases of matrix models are separated by singular domains (in the space of complex potentials), where no large $N$ limit exists \cite{d91, d92, bde00}. Removal of the ``divergent regions'' where $\re\, V(z) \to - \infty$ as $|z| \to \infty$ corresponds to holes in the complex plane, in which case one is led to decompose a generic integration path $\gamma$ on a homological basis of paths $\{ \gamma_1, \ldots, \gamma_s \}$ as\footnote{For a polynomial potential of degree $d$ there will be $d$ holes in the complex plane, in which case the dimension of the homological basis will be $s = d-1$.}
\be\label{gammacigammai}
\gamma = \sum_{k=1}^s \zeta_k\, \gamma_k,
\ee
\noindent
where we shall place $N_i$ eigenvalues on the path $\gamma_i$, with arbitrary distribution $\{ N_i \}$ but such that $\sum_{i=1}^s N_i = N$. As we shall make clear in the following, the coefficients $\zeta_k$ may be regarded as theta--parameters leading to different theta--vacua \cite{d92}. We may then define, with the appropriate symmetrizations,
\be
\widehat{Z} (N_1, \ldots, N_s) = \frac{1}{N_1! \cdots N_s!} \int_{\gamma_1} \prod_{i_1 = 1}^{N_1} \frac{\rmd\lambda_{i_1}}{2\pi} \dots \int_{\gamma_s} \prod_{i_s = 1}^{N_s} \frac{\rmd\lambda_{i_s}}{2\pi}\, \Delta^2 (\lambda)\, \rme^{- \frac{1}{g_s} \sum_{i=1}^N V(\lambda_i)}.
\ee
\noindent
For particular choices of the integration contours and particular choices of the filling fractions $\epsilon_i = \frac{N_i}{N}$, with $i = 1, \ldots, s$, the free energy $\log \widehat{Z} (N_1, \ldots, N_s)$ may have a perturbative large $N$ expansion\footnote{While in general there is no topological large $N$ expansion, there are of course some cases where this may be achieved: for instance in the case of degree $d$ polynomial potential one may choose as homological basis the $d-1$ steepest--descents paths which go through each of the $d-1$ critical points of the potential \cite{d91, d92}.}
\be
\log \widehat{Z} (N_1, \ldots, N_s) = \widehat{F} (N_1, \ldots, N_s) = \sum_{g=0}^{+\infty} N^{2-2g}\, F_g (t),
\ee
\noindent
where one finds the usual large--order behavior of $F_g \sim (2g)!$ rendering the topological $1/N$ expansion asymptotic. Nonperturbative effects associated to singularities in the complex Borel plane are interpreted as instanton configurations as follows.

Regarding $\widehat{Z} (N_1, \ldots, N_s)$ as the partition function associated to a specific topological sector, characterized by the filling fraction $\epsilon_1, \ldots, \epsilon_s$, it becomes natural to consider the general partition function where one sums over all possible ways of distributing the $N$ eigenvalues
\be\label{ZgammaZhat}
\CZ (\zeta_1, \ldots, \zeta_s)  = \sum_{\sum_{i=1}^s N_i = N}\, \zeta_1^{N_1} \cdots \zeta_s^{N_s}\, \widehat{Z} (N_1, \ldots, N_s),
\ee
\noindent
and where it is now clear that the $\zeta_k$ play the role of theta--parameters. This expression has been proposed by \cite{e08, m08, em08} to provide a \textit{nonperturbative} partition function for the matrix model. That (\ref{ZgammaZhat}) realizes such a nonperturbative completion of the theory is made clear by understanding how it encodes all possible multi--instanton corrections. It was pointed out in \cite{msw08} that if one is to consider the partition functions associated to two distinct topological sectors, with distinct fillings, $\{N_i\}$ and $\{N_i'\}$, one finds \cite{msw08}
\be
\frac{\widehat{Z} (N_1', \ldots, N_s')}{\widehat{Z} (N_1, \ldots, N_s)} \sim \rme^{-\frac{1}{g_s} \sum_{i=1}^s \left( N_i - N_i' \right) \frac{\partial F_0}{\partial t_i}}, \qquad t_i = g_s N_i,
\ee
\noindent
implying that once one selects a reference background, $\{N_i\}$ say, \textit{all} other sectors are different \textit{instanton} sectors of the matrix model. Let us consider the one--cut cubic matrix model in the following, in order to be a bit more concrete. The cubic potential $V(z) = \frac{1}{2} z^2 + \frac{g}{3} z^3$ has two critical points; the maximum located at $z=z_*$ and the metastable minimum $z=0$, as illustrated in Figure \ref{cubicveff}. There are two steepest--descent paths naturally associated to these critical points, $\gamma_0$ through $z=0$ and $\gamma_1$ through $z=z_*$. While the lowest energy configuration is associated to having all eigenvalues integrated along $\gamma_0$, this is an unstable configuration due to tunneling mediated by instanton configurations, which correspond to the integration of eigenvalues along $\gamma_1$ \cite{msw07}. In particular, the partition function in the one--instanton sector is given by \cite{msw07}
\be
Z_N^{(1)} = \frac{1}{2\pi}\, Z^{(0)}_{N-1} \int_{x \in \gamma_1} \rmd x \left\langle \det (x \mathbf{1} - M')^2 \right\rangle^{(0)}_{N-1}\, \rme^{-\frac{1}{g_s} V(x)}.
\ee
\noindent
Let us explain this expression. We removed one out of the $N$ eigenvalues in the cut, $x$, and we are integrating it over the non--trivial saddle--point, $\gamma_1$. The remaining $N-1$ eigenvalues are, of course, still integrated over the leading saddle, associated to $\gamma_0$, and $Z_N^{(0)}$ is the zero--instanton partition function evaluated around this standard saddle--point. Finally, $M'$ is an $(N-1) \times (N-1)$ hermitian matrix, all of its eigenvalues still integrated around the standard saddle--point in the zero--instanton correlation function. We refer the reader to \cite{msw07} for the details on the explicit computation of the quantity above. As it turns out this expression implies that, at leading order, the one--instanton contribution to the free energy is given by
\be
F^{(1)} = \frac{Z_N^{(1)}}{Z_N^{(0)}} = \frac{1}{2\pi}\, \frac{Z_{N-1}^{(0)}}{Z_N^{(0)}} \int_{x \in \gamma_1} \rmd x\, \left\langle \det (x \mathbf{1} - M')^2 \right\rangle^{(0)}_{N-1}\, \rme^{-\frac{1}{g_s} V(x)} \sim \rmi\, \rme^{-\frac{A}{g_s}},
\ee
\noindent
where the instanton action is \cite{msw07}
\be\label{instdiff}
A = V_{\mathrm{h,eff}}(x_0) - V_{\mathrm{h,eff}}(b) = \int_b^{x_0} \rmd z \, y(z), 
\ee
\noindent
with $b$ the endpoint of the single cut $\CC=[a,b]$. This formula has an obvious semiclassical interpretation, as the instanton action (\ref{instdiff}) is nothing but the height of the potential barrier under which instantons are tunneling. Furthermore, a configuration where $N_1$ eigenvalues are integrated along the contour $\gamma_1$, and $N_0 = N - N_1$ along $\gamma_0$, can be naturally regarded either as a two--cut solution with filling fractions $N_0$ and $N_1$, or as a $N_1$--instanton excitation above the reference one--cut solution \cite{msw08}. It has thus become clear that the general partition function (\ref{ZgammaZhat}) provides the nonperturbative completion of the matrix model since, by summing over all the filling fractions, it naturally encompasses all the multi--instanton configurations of the theory.

\FIGURE[ht]{
\label{cubicveff}
\centering
\psfrag{zst}{$z_*$}
\epsfig{file= 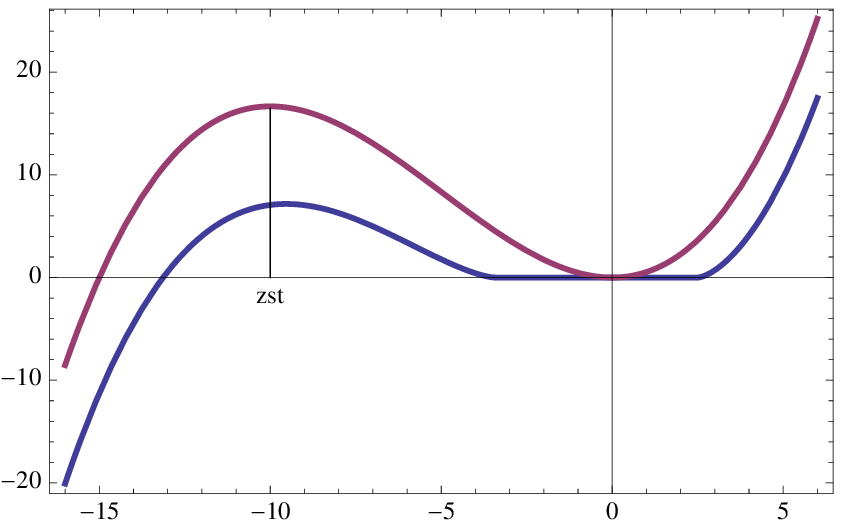, width=12cm, height=7cm}
\caption{The real part of the holomorphic effective potential for the one--cut cubic matrix model, in blue, and the simple cubic potential, in purple, for the values $g=0.1$ and $t=3$.}
}

What we shall see in the following is that, for our class of $c=1$ matrix models and topological strings, one also has to allow for generalized integration contours which contain copies of the leading saddle--point and which find themselves going through different sheets of the multi--valued holomorphic effective potential.

\subsection{Instantons and the Multi--Sheeted Effective Potential}

Let us now turn to the $c=1$ matrix models we are considering in this work. A quick glance at the real part of the holomorphic effective potentials for the three models, plotted in Figure \ref{veffs}, is enough to realize that there are no critical points of these potentials outside their single cuts. While this may not seem surprising for real and positive $g_s$, as the asymptotic expansions are Borel summable and there are no nonperturbative ambiguities in the reconstruction of the partition functions, we also know that when allowing for imaginary values of the string coupling our topological expansions become non--Borel summable, with instantons controlling large order. As such, and given that the discussion in the previous section cannot be applied, at least not in a straightforward fashion, in what follows we shall have to define the instanton actions for our models by exploiting the structure of the holomorphic effective potentials in the complex plane.

\FIGURE[ht]{
\label{veffs}
\centering
\epsfig{file=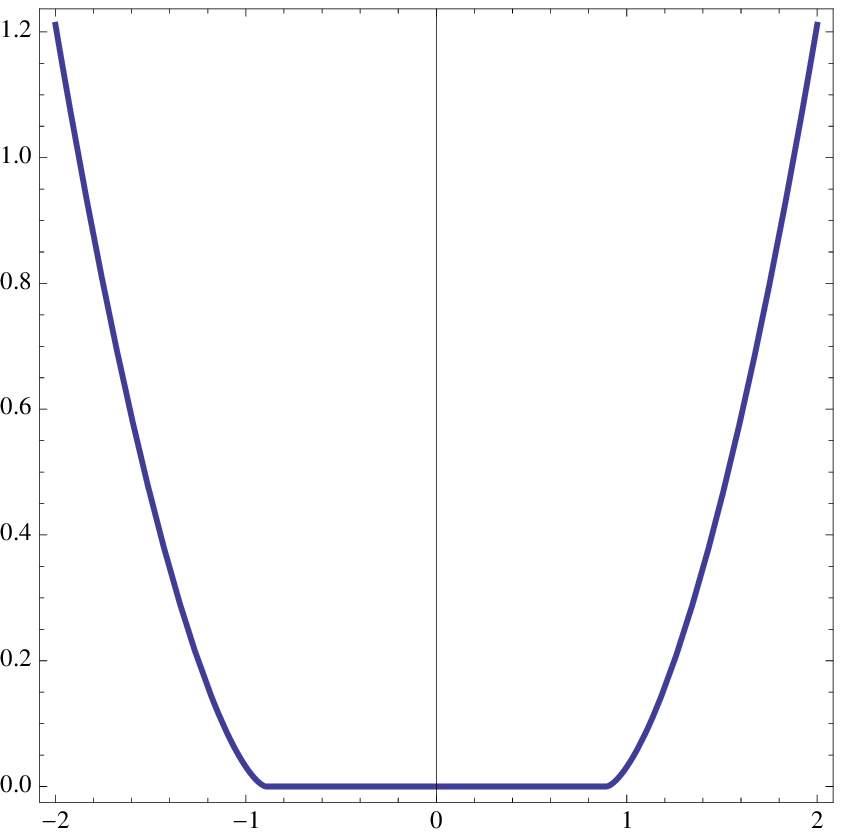, width=4.5cm, height=4.5cm}
$\quad$
\epsfig{file=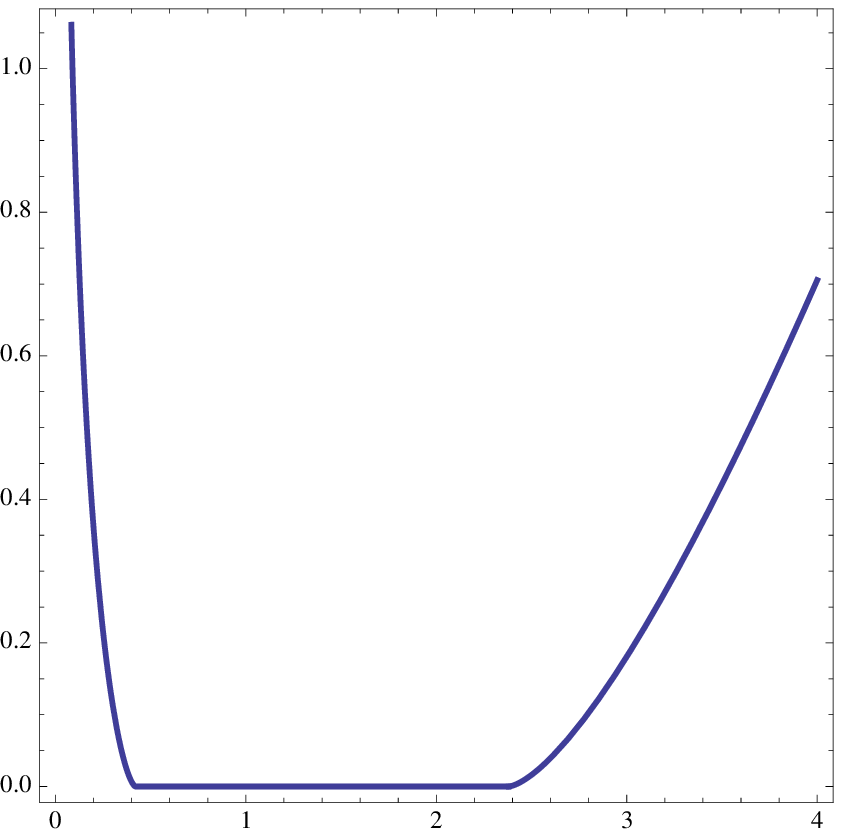, width=4.5cm, height=4.5cm}
$\quad$
\epsfig{file=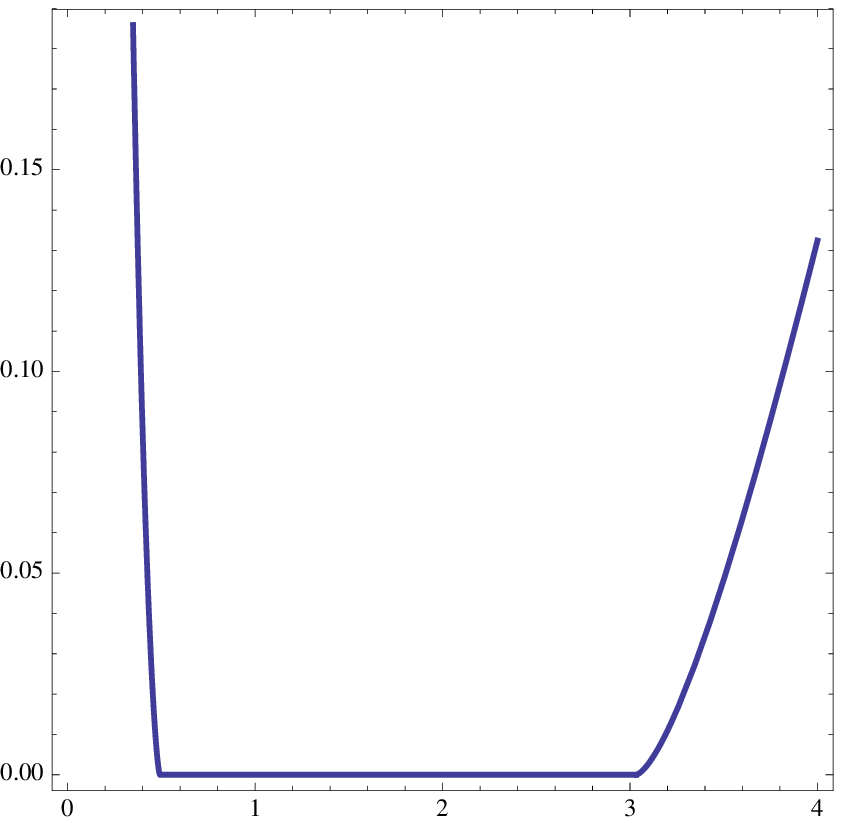, width=4.5cm, height=4.5cm}
\caption{The real part of holomorphic effective potentials for Gaussian, Penner and
Chern--Simons matrix models, from left to right, respectively, when $t=0.2$.}
}

Let us have a closer look at the holomorphic effective potentials for Gaussian, (\ref{veg}), Penner, (\ref{vep}), and Chern--Simons, (\ref{vecs}), matrix models. Due to the presence of either logarithmic or dilogarithmic functions these holomorphic effective potentials have a multi--sheeted branch structure in all examples, a clear feature from the structure of their Stokes lines (recall Figures \ref{gstokescurve}, \ref{pstokescurve} and \ref{csstokescurve}). As one takes the derivative of the effective potential, in order to obtain the spectral curve, all logarithmic or dilogarithmic sheets artificially collapse on top of each other. As such, the spectral curve cannot explicitly see this structure. One way to lift this artificial degeneracy and make the multi--sheeted structure of the effective potentials manifest, is to repeat what we did in section \ref{stogamma}, in the context of hyperasymptotic analysis.

Let us start with the Gaussian model. Akin to what we did for the ``Gamma partition--function'' it is useful to first change variables from $z$ to
\be
\rme^u = z + \sqrt{z^2-4t},
\ee
\noindent
in which case the Gaussian holomorphic effective potential gets written as
\be
V_{\mathrm{h;eff}}^{\mathrm{G}} (u) = \frac{1}{8} \rme^{2u} - 2 t^2 \rme^{-2u} - 2 t\, u + 2 t \log 2\sqrt{t}.
\ee
\noindent
In this new variable the critical points (or saddles) of the potential are located at\footnote{We use the standard definition of the logarithm, $\log z \equiv \log |z| + \rmi \arg z$, and make explicit its multi--sheeted structure by defining $\arg z$ up to $2\pi\rmi$, \textit{i.e.}, $\left. \arg z \right|_n \equiv \arg z + 2\pi\rmi n$ with $n \in \BZ$.}
\be
\rme^{2u} = 4t \qquad \Rightarrow \qquad u^{(n)}_{\pm} = \log (\pm 2\sqrt{t}) + \rmi \pi n, \qquad n \in \BZ,
\ee
\noindent
which is just the solution of $a,b = \rme^u$. The interesting property of the $u$ variable is that it lifts the aforementioned artificial degeneracy where all sheets have collapsed on top of each other, and insures that one identifies all possible critical points of the holomorphic effective potential in its fully unfolded multi--sheet domain. One furthermore selects $n=0$ for the reference saddle and identifies all saddles $m \not = 0$ as adjacent saddles to $n=0$.

In this new variable the multi--sheeted structure of the holomorphic effective potential is much simpler to visualize: one finds an infinite number of copies of the original single cut (with each distinct degenerate sheet parametrized by $n$), with endpoints $u^{(n)}_-$ and $u^{(n)}_+$, and where $V_{\mathrm{h;eff}}^{\mathrm{G}} (u)$ is constant on each replica of the cut. This, of course, has dramatic implications: as we proceed to evaluate the instanton action via $A = V_{\mathrm{h;eff}}(x_0) - V_{\mathrm{h;eff}}(b)$ the sheets are no longer degenerate and we shall find non--zero values every time we place an eigenvalue in another sheet. The placing of the eigenvalue is simple and totally analogous to the previous discussion of eigenvalue tunneling: one removes the single eigenvalue from the endpoint of the cut and places it at the starting point of the ``next'' cut, in the following sheet (in such a way that the spectral curve cannot ``see'' any difference in the configuration, only the holomorphic effective potential can), an idea first suggested in \cite{m05}. We conclude that the multi--instanton action (the singulant, in the hyperasymptotic language we used earlier, \textit{e.g.}, in section \ref{stogamma}) is given by the difference between the holomorphic effective potential evaluated on the principal sheet (corresponding to the choice $n=0$ and denoted with a $\star$ in the following) and its value on a generic sheet,
\be\label{semiG}
A^{\mathrm{G}}_{n} = V^{\mathrm{G}}_{\mathrm{h;eff}} ( u^{(n)}_+ ) - V^{\mathrm{G}}_{\mathrm{h;eff}} ( u^{\star}_+ ) = - 2 \pi \rmi n t = \frac{1}{2} \oint_{\gamma_n} \rmd z\, y(z).
\ee
\noindent
In the last equality we compute the instanton action as the integral of $y(z)\, \rmd z$ along $\gamma_n$, a non--contractible contour encircling the eigenvalue cut $n$ times\footnote{Here, $\gamma_1$ is the contour used to define the filling fraction in matrix models, \textit{i.e.}, the $A$--cycle. The factor of $\frac{1}{2}$ insures that the action describes moving an eigenvalue from the endpoint of the cut to the \textit{beginning} of the ``next'' cut in the multi--sheeted structure, \textit{i.e.}, that we keep the standard eigenvalue tunneling picture.}. Analogously, the instanton action may be regarded as half the shift due to the additive monodromy of the logarithm present in $V^{\rm G}_{\mathrm{h;eff}(x)}$ (recall equation (\ref{veg})). The action (\ref{semiG}) we have just obtained from a semiclassical viewpoint coincides with the one previously obtained with either Borel or Stokes analysis. It is also important to notice that the definition we are now suggesting for the instanton action is in perfect analogy with the one that we have previously described when addressing the semiclassical derivation of the Schwinger pair--production effect, in section \ref{sec:se}. With an appropriate identification of the parameters, one may identify the semiclassical Schwinger effective action (\ref{ac}) with the matrix model effective potential (\ref{veg}). As such, in here one may still interpret the instantons as an eigenvalue tunneling process, with the sub--barrier motion consisting of multiple windings in the complex plane around the eigenvalue cut, as in the discussion of section \ref{sec:se}.

As we move to the Penner model the holomorphic effective potential becomes more intricate, as displayed in (\ref{vep}), and we shall focus on obtaining the instanton action straight out of the monodromy shift of (\ref{vep}) around the cut $\CC=[a,b]$. There are now contributions from more than one logarithmic branch cut and one needs to properly consider them all. In this case, there are two possible ways to choose the contour: one may encircle the eigenvalue cut \textit{without} crossing the $\log z$ cut in (\ref{vep}), in which case one is forced to cross the other two logarithmic branch cuts; or one may encircle the eigenvalue cut \textit{crossing} the $\log z$ cut. In the first option, clockwise winding $n$ times around the cut $\CC=[a,b]$ one will cross two logarithmic branch cuts and obtain the global shift $-(2t+1) 2 \pi \rmi n - 2 \pi \rmi n = - 4 \pi \rmi n \left( t+1 \right)$\footnote{The reader may be puzzled by the fact the the integral of the curve around the cut, \textit{i.e.}, the $A$--cycle $\oint_A \rmd z\, y(z)$, yields $2(t+1)$ instead of $2t$ (as one would have expected from the normalization of the genus--zero resolvent at infinity, $\omega_0 (z) \sim \frac{1}{z} + \cdots$ as $z \to \infty$). The reason for this is that, when deforming the contour back from infinity to the cut $\CC$, we still have to pick the residues at $z=0$. This unusual feature is, of course, due to the fact that we are not dealing with a polynomial potential.}. This immediately yield the multi--instanton action
\be
A^{\mathrm{P}}_{n} = - 2 \pi \rmi \left( t+1 \right) n = \frac{1}{2} \oint_{\gamma_n} \rmd z\, y(z),
\ee
\noindent
a result which agrees with the one we have previously obtained via Borel and Stokes analysis. If, instead, one chooses the second option, which involves crossing the $\log z$ cut, this will naturally yield an additional contribution of $2 \pi \rmi m$ (a result we expected to find from Borel or Stokes analysis in any case), but it is not entirely clear to us how this relates to a cycle of the curve.

Finally, we need to confirm the validity of our proposal within the Chern--Simons matrix model. In this case, the holomorphic effective potential (\ref{vecs}) is rather complex, involving both logarithmic and dilogarithmic branch cuts. As such, we shall restrain to computing the instanton action from the monodromy around the cut $\CC=[a,b]$. If the cut $\CC = [a,b]$ is placed on the positive real axis, we may choose our contour $\gamma_n$ such that $\re\, (z) \geq 0$ when contouring the cut, in which case one also finds $\re\, (\xi) > 0$ and both $\log z$ and $\log \xi$ will have no monodromy. We will then be left with the contributions of $\log \left( 1-\xi \right)$ and $\log \left( 1-\rme^{-t} \xi \right)$. Since both $\rme^{-t} \xi$ and $\xi$ will be bigger than one on the right hand side of the cut, on the real axis, then both these logarithms will produce the same shift of $2\pi\rmi$, at each winding. Turning to the dilogarithmic dependence, we find a more complicated monodromy structure and we refer the reader to, \textit{e.g.}, \cite{m03, v07} for a more in depth analysis, or to appendix \ref{ap1} for a brief review of this topic. The dilogarithm $\mathrm{Li}_2 (z)$ has a branch cut starting at $z=1$, and we choose it to be on the positive real axis. However, as one moves from the principal sheet to a generic sheet, by crossing the principal branch cut, a new branch cut will emerge from $z=0$ to the right. As such, we shall need two distinct integers in order to specify the value of $\mathrm{Li}_2 (z)$ on a generic sheet, in terms of its value on the principal sheet, $\mathrm{Li}_2^\star (z)$. This turns out to be \cite{m03, v07}
\be
\mathrm{Li}_2 (z) = \mathrm{Li}_2^\star (z) + 2 \pi \rmi n \log (z) - 4 \pi^2 k n, \qquad n \in \BN_+, \quad k \in \BZ.
\ee
\noindent
In here, the integer $n$ counts how many times we wound clockwise in the complex plane, by crossing the principal branch cut, while the integer $k$ counts how many times we crossed the ``hidden'' branch cut. Putting it all together, it follows
\bea
\left. V_{\mathrm{h;eff}} \right|_{\mathrm{CS}} &=& \left. V_{\mathrm{h;eff}}^\star \right|_{\mathrm{CS}} + 4 \pi \rmi n \log \left( \rme^{-t} \xi \right) - 8 \pi^2 n k_1 - 4 \pi \rmi n \log \xi + 8 \pi^2 n k_2 = \nonumber \\
&=& \left. V_{\mathrm{h;eff}}^\star \right|_{\mathrm{CS}} - 4 \pi \rmi t n - 8 \pi^2 n \left( k_1 - k_2 \right).
\eea
\noindent
Setting $k=k_1-k_2$, with $k_1$ and $k_2$ being the two, \textit{a priori} different, windings around the ``hidden'' branch cut of the two dilogarithms in (\ref{vecs}), one may write
\be
\oint_{\gamma_{n,k}} \rmd z\, y(z) = - 4 \pi \rmi t n - 8 \pi^2 n k
\ee
\noindent
where $\gamma_{n,k}$ is a contour winding around the cut $\CC$ until it reaches the $(n,k)$ sheet. Again, on each $(n,k)$ sheet we find a copy of $\CC$, with the effective potential being  constant on each replica of the cut. The instanton action is thus given by
\be
A^{\mathrm{CS}}_{n,k} = - 2 \pi \rmi t n - 4 \pi^2 n k = \frac{1}{2} \oint_{\gamma_{n,k}} \rmd z\, y(z).
\ee
\noindent
Once again the instanton action we have obtained with our proposed semiclassical reasoning precisely matches the one we previously obtained from Borel or Stokes analysis.

It is interesting to note that in all cases above the instanton action is essentially given by the integral of the spectral curve along the $A$--cycle of the single cut, while in the cases considered in \cite{msw07, msw08} the instanton action was given by the integral of the spectral curve along the $B$--cycle of the cut(s). This complete our matrix model derivation of the instanton action. In the following we focus on the spacetime interpretation of the nonperturbative effects we have discussed do far.

\subsection{Spacetime D--Instanton Interpretation}

Nonperturbative effects in hermitian matrix models have been realized in terms of D--brane instanton effects in the holographically dual minimal models with $c<1$. In particular, in \cite{akk03} matrix model instantons, as described by eigenvalue tunneling, have been shown to match the disk contribution of ZZ branes. An analogous situation takes place in the case of $c=1$ where the $Z^{\mathrm{ZZ}}_{(k,1)}$ disk contribution (\ref{zz1}) coincides with the instanton action obtained from the MQM nonperturbative integral formula \cite{gk90, gm90, a04b}. In this case the singularities of the curve leading to the ZZ brane disk amplitude are not pinched cycles, but instead are points where the curve has self--intersections, as described in \cite{a04b}.

As usual, let us first analyze the Gaussian matrix model, or, equivalently, the $c=1$ string at self--dual radius. We have already observed that the MQM spectral curve is just an infinite covering of the hyperboloid defining the Gaussian spectral curve and identified correlators in the Gaussian matrix model with open amplitudes with FZZT boundary conditions. The natural next step is to identify the instanton action (\ref{semiG}) with the ZZ brane partition function in the dual $c=1$ theory. Moreover, considering the relation (\ref{rela}) between FZZT and ZZ boundary states, one may write the instanton action as the difference between two FZZT branes located at the branch point $u^\star_+$, on the principal sheet, and its replica $u^{(n)}_+$, on the $n$ sheet,
\be
A^{c=1}_n \left( = \frac{A^{\mathrm{G}}_n}{\rmi \bar{g}_s} \right) = Z^{\mathrm{FZZT}} (u^{(n)}_+) - Z^{\mathrm{FZZT}} (u^\star_+) = Z^{\mathrm{ZZ}} = \frac{1}{2\rmi\bar{g}_s} \oint_{\gamma_n} \rmd z\, y(z) = 2 \pi n \mu. 
\ee
\noindent
A very similar story goes through for the Penner model, thus completing the identification of the (double--scaled) Gaussian and Penner matrix model nonperturbative effects as ZZ brane D--instantons, in the dual $c=1$ string theory.

Let us now turn to the case of the Chern--Simons matrix model where, much as in the local curve backgrounds studied in \cite{m06}, the instanton action may be interpreted, in the dual large $N$ description on the resolved conifold, in terms of toric branes. In this case, the disk amplitude with (mirror of) toric--brane boundary conditions may be written, in terms of the mirror curve\footnote{For backgrounds with a matrix model dual, this mirror curve precisely coincides with the spectral curve.}, as \cite{av00, akv01}
\be
A^{(0)}_1 (x) = \int^x \rmd s\, y(s).
\ee
\noindent
\noindent
It thus follows that one may write the instanton action as the difference of two disk amplitudes; simply set
\be\label{toric}
A^{\mathrm{conif}}_{(n,k)} \left( = \frac{A^{\mathrm{CS}}_{(n,k)}}{\rmi\bar{g}_s} \right) = \frac{\rmi}{\bar{g}_s} A^{(0)}_1 (u_+^\star) - \frac{\rmi}{\bar{g}_s} A^{(0)}_1 (u_+^{(n,k)}) = -\frac{\rmi}{2 \bar{g}_s} \oint_{\gamma_{n,k}} \rmd z\, y(z) = \frac{2 \pi n}{\bar{g}_s} \left( t + 2 \pi \rmi k \right),
\ee
\noindent
where the toric branes have been place at the endpoint $u_+^\star$ and at its copy, $u_+^{(n,k)}$, on the sheet labelled by $n$ and $k$ (recall our previous discussion). The instanton action is in this way given by the tension of the domain wall interpolating in between the two branes (see as well \cite{msw07}).

At this point, two comments are in order. First, when we discussed the $c=1$ double--scaling limit we also checked that the spectral curve---alongside with all the open correlators---of the Chern--Simons model collapse, at the conifold point, to their corresponding $c=1$ values. In particular, this implies that in this limit also the instanton action reduces to the $c=1$ one which indicates, as also observed in \cite{m06}, that in the double--scaling limit nonperturbative effects due to toric branes reduce to the ones due to Liouville branes in minimal models. The second observation we want to make is that the instanton effects leading to the above action (\ref{toric}) involve vector multiplet moduli. This can be realized by simply noting that the instanton action (\ref{toric}) is given by the integral of the one--form $y\, \rmd z$ on the contour $\gamma$, which are, respectively, reductions of the complex structure three--form $\Omega$ and of the three--cycle $\Gamma$ for a local CY,
\be\label{generalA}
A^{\mathrm{conif}}_{(n,k)} = \rmi \int_\Gamma \Omega, \qquad \Gamma = C^\star_+ - C^{(n,k)}_+, 
\ee
\noindent
where the two--cycles $C_+^\star$ and $C_+^{(n,k)}$ are given by line--bundles, with base points $u^\star_+$ and $u^{(n,k)}_+$ on the spectral curve. The instanton action may in this way be identified with the tension of the domain wall interpolating in--between the two branes (see as well \cite{msw07}).

So far we provided a spacetime interpretation of nonperturbative effects in $c=1$ matrix models and topological strings in terms of Liouville and toric branes. However, there is a further, obvious, spacetime interpretation of the instanton expansion as due to BPS particle production, via the Schwinger effect that we have discussed earlier in this paper. Furthermore, it is possible to relate these particle production effects to spacetime D--brane instantons via a compactification to three dimensions, followed by a $T$--duality \cite{ov96}. Recently, in \cite{csu09}, this connection has been exploited in order to study the continuity, across walls of marginal stability, of nonperturbative effects in type II compactifications.

\section{Trans--Series and the Toda Equation}

Another approach to the calculation of instanton corrections to the free energy of matrix models and topological strings, and first developed away from criticality in \cite{m08}, deals with trans--series solutions in the orthogonal polynomial framework, rather than the spectral geometry as in the preceding section. In this section we wish to learn what this approach has to say on what considers Gaussian, Penner and Chern--Simons matrix models; in particular, we wish to confirm our instanton results in yet a novel setting. Let us begin by briefly reviewing the ideas behind the trans--series approach, and then apply it within our interests.

Let us first recall that in the orthogonal polynomial formalism one may compute the partition function via (\ref{zop}) once one knows the recursion coefficients $\{ r_n \}$. Making use of the definition (\ref{zop}) it is not too hard to obtain
\be\label{zzz}
\frac{Z_{N+1} Z_{N-1}}{Z_N^2} = r_N.
\ee
\noindent
In the continuum $N \to \infty$ limit the coefficients $r_n$ become a function $r_n \to R(x,g_s)$ of the variable $x = t \frac{n}{N} \in [0,t]$, where the function $R(x,g_s)$ can be determined by solving the so--called pre--string equation, a finite difference equation obtained from the continuum limit of the recursion for the coefficients $\{ r_n \}$ \cite{biz80}. Analogously, the continuum limit of equation (\ref{zzz}) produces the following Toda--like equation for the free energy, $F_N=\log Z_N\to F(t,g_s)$,
\be\label{toda}
\exp \left( F(t+g_s,g_s) - 2 F(t,g_s) + F(t-g_s,g_s) \right) = R(t,g_s).
\ee
\noindent
Given a solution to the recursion coefficients $R(x,g_s)$, this equation then determines the free energy of our model as a solution to the Toda hierarchy.

Let us now consider trans--series solutions (in the sense of exponential asymptotics) to the equations above\footnote{Notice that these trans--series solutions will only be valid in a specific region of the complex plane; as we change sectors and cross a Stokes line the asymptotics will change. This change will be given by a shift in the nonperturbative ambiguity parameter, $C$, as $C \to C+S$ with $S$ the Stokes multiplier.}. For that, consider the trans--series \textit{ans\"atz} for the recursion coefficients
\be\label{rtransseries}
R(x,g_s) = \sum_{\ell=0}^{+\infty} C^\ell R^{(\ell)} (x,g_s),
\ee
\noindent
with the zero--instanton contribution given by
\be \label{r0inst}
R^{(0)} (x,g_s) = \sum_{n=0}^{+\infty} g_s^{2n} R^{(0)}_{2n} (x)
\ee
\noindent
and the $\ell$--instanton contributions given by
\be\label{rellinst}
R^{(\ell)} (x,g_s) = R^{(\ell)}_1 (x)\, \rme^{- \frac{\ell A(x)}{g_s}} \left( 1 + \sum_{n=1}^{+\infty} g_s^n\, R^{(\ell)}_{n+1} (x) \right), \qquad \ell \ge 1.
\ee
\noindent
Plugging this \textit{ans\"atz} into the pre--string equations one may, in principle, determine recursively both the instanton action $A$ and all the loop terms $R^{(\ell)}_{n} (x)$. Once this is done, one finally plugs the trans--series solution for $R(t,g_s)$ on the right--hand side of equation (\ref{toda}) and solves it with a trans--series \textit{ans\"atz} for the free energy
\be\label{ftransseries}
F(t,g_s) = \sum_{\ell=0}^{+\infty} C^\ell F^{(\ell)} (t,g_s),
\ee
\noindent
where
\be\label{f0inst}
F^{(0)}(t,g_s) = \sum_{g=0}^{+\infty} g_s^{2g-2} F^{(0)}_g (t),
\ee
\noindent
and
\be\label{fellinst}
F^{(\ell)}(t,g_s) = F^{(\ell)}_1 (t)\, \rme^{- \frac{\ell A(t)}{g_s}} \left( 1 + \sum_{n=1}^{+\infty} g_s^n\, F^{(\ell)}_{n+1} (t) \right), \qquad \ell \ge 1.
\ee
\noindent
Again, at least in principle, all the coefficients $F^{(\ell)}_{n} (t)$ may be determined recursively.

In \cite{m08} this formalism has been applied and checked against a large--order analysis in several matrix models. Remarkably, this method appears to work beyond the context of matrix models: for instance, in \cite{m08}, the full instanton series has been obtained for the case of Hurwitz theory, which is also controlled by a Toda--like equation.

\subsection{The Trans--Series Approach for $c=1$ Matrix Models}

Let us now try to apply the trans--series method to solve our $c=1$ models. As we have just learned, the first thing to do is to look for a trans--series solution to the pre--string equation. But it so happens that it is a common feature of all our $c=1$ models that the recursion relations for the coefficients $\{ r_n \}$ may be solved exactly, \textit{without} a genus expansion. In particular, for Gaussian, Penner and Chern--Simons models we find
\bea
r^{\mathrm{G}}_n = g_s n \quad &\to& \quad R^{\mathrm{G}} (x,g_s) = x, \\
r^{\mathrm{P}}_n = g_s n \left( 1+g_s n \right) \quad &\to& \quad R^{\mathrm{P}} (x,g_s) = x \left( 1+x \right), \\
r^{\mathrm{CS}}_n = q^{3n} \left( q^n-1 \right) \quad {\mathrm{with}} \quad q=\rme^{g_s} \quad &\to& \quad R^{\mathrm{CS}} (x,g_s) = \rme^{3x} \left( \rme^{x}-1 \right).
\eea
\noindent
Since we have exact solutions for the functions $R(x,g_s) \equiv R(x)$, without a genus expansion, there is clearly no asymptotics and thus no trans--series expansion for $R(x)$. Thus, for all three cases that we are considering, we only have to worry about the trans--series \textit{ans\"atz} (\ref{fellinst}) and plug into the homogeneous Toda equation
\be
F(t+g_s) - 2 F(t) + F(t-g_s) = 0.
\ee
\noindent
This is a rather interesting point, also implying that all Borel poles are controlled by the Toda equation (and not by the pre--string equations). In hindsight this is not so surprising, as at criticality all our examples are in the universality class of the $c=1$ string, and it is precisely the case that the partition function of $c=1$ string theory is a $\tau$--function of the Toda hierarchy, satisfying the Toda equation \cite{dmp92}. In some sense, the above Toda equation plays a role very analogous to the one played by the Painlev\'e I equation in $c=0$ string theory, and which was also studied in connection to the large--order behavior of topological strings in \cite{msw07, msw08}.

Let us start with the one--instanton sector. By plugging into the homogeneous Toda equation the $\ell=1$ term in (\ref{fellinst}), at first order in $g_s$ one obtains
\be
4 F^{(1)}_1 (t)\, \rme^{- \frac{A(t)}{g_s}}\, \sinh^2 \left( \frac{A'(t)}{2} \right) = 0, 
\ee
\noindent
and setting $F^{(1)}_1 (t) \not = 0$ it follows
\be
\sinh \left( \frac{A'(t)}{2} \right) = 0 \qquad \Leftrightarrow \qquad \frac{A'(t)}{2} = \rmi \pi k, \quad k \in \BZ.
\ee
\noindent
It is quite remarkable that the solution to the above equation already includes all $k$--instanton actions
\be
A_k (t) = 2\pi\rmi t\, k + \alpha(k),
\ee
\noindent
where we have also allowed the integration constant to depend on $k$. Making further use of the trans--series \textit{ans\"atz} we may compute
\be
F^{(k)}(t \pm g_s,g_s) = F^{(k)}_1 (t)\, \rme^{-\frac{A_k(t)}{g_s}} \left( 1 + \sum_{n=1}^{+\infty} g_s^n\, \CF_{(n+1)}^{(k)} \right),
\ee
\noindent
where, for example,
\bea
\CF_{(2)}^{(k)} &=& F^{(k)}_2 (t) \pm \frac{\partial_t F^{(k)}_1}{F^{(k)}_1} (t), \\
\CF_{(3)}^{(k)} &=& F^{(k)}_3 (t) \pm \frac{F^{(k)}_2 \partial_t F^{(k)}_1}{F^{(k)}_1} (t) \pm \partial_t F^{(k)}_2 (t) + \frac{1}{2} \frac{\partial_t^2 F^{(k)}_1}{F^{(k)}_1} (t).
\eea
\noindent
We now insert these expressions into the Toda equation, and solve it perturbatively in $g_s$. At second order in the string coupling it follows,
\be
\partial_t^2 F^{(k)}_1(t)=0 \quad \Rightarrow \quad F^{(k)}_1 (t) = \Phi_1(k) + \Phi_2(k)\, t.
\ee
\noindent
In fact, solving the infinite chain of differential equations one perturbatively obtains, in a recursive fashion, one is always led to second order differential equations. All the higher order terms are then fixed to be
\be
F^{(k)}_{n+1} (t) = - \frac{\Phi^{(n+1)}_1(k)}{\Phi_2(k)\, F^{(k)}_1 (t)} + \Phi^{(n+1)}_2(k), \qquad n \ge 1. 
\ee
\noindent
The $k$--instanton contribution in (\ref{fellinst}) is thus given by
\bea
F^{(k)} (t,g_s) &=& \left( \Phi_1(k) + \Phi_2(k)\, t \right) \rme^{ -\frac{2\pi\rmi t\, k + \alpha(k)}{g_s}} \times \nonumber \\
&& \times \left( 1 +  \sum_{n=1}^{+\infty} g_s^n \left( - \frac{\Phi^{(n+1)}_1(k)}{\Phi_2(k) \left( \Phi_1(k) + \Phi_2(k)\, t \right)} + \Phi^{(n+1)}_2(k) \right) \right).
\label{gene1}
\eea

It is now simple to see that the instanton series of our three models, that we have computed either via Borel or Stokes analysis, (\ref{gdisc}), (\ref{pendisc}) and (\ref{csdisc}), are all solutions to the homogeneous Toda equation\footnote{Because the Toda equation is linear, linear combinations of (\ref{gene1}) are also a solution to the problem, in particular linear combinations where $\alpha$ also depends on an integer $m$ and we sum over $m \in \BZ$ as in the conifold.}. This confirms, within the trans--series setting, the validity of our results. However, it is not equally clear how to flow in the other direction, \textit{i.e.}, how to obtain our results, (\ref{gdisc}), (\ref{pendisc}) and (\ref{csdisc}), \textit{starting} from the trans--series formalism. In particular, it is not obvious to us how to provide enough boundary conditions in order to fix, for each distinct case, the integration constants of the general solution (\ref{gene1}). A natural boundary condition is to impose matching to the $c=1$ solution, in the double--scaling limit \cite{kkk01}. But this cannot be quite enough off--criticality: in fact all our \textit{distinct} examples are satisfying the \textit{same} difference equation with the \textit{same} aforementioned boundary condition. Thus, one is forced to demand, as a final boundary condition and in order to obtain a unique solution in each case, a comparison with the large--order behavior of perturbation theory, in each distinct model we wish to study. So, at least for these models, Borel or saddle--point methods seem more powerful and efficient ways to proceed.

\subsection{A Comment on Parametric Resurgence}

In many cases it is not possible to resum the Borel transform of an asymptotic series, or even to explicitly compute this Borel transform, and thus one cannot locate the singularities in the Borel complex plane (some of which, in particular, control the large--order behavior of the perturbation theory). However, it may be the case that, even if we are not able to resum the Borel transform, we may know that the asymptotic series arises as a solution to a finite difference equation (where we are, of course, interested in the example of the Toda equation). In this case we may still obtain some interesting information, as shown in \cite{s07}. Let us quickly apply \cite{s07} to our problem.

Consider the asymptotic series
\be
F(t,g_s) \sim \sum_{n=0}^{+\infty} g_s^{2n} F_n (t),
\ee
\noindent
which we take as a perturbative solution to the finite difference equation
\be
F (t+g_s) - 2 F(t) + F(t-g_s) = g_s^2\, G(t) \qquad \Leftrightarrow \qquad 4 \sinh^2 \left( \frac{g_s}{2} \frac{\partial}{\partial t} \right) F(t) = g_s^2\, G(t).
\ee
\noindent
Then, it is possible to formally solve this differential equation \cite{s07}, with the formal solution being expressed in terms of the power--series coefficients of the function
\be
H(x) = - 1 + \frac{x^2}{4 \sinh^2 \left( \frac{x}{2} \right)} = \sum_{n=0}^{+\infty} H_n\, x^{2n+2}.
\ee
\noindent
Furthermore, the poles of the Borel transform $\CB[F] (\xi)$ may be related to the poles of the Borel transform of $H(x)$. In particular \cite{s07} this implies that in this case the Borel poles are located at $2\pi\rmi n t$, with $n \in \BZ$. It is indeed the case that the instanton actions for the three matrix models we have studied in this paper are of this kind.

\section{Conclusions and Outlook}

In this paper we have addressed the nonperturbative structure of topological strings and $c=1$ matrix models, focusing on Gaussian, Penner and Chern--Simons matrix models together with their holographic duals, $c=1$ minimal strings and topological strings on the resolved conifold. Making use of either Borel or Stokes analysis, we have uncovered the nature of instanton effects in these models, and have further explored the relation of these nonperturbative phenomena to the large--order behavior of the $1/N$ expansion. While this builds up on previous work along the same direction \cite{msw07, m08, msw08, em08}, clarifying the discussion for a big class of models, we believe there is still much work to be done. In particular, let us end by listing several issues raised in this paper which we believe deserve further and immediate investigation (in no particular order):

\begin{itemize}

\item While our results were checked in the trans--series formalism, we have seen that it is not obvious how to obtain these results for the multi--instanton series starting straight from within the trans--series set up. In particular, the choice of boundary conditions is not completely clear. It would be interesting to further explore and better understand the trans--series {\textit{ans\"atz}} in this context, possibly solving the questions we have just mentioned.

\item In the Gaussian and Penner models we have explicitly shown that instanton effects may be understood as Stokes phenomena for the logarithm of the $G_2 (z)$ Barnes function. However, we could not say much along these lines for the Chern--Simons model, as we do not know of any appropriate hyperasymptotic framework for the quantum Barnes function, $G_q (z)$. It would be very interesting to study the hyperasymptotics of $G_q (z)$ and show that Stokes phenomena in this case is also related to the instanton effects of the Chern--Simons model.

\item In our matrix model derivation of the instanton effects, in terms of eigenvalue tunneling, we have obtained the instanton action expressed in terms of the spectral curve (a cycle of the $y\, \rmd z$ one--form). It would be rather interesting to extend this calculation in order to contemplate loop corrections. Indeed, in \cite{msw07, msw08}, higher loop terms around the multi--instanton configurations were computed, in terms of matrix model open correlators. Extending that calculation to the present set up would be very interesting, since these correlators can be computed entirely in terms of the spectral curve and, as such, this formalism could be extended to other topological string scenarios where a dual matrix model description is not available. In particular, this could allow for a direct understanding---from a spectral geometry point of view---of why the loop expansion around an one/multi--instanton configuration truncates in all our examples. Furthermore, this would also provide an explanation of the instanton expansion in terms of open string amplitudes with either Liouville or toric boundary conditions, for the $c=1$ string and the resolved conifold, respectively.

\item Much of our Borel analysis was very much related to the existence of a GV integral representation for the free energies of our models. Since also on a generic CY background the topological string free energy admits a GV integral representation, one is led to wonder if our approach may be applied to other, more general cases. Recall that on a general CY threefold $\CX$, the topological string free energy is given in terms of GV integer invariants by the expansion
\be\label{sct}
F_\CX (g_s) = \sum_{r=0}^{+\infty} \sum_{d_i=1}^{+\infty} n_r^{(d_i)} (\CX) \sum_{m \in \BZ} \int_0^{+\infty} \frac{\rmd s}{s} \left( 2 \sin \frac{s}{2} \right)^{2r-2}\, \rme^{- \frac{2 \pi s}{g_s} \left( d \cdot t + \rmi m \right)}.
\ee
\noindent
When $r=0$ the zeros of the sine will be poles of the integrand and this contribution to the total sum will, as in the case of the resolved conifold which we address in the paper, yield a nonperturbative contribution to the free energy. However, for generic CY backgrounds, one will also have to consider the summation over the K\"ahler classes $\{d_i\}$, as weighted by $n_0^{(d_i)}$. In this way, one is led to write (we neglect the pole at zero)
\be\label{gdiscX}
{\rm Disc}~\widetilde{F}_{\CX} (\bar{g}_s) = - \frac{\rmi}{2\pi\bar{g}_s} \sum_{d_i=1}^{+\infty} n_0^{(d_i)} (\CX) \sum_{n=1}^{+\infty} \sum_{m \in \BZ} \left( \frac{\left( 2\pi \right)^2 \left( d \cdot t + \rmi m \right)}{n} + \frac{\bar{g}_s}{n^2} \right) \rme^{-\frac{\left( 2\pi \right)^2 n}{\bar{g}_s} \left( d \cdot t + \rmi m \right)}.
\ee
\noindent
Higher terms with $r>0$ in (\ref{sct}) have no poles in the complex plane and will only contribute to the nonperturbative corrections through the residues at infinity (recall (\ref{disc}) and our discussion in the appendix). What role they might play is beyond the scope of our analysis.

It seems likely that nonperturbative corrections obtained from the GV representation actually provide the \textit{full} nonperturbative corrections to the topological string free energy, in those cases where the number of GV invariants $n_r^{(d_i)} (\CX)$ is \textit{finite} (as in our example of the resolved conifold). Indeed, in these cases (\ref{gdiscX}) may provide the complete tower of nonperturbative corrections to the topological string free energy, as (\ref{sct}) is basically given by a finite sum, \textit{not} a power series expansion. In particular, the number of $n_0^{(d_i)}$ invariants is finite for non--singular curves of any genus \cite{bp00}, for rational curves with nodal singularities \cite{bkl99}, and for the configurations studied in \cite{klm05} (among which are the CY threefolds which are $A_k$--type ALE spaces, times $\BC$). Further notice that in (\ref{gdiscX}) the loop expansion around multi--instanton configurations truncates. In this case, and if indeed (\ref{gdiscX}) turns out to be the full answer for backgrounds with a finite number of GV invariants, then it must also be the case that these backgrounds will display no non--trivial large--order behavior in their multi--instanton sectors. Finally, observe that the $A_k$--type ALE backgrounds have also been studied in the context of the OSV conjecture \cite{osv04}, in \cite{ajs05}. The OSV conjecture \cite{osv04} relates the topological string partition function to the partition function of a configuration of branes in type II string theory, giving rise to a four--dimensional BPS black hole. The brane partition function was further suggested to provide a nonperturbative completion of topological string theory. For $A_k$--type ALE spaces, times $\BC$, it would thus be interesting to compare OSV and Schwinger completions.

\item When the number of GV invariants $n_r^{(d_i)} (\CX)$ is infinite, it seems very unlikely that the GV integral representation can still provide the full set of nonperturbative corrections to the topological string free energy. Indeed, it is now the case that $n_0^{(d)} \sim \rme^d$, for large $d$, and it seems to be the case that (\ref{gdiscX}) cannot provide the complete nonperturbative information. A particularly interesting example to further explore this issue is that of the local curve, which indeed has an infinite number of GV invariants. The good news is that this background has been extensively studied in \cite{msw07}, with instanton configurations identified and checked against the large--order behavior. As such, it would be extremely interesting to analyze the relation between the nonperturbative corrections arising due to the GV representation, with the ones derived in \cite{msw07}. In particular, the analysis of \cite{msw07, msw08} seems to indicate that, for the local curve, the loop expansion around multi--instanton configurations will \textit{not} truncate, which is to say that (\ref{gdiscX}) cannot be the full correct answer. On the other hand, and as we have already remarked, the local curve is in the universality class of 2d gravity, with $c=0$, and as such it should better be understood as a first step to understand backgrounds with an infinite number of GV invariants. A second natural step in this direction would be looking at the case of local $\BP^2$, a $c=1$ toric geometry with an infinite number of GV invariants. Indeed, the free energy of this model can be computed very efficiently to all genus, by means of direct integration of the holomorphic anomaly equations as shown in \cite{hkr08}, and would thus provide for a natural testing ground for our aforementioned questions.

\item Besides an explicit check against the large--order behavior of the theory, another way to test the Schwinger completion of topological string theory, for generic backgrounds, would be to study the modular properties of its nonperturbative free energy. In fact, it is expected that modular invariance may be recovered at the nonperturbative level \cite{em08}, an important issue also in the context of large $N$ dualities. Clearly, for the case of the resolved conifold, which we studied in this paper, there are no constraints arising from modularity since the moduli space is trivial, with the mirror 
geometry having genus zero, but this will not be the case for, \textit{e.g.}, local $\BP^2$, whose mirror geometry has a spectral curve of genus one. Notice that backgrounds with a finite number of GV invariants seem not to give rise to mirror geometries with spectral curves of genus one and, as such, modularity should become an issue precisely when (\ref{gdiscX}) ceases to be the full correct answer. In particular, modularity may play a key role in order to understand exactly what type of information is (\ref{gdiscX}) missing in general backgrounds, and these issues should be addressed in future work.

\end{itemize}

\acknowledgments
We would like to thank Jacopo Belfi, Andrea Brini, Luca Griguolo, Jos\'e Mour\~ao, Nicolas Orantin, Christoffer Petersson, Domenico Seminara, Jorge Drumond Silva, Angel Uranga, Marcel Vonk and, specially, Marcos Mari\~no, for useful discussions, comments and/or correspondence. RS would like to thank CERN TH--Division for hospitality, where a part of this work was conducted.

\newpage

\appendix

\section{The Polylogarithm: Branch Points and Monodromy}\label{ap1}

This appendix is devoted to the study of the polylogarithm, with emphasis towards its branch points and monodromy. This function is defined by
\be
\mathrm{Li}_{p} (z) = \sum_{n=1}^{+\infty} \frac{z^n}{n^p},
\ee
\noindent
and on its principal sheet it has a branch point at $z=1$ leading, by convention, to a branch cut discontinuity in the complex $z$ plane running from $1$ to infinity. As one starts ``exploring'' the multi--sheeted structure of the polylogarithm and moves off its principal sheet, one finds that there exists another branch point, at $z=0$. In this case, the resulting monodromy group will be generated by two elements, acting on the covering space of the bouquet $S^{1}\vee S^{1}$ of homotopy classes of loops in $\BC \setminus \{ 0,1 \}$, passing around the branch points $z=0$ or $z=1$. For further details, we refer the reader to the very thorough explanations that can be found in, \textit{e.g.}, \cite{v07, m03}.

For our purposes in this paper, a simple analysis in terms of explicit topological language will suffice. Let $m_1$ represent the homotopy class of all loops based at some point $z$ in $\BC$, which wind once, clockwise around the branch point at $z=1$. The action of $m_1$ on the polylogarithm has the effect of carrying this function from one sheet to the next. It was shown in \cite{v07} that one may write
\be
m_1 \cdot \mathrm{Li}_s (z) = \mathrm{Li}_s (z) - \Delta_1,
\ee
\noindent
where $\Delta_1$ is a function, whose specific form is not important at the moment, but which includes a logarithm with a branch point at $z=0$. This implies that, after acting once with $m_1$, one finds oneself on a sheet which has a branch cut discontinuity running from $0$ to minus infinity. If we now let $m_0$ represent the homotopy class of all loops based at some point $z$ in $\BC$, which wind once, clockwise around this new branch point at $z=0$, its action on the logarithm is the familiar one:
\be
m_0 \cdot \log z = \log z + 2\pi\rmi.
\ee
\noindent
Now, because the principal sheet of the polylogarithm has no branch point at $z=0$, it simply follows
\be
m_0 \cdot \mathrm{Li}_s (z) = \mathrm{Li}_s (z).
\ee
\noindent
If one now winds with $m_1$ in the opposite direction, one is led to write instead
\be
m_1^{-1} \cdot \mathrm{Li}_s (z) = \mathrm{Li}_s (z) - \Delta_{-1},
\ee
\noindent
where again $\Delta_{-1}$ is a function we shall leave unspecified; see \cite{v07} for details. If $m_1$ is to be properly considered the group--theoretic inverse of $m_1$ it better be the case that $m_1 \cdot m_1^{-1} = 1 = m_1^{-1} \cdot m_1$, when acting on $\mathrm{Li}_s (z)$. This immediately implies, \textit{e.g.}, $m_1^{-1} \Delta_1 = - \Delta_{-1}$, where we recall that $\Delta_1$ includes the standard logarithmic branch cut discontinuity starting off at $z=0$. This relation thus seems odd, as the logarithm has no branch point at $z=1$ and there should be nothing to wind around. This is a subtle point, further explained in \cite{v07}, and it should be stressed that it is the joining of polylogarithmic and logarithmic cuts that causes this effect. One way to capture the idea of there being no obstruction for the logarithm at $z=1$ is \cite{v07} to define group elements $g_1 = m_1 \cdot m_0^{-1}$ and $g_0 = m_0$, such that $g_1 \cdot \log z = \log z$ and
\bea
g_0 \cdot \mathrm{Li}_s (z) &=& \mathrm{Li}_s (z), \\
g_1 \cdot \mathrm{Li}_s (z) &=& \mathrm{Li}_s (z) - \Delta_1.
\eea
\noindent
We shall explore this monodromy group for the dilogarithm in the following.

The free combinations of powers of the two generators $g_0$ and $g_1$ form the monodromy group of the polylogarithm. If $s$ is a positive integer, this monodromy group has a finite--dimensional representation with dimension $s+1$. A particularly well known case is the dilogarithm $s=2$, also further discussed in \cite{m03}. In this case, the monodromy group is the discrete Heisenberg group \cite{v07}. In particular, one finds
\be
\Delta_n = 2\pi\rmi \left( \log z + 2\pi\rmi \left(n-1\right) \right).
\ee
\noindent
As such, repeated applications of $g_0$ and $g_1$ will only result in linear combinations of the dilogarithm $\mathrm{Li}_2 (z)$, the logarithm $\log z$, and the identity operator. Indeed, one could further take each of these three elements as a basis of a three--dimensional vector space, $\mathbf{e}_1 = 4\pi^2$, $\mathbf{e}_2 = -2\pi\rmi\, \log z$ and $\mathbf{e}_3 = \mathrm{Li}_2 (z)$, in which case the matrix representation of the monodromy group would become
\be
g_1 = \left[
\begin{array}{ccc}
1 & 0 & 0\\
0 & 1 & 1\\
0 & 0 & 1
\end{array}
\right] \qquad \mathrm{and} \qquad g_0 = \left[
\begin{array}{ccc}
1 & 1 & 0\\
0 & 1 & 0\\
0 & 0 & 1
\end{array}
\right].
\ee
\noindent
These two matrices are in fact the generators of the discrete Heisenberg group $\CH_3 (\BZ)$, see \cite{v07} for full details on this discussion.

In this paper we are interested in the following action (see \cite{v07} for any missing details)
\bea
g_0^k \cdot g_1^n \cdot \mathrm{Li}_2 (z) &=& g_0^k \cdot \left( \mathrm{Li}_2 (z) - n\, \Delta_1 \right) = \nonumber \\
&=& \mathrm{Li}_2(z) - n\, \Delta_{k+1} = \mathrm{Li}_2 (z) - 2\pi\rmi\, n\, \log z + 4 \pi^2\, k\, n,
\eea
\noindent
and we make use of this result in the main body of the paper.

\section{Dispersion Relation for Topological Strings}

In this appendix we address the Cauchy dispersion relation (\ref{disc}) for the case of general topological string theories on a CY threefold $\CX$, whose free energy is given in terms of GV integer invariants by the expansion (\ref{GVexpansion}), and which we recall in here as
\be
F_\CX (g_s) = \sum_{r=0}^{+\infty}\, \sum_{d_i=1}^{+\infty} n_r^{(d_i)} (\CX) \sum_{n=1}^{+\infty} \frac{1}{n} \left( 2 \sin \frac{ng_s}{2} \right)^{2r-2}\, \rme^{- 2 \pi n\, d \cdot t}.
\ee
\noindent
In particular, we wish to evaluate
\be\label{contributionatinfinity}
\oint_{(\infty)} \frac{\rmd w}{2\pi\rmi}\, \frac{F_\CX (w)}{w-g},
\ee
\noindent
and show that this vanishes in the case of the resolved conifold, a result we have used in the main body of the paper. Notice that one cannot compute the pole at infinity straight: infinity is an essential singularity of the integrand in the GV representation and, as such, residue calculus does not apply. Nonetheless, for $r=0$ one may decompose the contour $\mathcal{C}_\infty$ into the sum of two contours in upper and lower hemispheres, $\mathcal{C}_+$ and $\mathcal{C}_-$, respectively, 
\be
\oint_{\mathcal{C}_\infty} \frac{\rmd w}{2\pi\rmi}\, \frac{1}{w-g} \frac{1}{\left( 2 \sin \frac{n w}{2} \right)^{2} } = \int_{\mathcal{C}_+} \frac{\rmd w}{2\pi\rmi}\, \frac{\rme^{\rmi n w}}{\left( w-g \right) \left( 1 - \rme^{\rmi n w} \right)^2} + \int_{\mathcal{C}_-} \frac{\rmd w}{2\pi\rmi}\, \frac{\rme^{- \rmi n w}}{\left( w-g \right) \left( 1 - \rme^{- \rmi n w} \right)^2},
\ee
\noindent
and then use Jordan's lemma---applicable as
\be
\lim_{R \to + \infty} \max_{\theta \in [0,\pi]} \left| \frac{1}{\left( R \rme^{\rmi \theta} - g \right) \left( 1 - \rme^{\rmi n R \rme^{\rmi \theta}} \right)^2} \right| = 0
\ee
\noindent
in the upper hemisphere, and analogously in the lower---in order to find that this implies that the integral vanishes at infinity, and it thus follows that for the resolved conifold, where only $n_0^{(1)}=1$ is non--zero, the (\ref{contributionatinfinity}) contribution indeed vanishes. For more complicated CY threefolds with $r \ge 1$ it seems rather likely that there will be a Cauchy contribution at infinity, and a complete analysis of this situation is beyond the scope of the present work.

\newpage


\bibliographystyle{plain}

\end{document}